\documentclass[twocolumn, 12pt]{aastex63}
\usepackage{float}
\usepackage{hyperref}
\usepackage{amsmath}
\received{June 6, 2020}
\revised{Oct 26, 2020}
\accepted{Nov 13, 2020}
\submitjournal{APJL}

\newcommand{\ceo}{C$^{18}$O}
\newcommand{\tco}{$^{13}$CO}
\newcommand{\htco}{H$_{2}$CO}
\newcommand{\htcn}{H$^{13}$CN}

\newcommand{\cso}{C$^{17}$O}
\newcommand{\htcop}{H$^{13}$CO$^+$}
\newcommand{\co}{$^{12}$CO}
\newcommand{\sio}{SiO}
\newcommand{\sot}{SO$_{2}$}
\newcommand{\lsot}{SO$_{2}$~(J~=~13$_{2,12}\rightarrow$12$_{1,11}$)}
\newcommand{\lco}{\co~(J~=~3$\rightarrow$2)}  
\newcommand{\lhtcn}{\htcn~(J~=~4$\rightarrow$3)}  
\newcommand{\lcso}{\cso~(J~=~3$\rightarrow$2)}  
\newcommand{\lhtcop}{\htcop~(J~=~4$\rightarrow$3)}  
\newcommand{\lsio}{\sio~(J~=~7$\rightarrow$6)}  
\renewcommand{\arcsec}{$^{\prime\prime}$}
\renewcommand{\arcmin}{$^{\prime}$}
\renewcommand{\deg}{\degr}
\renewcommand{\micron}{$\mu$m}
\newcommand{\pdspy}{\textit{pdspy}}

\newcommand{\ab}{$\sim$}
\newcommand{\kms}{km~s$^{-1}$}

\newcommand{\tbol}{T$_{bol}$}
\newcommand{\rsun}{R$_{\odot}$}
\newcommand{\mstar}{M$_{*}$}
\newcommand{\solm}{M$_{\odot}$}
\newcommand{\msun}{M$_{\odot}$}
\newcommand{\lsun}{L$_{\odot}$}
\newcommand{\mdot}{$\dot{M}$}

\newcommand{\csobeam}{0\farcs21$\times$0\farcs13}
\newcommand{\siobeam}{0\farcs85$\times$0\farcs52}
\newcommand{\cobeam}{0\farcs19$\times$0\farcs11}
\newcommand{\htcopbeam}{0\farcs85$\times$0\farcs52}
\newcommand{\htcnbeam}{0\farcs22$\times$0\farcs14}
\newcommand{\contbeam}{0\farcs11$\times$0\farcs05}
\newcommand{\tttfbeam}{0\farcs21$\times$0\farcs13}

\begin{document}
\nocite{*}
\title{Kinematic Analysis of a Protostellar Multiple System: Measuring the Protostar Masses and Assessing Gravitational Instability in the Disks of L1448 IRS3B and L1448 IRS3A}
\shorttitle{L1448 IRS3B}
\shortauthors{Reynolds et al.}

\author{Nickalas K. Reynolds}
\affiliation{Homer L. Dodge Department of Physics and Astronomy, University of Oklahoma, 440 W. Brooks Street, Norman, OK 73019, USA}
\author{John J. Tobin}
\affiliation{National Radio Astronomy Observatory, 520 Edgemont Rd. Charlottesville, VA 22901}
\author{Patrick Sheehan}
\affiliation{National Radio Astronomy Observatory, 520 Edgemont Rd. Charlottesville, VA 22901}
\affiliation{Department of Physics and Astronomy, Northwestern University, 2145 Sheridan Road, Evanston, IL 60208, USA}
\author{Sarah I. Sadavoy}
\affiliation{Department of Physics, Engineering Physics \& Astronomy, Queen's University, Kingston, Ontario, Canada}
\author{Kaitlin M. Kratter}
\affiliation{University of Arizona, Steward Observatory, Tucson, AZ 85721}
\author{Zhi-Yun Li}
\affiliation{Department of Astronomy, University of Virginia, Charlottesville, VA 22903}
\author{Claire J. Chandler}
\affiliation{National Radio Astronomy Observatory, P.O. Box O, Socorro, NM 87801}
\author{Dominique Segura-Cox}
\affiliation{Department of Astronomy, University of Illinois, Urbana, IL 61801}
\author{Leslie W. Looney}
\affiliation{Department of Astronomy, University of Illinois, Urbana, IL 61801}
\author{Michael M. Dunham}
\affiliation{Department of Physics, State University of New York Fredonia, Fredonia, New York 14063, USA}
\affiliation{Harvard-Smithsonian Center for Astrophysics, 60 Garden St, MS 78, Cambridge, MA 02138}

\begin{abstract}
We present new Atacama Large Millimeter/submillimeter Array (ALMA) observations towards a compact (\ab230~au separation) triple protostar system, L1448 IRS3B, at 879~\micron\space with \contbeam~resolution. Spiral arm structure within the circum-multiple disk is well resolved in dust continuum toward IRS3B, and we detect the known wide (\ab2300~au) companion, IRS3A, also resolving possible spiral substructure. Using dense gas tracers, \lcso, \lhtcop, and \lhtcn, we resolve the Keplerian rotation for both the circum-triple disk in IRS3B and the disk around IRS3A. Furthermore, we use the molecular line kinematic data and radiative transfer modeling of the molecular line emission to confirm that the disks are in Keplerian rotation with fitted masses of \added{$1.19^{+0.13}_{-0.07}$\solm\space for IRS3B-ab, $1.51^{+0.06}_{-0.07}$~\solm\space for IRS3A}, and place an upper limit on the central protostar mass for the tertiary IRS3B-c of 0.2~\solm. We measure the mass of the fragmenting disk of IRS3B to be \ab0.29~\solm\space from the dust continuum emission of the circum-multiple disk and estimate the mass of the clump surrounding IRS3B-c to be 0.07~\solm. We also find that the disk around IRS3A has a mass of \ab0.04~\solm. By analyzing the Toomre~Q parameter, we find the IRS3A circumstellar disk is gravitationally stable (Q$>$5), while the IRS3B disk is consistent with a gravitationally unstable disk (Q$<$1) between the radii \ab200-500~au. This coincides with the location of the spiral arms and the tertiary companion IRS3B-c, supporting the hypothesis that IRS3B-c was formed in situ via fragmentation of a gravitationally unstable disk.
\end{abstract}

\section{Introduction}\label{sec:intro}
Star formation takes place in dense cores within molecular clouds \citep{1987ARAA..25...23S}, that are generally found within filamentary structures \citep{2014prpl.conf...27A}. The Perseus Molecular Cloud, in particular, hosts a plethora of young stellar objects \citep[YSOs;][]{2014ApJ...787L..18S,2009ApJ...692..973E} and is nearby \citep[d\ab288$\pm$22~pc; e.g.,][]{2018arXiv180803499O, 2019ApJ...879..125Z}, making its protostellar population ideal for high-spatial resolution studies. By observing these YSOs during the early stages of star formation, we can learn about how cores collapse and evolve into protostellar and/or proto-multiple systems, and how their disks may form into proto-planetary systems.

Protostellar systems have been classified into several groups following an evolutionary sequence: Class 0, the youngest and most embedded objects characterized by low L$_{bol}$/L$_{submm}$ \citep[$<5\times10^{-3}$; ][]{1993ApJ...406..122A} and \tbol\space$\le$70~K, Class I sources which are still enshrouded by an envelope that is less dense than the Class 0 envelope, with T$_{bol}<=650$~K, Flat Spectrum sources, which are a  transition phase between Class I and Class II, and Class II objects, which have shed their envelope and consist of a pre-main sequence star (pre-MS) and a protoplanetary disk. Most stellar mass build-up is expected to occur during the Class 0 and Class I phases \citep[$<5\times10^{5}$~yr; e.g.][]{2018arXiv180711262K,1987IAUS..115....1L}, because by the time the system has evolved to the Class II stage, most of the mass of the envelope has been either accreted onto the disk/protostar or blown away by outflows \citep[][]{2006ApJ...646.1070A, 2014ApJ...784...61O}.

Studies of multiplicity in field stars have observed multiplicity fractions of 63\% for nearby stars \citep[][]{1962AJ.....67R.590W}, 44-72\% for Sun-like stars \citep[][]{1983ARAA..21..343A, 2010ApJS..190....1R}, 50\% for F-G type nearby stars \citep[][]{1991AA...248..485D}, 84\% for A-type stars \citep[][]{2017ApJS..230...15M}, and 60\% for pre-MS stars \citep[][]{1994ARAA..32..465M}. These studies demonstrate the high frequency of stellar multiples and motivates the need for further multiplicity surveys toward young stars to understand their formation mechanisms.

Current theories suggest four favored pathways for forming multiple systems: turbulent fragmentation \citep[on scales \ab1000s of au; e.g.][]{2004ApJ...617..559P, 2004ApJ...600..769F}, thermal fragmentation \citep[on scales \ab1000s of au; e.g.][]{2010ApJ...725.1485O, 2013ApJ...764..136B}, gravitational instabilities within disks \citep[on scales \ab100s of au; e.g.][]{1989ApJ...347..959A, 2009MNRAS.392..413S, 2010ApJ...708.1585K}, and/or loose dynamical capture of cores \citep[\ab10$^{4-5}$~au scales][]{2002MNRAS.336..705B, 2019ApJ...887..232L}. Additionally, stellar multiples may evolve via multi-body dynamical interactions which can alter their hierarchies early in the star formation process \citep{2002MNRAS.336..705B, 2010MNRAS.404..721M, 2012Natur.492..221R}. In order to fully understand star formation and multiple-star formation, it is important to target the youngest systems to characterize the initial conditions.

The VLA Nascent Disk and Multiplicity (VANDAM) survey \citep{2016ApJ...818...73T} targeted all known protostars down to 20~au scales within the Perseus Molecular Cloud using the Karl G. Jansky Very Large Array (VLA) to better characterize protostellar multiplicity. They found the multiplicity fraction (MF) of Class 0 protostars to be \ab57\% (15-10,000~au scales) and \ab28\% for close companions (15-1,000~au scales), while, for Class I protostars, the MF for companions (15-10,000~au scales) is 23\% and 27\% for close companions (15-1,000~au scales). This empirical distinction in MF motivates the need to observe Class 0 protostars to resolve the dynamics before the systems evolve. It was during this survey that the multiplicity of L1448 IRS3B, a compact (\ab230~au) triple system, was discovered. \citet{2016Natur.538..483T} observed this source at 1.3~mm, resolving spiral arms, kinematic rotation signatures in \ceo, \tco, and \htco, with strong outflows originating from the IRS3B system. 

L1448 IRS3B has a hierarchical configuration, which features an inner binary (separation 0\farcs25$\approx$75~au, denoted -a and -b, respectively) and an embedded tertiary (separation 0\farcs8$\approx$230~au, denoted -c). The IRS3B-c source is deeply embedded within a clump positioned within the IRS3B disk, thus we reference the still forming protostar as IRS3B-c and the observed compact emission as a ``clump'' around IRS3B-c. \citet{2016ApJ...818...73T}\space found evidence for Keplerian rotation around the disks of IRS3B and IRS3A. They also found that the circum-triple disk was likely gravitationally unstable.

\added{Theory suggests that during stellar mass assembly via disk accretion, fragmentation via gravitational instability (hereafter GI) may occur if the disk is sufficiently massive, cold, and rapidly accreting\citep{1989ApJ...347..959A, 1999ApJ...525..330Y, 2010ApJ...710.1375K}. Due to the scales of fragmentation, and on-going infall, fragments likely turn into stellar or brown dwarf mass companions, and GI is a favored pathway for the formation of compact multi-systems ($\lesssim$100~au). Since observations show that the youngest systems, like L1448 IRS3B, have higher disk masses than their more evolved counterparts \citep{2020ApJ...890..130T}, we would also expect observational signatures of disk instability and fragmentation to be most prevalent in the Class 0 stage.}

\deleted{Theory suggests that during stellar mass assembly, when the disk is forming via conservation of angular momentum \citep{1976ApJ...210..377U, 1984ApJ...286..529T, 2014prpl.conf..173L}, gravitational fragmentation should occur if the disk is massive \citep{1989ApJ...347..959A, 1999ApJ...525..330Y, 2010ApJ...710.1375K}. These fragmentation events, if large enough, can form stellar multiples and is a possible pathway for the formation of compact (\ab100s of au scales) multiple systems. Since observations show that the youngest systems, like L1448 IRS3B, have higher disk masses than their more evolved counterparts \citep{2020ApJ...890..130T}, we would also expect disk fragmentation to be the most common mechanism in the Class 0 stage.}

The wide and compact proto-multiple configurations of IRS3A and IRS3B contained within a single system provides a test bed for multiple star formation pathways to determine which theories best describe this system. Here we detail \added{our follow-up observations to \citet{2016Natur.538..483T} of L1448 IRS3B with the Atacama Large Millimeter/submillimeter Array (ALMA) in Band 7, with $2\times$~higher resolution and $6\times$~higher sensitivity. We resolve the kinematics toward both IRS3B and IRS3A with much higher fidelity that the previous observations, enabling us to characterize the nature of the rotation in the disks, measure the protostar masses, and characterize the stability of both disks.} We show our observations of this system and describe the data reduction techniques in Section~\ref{sec:obs}, we discuss our empirical results and our use of molecular lines in Section~\ref{sec:results}, we \deleted{then refine}\added{ further analyze} the molecular line \deleted{s used to trace disk kinematics}\added{kinematics} in Section~\ref{sec:keprotation}, we further detail our models and the results in Section~\ref{sec:kmodelresults}, and we interpret our findings in Section~\ref{sec:discussion}, where we discuss the implications of our empirical and model results and future endeavors. 

\section{Observations}\label{sec:obs}
\explain{Bulk of observation explanations was moved to appendix. Particularly the content surrounding self calibration.}
We observed L1448 IRS3B with ALMA in Band 7 (879~\micron) during Cycle 4 in two configurations, an extended (C40-6) and a compact (C40-3) configuration in order to fully recover the total flux out to \ab5\arcsec\space angular scales in addition to resolving the structure in the disk. C40-6, was used on 2016 October 1~and~4 with 45 antennas. The baselines ranged from 15 to 3200~meters, for a total of 4495 seconds on source (8052 seconds total) for both executions. C40-3, was used on 19 December 2016 with 41 antennas. The baselines covered 15 to 490 meters for a total of 1335 seconds on source (3098 seconds total).

The complex gain calibrator was J0336$+$3218, the bandpass calibrator was J0237$+$2848, and the flux calibrator was the monitored quasar J0238$+$1636. The observations were centered on IRS3B. IRS3A, the wide companion, is detected further out in the primary beam with a beam efficiency \ab60\%).We summarize the observations in Tables~\ref{table:obssummary1}~and~\ref{table:obssummary2} and further detail our observations and reductions in Appendix~\ref{sec:appobs}.

It should also be noted there is possible line blending of \lhtcn\space and \lsot\space\citep{1997Icar..130..355L} (Table~\ref{table:obssummary2}). The \sot\space line has an Einstein A coefficient of 2.4$\times10^{-4}$~s$^{-1}$ with an upper level energy of 93~K, demonstrating the transition line strength could be strong. \sot\space provides another shock tracer which could be present toward the protostars. We label \htcn\space and \sot\space together for the rest of this analysis to emphasize the possible line blending of these molecular lines. Additionally, the \co\space and \sio\space emission primarily trace outflowing material and analysis of these data is beyond the scope of this paper, but the integrated intensity maps of select velocity ranges are shown in \deleted{in} Appendix~\ref{sec:coemission}~and~\ref{sec:sioemission}. The results of this analysis are summarized for each of the sources in Table~\ref{table:obssummary3}.

\section{Results}\label{sec:results}
\subsection{879~\micron~Dust Continuum}\label{sec:dcont}
The observations contain the known wide-binary system L1448 IRS3A and L1448 IRS3B and strongly detect continuum disks towards each protostellar system (Figures~\ref{fig:contimage}~and~\ref{fig:zoomincont}). We resolve the extended disk surrounding IRS3A (Briggs robust weight $=0.5$: Figure~\ref{fig:contimage}, superuniform: Figure~\ref{fig:widesuperuniform})\deleted{ and marginally resolve possible spiral arm substructure}. 

\subsubsection{IRS3B}
We resolve the extended circum-multiple disk of IRS3B and the spiral arm structure that extends asymmetrically to \ab600~au North-South  \added{in diameter}. Figure~\ref{fig:zoomincont}\space shows a zoom in on the IRS3B circumstellar disk, \deleted{showing}\added{exhibiting} clear substructure\deleted{s}. Furthermore, we observe the three distinct continuum sources within the disk of IRS3B as identified by \citet{2016Natur.538..483T}, but with our superior resolution and sensitivity (\ab2$\times$\space higher), our observations are able to marginally resolve smaller-scale detail closer to the inner pair of sources, IRS3B-a and -b (Figure~\ref{fig:zoomincont}). We now constrain the origin point of the two spiral arm structures. Looking towards IRS3B-ab, we notice a decline in the disk continuum surface brightness in the inner region, north-eastward of IRS3B-ab. We also observe a ``clump'' \ab50~au East of IRS3B-b. However, given that this feature is located with apparent symmetry to IRS3B-a, it is possible that the two features (``clump'' and IRS3B-a) are a part of an inner disk structure as there appears a slight deficit of emission located between them (``deficit''), while IRS3B-b is just outside of the inner region. \deleted{While this so-called ``clump'' could just be a surface brightness variation in the spiral substructure, the source was observed in \citet{2016ApJ...818...73T}\space at longer wavelengths.}

\subsubsection{IRS3B-ab}
To best determine the position angle and inclination of the circum-multiple disk, we first have to remove the tertiary source that is embedded within the disk using the \textit{imfit} task in CASA by fitting two 2-D Gaussians with a constant emission offset (detailed fully in Appendix~\ref{sec:tertsub}). We fit the semi-major and semi-minor axis of the IRS3B-ab disk with a 2-D Gaussian using the task \textit{imfit} in CASA. To fit the general shape of the disk and not fit the shape of the spiral arms, we smooth the underlying disk structure (taper the uv visibilities at 500~k$\lambda$ during deconvolution using the CASA \textit{clean} task), yielding more appropriate image for single 2D Gaussian fitting. 

From this fit, we recovered the disk size, inclination, and position angle, which are summarized in Table~\ref{table:obssummary3}. The protostellar disk of IRS3B has a deconvolved major axis and minor axis FWHM of \added{1\farcs73$\pm$0\farcs05\space and 1\farcs22$\pm$0\farcs04\space} (497$\pm$17~au\space $\times$\space 351$\pm$12~au), respectively\deleted{, corresponding to an average continuum disk radius of \ab248.5~au (assuming the disk is symmetric)}. This corresponds to an inclination angle of 45.0\deg$^{+2.2}_{-2.2}$\space assuming the disk is symmetric and geometrically thin, where an inclination angle of 0\deg\space corresponds to a face-on disk. We estimate the inclination angle uncertainty to be as much as 25\% \space by considering the south-east side of the disk as asymmetric and more extended. The position angle of the disk corresponds to 28$\pm$4\deg\space East-of-North. 

\subsubsection{IRS3B-c}
In the process of removing the clump around the tertiary companion IRS3B-c, we construct a model image of this clump that can be analyzed through the same methods. We recover a deconvolved major axis and minor axis FWHM of 0\farcs28$\pm$0\farcs05\space and 0\farcs25$\pm$0\farcs04\space (80$\pm$17~au\space $\times$\space 71$\pm$12~au), respectively, corresponding to a radius \added{\ab40~au (assuming the disk is symmetric)}. This corresponds to an inclination angle of 27.0\deg$^{+19}_{-19}$\space and we fit a position angle of 21$\pm$1\deg\space East-of-North. We note the inclination estimates for IRS3B-c may not be realistic since the internal structure of the source (oblate, spherical, etc.) cannot be constrained from these observations, thus the reported angles are assuming a flat, circular internal structure, similar to a disk.

\subsubsection{IRS3A}
The protostellar disk of IRS3A has a FWHM radius of \ab100~au and has a deconvolved major axis and minor axis of \added{0\farcs69$^{+0.01}_{-0.01}$\space and 0\farcs25$^{+0.1}_{-0.1}$}\space(197$\pm$3~au $\times$\space 72$\pm$3~au), respectively. This corresponds to an inclination angle of 68.6\added{$\pm$1.2}\deg\space assuming the disk is axially symmetric. The position angle of the disk corresponds to 133$\pm$1\deg\space East-of-North. \added{We marginally resolve two emission deficits one beamwidth off IRS3A, along the major axis of the disk. The potential spirals appear to originate along the minor axis of the disk; however, due to the reconstructed beam elongation along the minor axis of the disk, we cannot fully resolve the substructure of the disk around IRS3A}\added{, limiting the characterization that we can perform on it}.

\subsection{Disk Masses}\label{sec:diskmass}
The traditional way to estimate the disk mass is via the dust component which dominates the disk continuum emission at millimeter wavelengths. If we make the assumption that the disk is isothermal, optically thin, \added{without scattering}, and the dust and gas are well mixed, then we can derive the disk mass from the equation:

\begin{equation}\label{eq:dustmasseq}
    M_{dust} = \frac{D^2 F_{\lambda}}{\kappa_{\lambda}B_{\lambda}(T_{dust})}
\end{equation}
where $D$ is the distance to the region (288~pc), $F_{\lambda}$\space is the flux density, $\kappa_{\lambda}$\space is the dust opacity, $B_{\lambda}$\space is the Planck function for a dust temperature, and $T_{dust}$\space is taken to be the average temperature of a typical protostar disk. The $\kappa_{\lambda}$\space at $\lambda$ = 1.3~mm was adopted from dust opacity models with value of 0.899~cm$^2$~g$^{-1}$, typical of dense cores with thin icy-mantles \citep{1994AA...291..943O}. We then appropriately scale the opacity:

\begin{equation}
    \kappa_{0.879 mm} = \kappa_{1.3 mm}\times\left(\frac{1.3 mm}{0.879 mm}\right)^{\beta}
\end{equation}
assuming $\beta$=1.78. We note that $\beta$\space values typical for protostars range from 1-1.8 \citep{2009ApJ...696..841K, 2013PhDT.......434S}\deleted{, we adopt the upper bound for the opacity which, in turn, represents a lower bound for the derived mass. We also note that the choice of $\kappa$\space significantly affects the derived mass}. \deleted{We adopt a $\kappa_{1.3mm}$\space value more typical of dense cores (0.899~cm$^{2}$~g$^{-1}$). However,} If we assume significant grain growth has occurred, typical of more evolved protoplanetary disks like that of \citet{2009ApJ...700.1502A}, we would then adopt a $\kappa_{0.899\mu m}\approx3.5$~cm$^{2}$~g$^{-1}$\space and $\beta$=1, which would lower our reported masses by a factor of 2.

\added{The assumed luminosities} of the sources are  13.0~\lsun\space and 14.4~\lsun\space for IRS3B and IRS3A at a distance of 300~pc, respectively \citep[8.3~\lsun\space and 9.2~\lsun\space for IRS3B and IRS3A, respectively at 230~pc; ][]{2016ApJ...818...73T}. We note \deleted{found}\added{that} in the literature there are several luminosity values for IRS3B, differing from our adopted value by a factor of a few. Reconciling this is outside of the scope of this paper, but the difference could arise from source confusion in the crowded field and differences in SED modeling.

We adopt a $T_{dust}\approx\added{40}~K$~for the IRS3B \deleted{source}\added{disk dust temperatures} from the equation $T_{dust}=30~K\times\left(L_{*} / L_{\odot}\right)^{1/4}$, which is comparable to temperatures derived from protostellar models \citep[43~K:][]{2013ApJ...771...48T} and larger than temperatures assumed for the more evolved protoplanetary disks \citep[25~K:][]{2013ApJ...771..129A}. The compact clump around IRS3B-c has a peak brightness temperature of 55~K. Thus we adopt a T$_{dust}$ = 55~K since the emission may be optically thick ($T_{dust}$\ab $T_{B}$). We determine the peak brightness temperature of this clump by first converting the dust continuum image from Jy into K via the Rayleigh-Jean's Law\footnote{\citep[T = 1.222$\times10^3\frac{I~mJy~beam^{-1}}{(\nu~GHz)^2(\theta_{major}~arcsec)(\theta_{minor}~arcsec)}$~K, ][]{2009tra..book.....W}}. \deleted{Similarly, }We adopt a $T_{dust}$ = 51~K~ for the IRS3A source.

If we assume the canonical ISM gas-to-dust mass ratio of 100:1 \citep{1978ApJ...224..132B}, we estimate the total mass of the IRS3B-ab disk (IRS3B-c subtracted) to be 0.29~\solm\space for $\kappa_{0.879~mm}=$1.80~cm$^2$~g$^{-1}$, $T_{dust}\approx\added{40}~K$~\citep{2019ApJ...886....6T}, and $F_{\lambda}\approx1.51~Jy$. We note that the dust to gas ratio \deleted{may} is expected to decrease as disks evolved from Class 0 to Class II \citep{2014ApJ...788...59W}, but for such a young disk, we expect it to still be gas rich and therefore have a gas to dust ratio more comparable with the ISM. We estimate 0.07~\solm\space to be associated with the circumstellar dust around IRS3B-c, from this analysis, for a T$_{dust}$ = 55~K. We perform the same analysis towards IRS3A and arrive at a disk mass estimate of 0.04~\solm, for a $T_{dust}$ = \added{51}~K and $F_{\lambda}\approx0.19~Jy$.

The dust around the tertiary source, IRS3B-c, is compact and it is the highest peak intensity source in the system, and thus the optical depth needs to be constrained. An optically thick disk will be more massive than what we calculate while an optically thin disk will be more closely aligned with our estimates. We calculate the average deprojected, cumulative surface density from the mass and radius provided in Table~\ref{table:obssummary3}, and determine the optical depth via \deleted{$\tau_{0.879 mm}=\kappa_{0.879 mm}\Sigma=\frac{D^2 F_{\lambda}}{\pi R_{disk}^{2}B_{\lambda}(T_{dust})}$\citep{2016Natur.538..483T}.}\explain{Turned into a proper equation for better readability.}
\begin{align*}
\tau_{0.879~mm} &= \kappa_{0.879 mm}\Sigma \\
 &=\frac{D^2 F_{\lambda}}{\pi R_{disk}^{2}B_{\lambda}(T_{dust})}
\end{align*}
from \citep{2016Natur.538..483T}. The dust surrounding the tertiary source has an average dust surface density ($\Sigma$) of \ab2.6~g~cm$^{-2}$\space and an optical depth ($\tau$) of \ab2.14, indicative of being optically thick, while IRS3B-ab (IRS3B-c clump subtracted) is not optically thick if we assume dust is equally distributed throughout the disk with an average dust surface density of \ab0.17~g~cm$^{-2}$\space and an optical depth of 0.34. However, since spiral structure is present, these regions of concentrated dust particles are likely much more dense. L1448 IRS3A has an average dust surface density of 0.32~g~cm$^{-2}$\space and an optical depth of 0.57. Optically thick emission indicates that our dust continuum mass estimates are likely lower limits for the mass enclosed in the clump surrounding IRS3B-c, while the IRS3B-ab circum-multiple disk and the IRS3A circumstellar disk are probably optically thin except for the inner regions. 

\added{
An effect that could impact our measurements of disk masses and surface densities is scattering. Scattering reduces the emission of optically thick regions of the disk to appear optically thin, thus underestimating the optical depth. \citet{2019ApJ...877L..18Z}, showed that in the lower limit of extended ($>100$~au) disks, this effect underestimates the disk masses by a factor of 2. However, towards the inner regions, this effect might be enhanced to factors $>10$. \citet{2020ApJ...892..136S} show that for wavelengths \ab870~\micron\space and 100~\micron \space size particles, only a $\Sigma\approx3.2$~(g~cm$^{-2}$) is needed for the particles to be optically thick. Thus our masses could be several factors higher.
}

\subsection{Molecular Line Kinematics}\label{sec:kinematics}
Additionally, we observe a number of molecular lines (\co, \sio, \htcop, \htcn/\sot, \cso) towards IRS3B and IRS3A to resolve outflows, envelope, and disk kinematics, with the goal of disentangling the dynamics of the systems. We summarize the observations of each of the molecules below and provide a more rigorous analysis towards molecules tracing disk kinematics. While outflows are important for the evolution and characterization of YSOs, the analysis of these complex structures is beyond the scope of this paper because we are focused on the disk and envelope. We find \co\space and \sio\space emission primarily traces outflows, \htcop\space emission traces the inner envelope, \htcn/\sot\space emission traces energetic gas which can take the form of outflow launch locations or inner disk rotations, and \cso\space primarily traces the disk. Non-disk/envelope tracing molecular lines (\co\space and \sio) are discussed in Appendix~\ref{sec:outflow}.

We construct moment 0 maps, which integrate the data cube over the frequency axis, to reduce the 3D nature of datacubes to 2D images. These images show spatial locations of strong emission and deficits. To help preserve some frequency information from the datacubes, we integrated at specified velocities to separate the various kinematics in these systems. However, when integrating over any velocity ranges, we do not preserve the full velocity information of the emission, thus we provide \deleted{a spectral profile of \cso\space emission towards IRS3B-c which exhibits absorption and appears as deficits in the moment maps (see Appendix~\ref{sec:spectra}).}\added{provide spectral profiles of \cso\space emission toward the IRS3B-ab, IRS3B-c, and IRS3A sources in Appendix~\ref{sec:spectra}.}

\subsubsection{\cso\space Line Emission}\label{sec:csoemission}
The \cso\space emission (Figure~\ref{fig:irs3bc17omoment}, \ref{fig:irs3ac17omoment}, and~\ref{fig:irs3abc17omoment1}) appears to trace the gas kinematics within the circumstellar disks because the emission is \added{largely confined to the scales}\deleted{on the scale} of the continuum disks for both IRS3B and IRS3A, appears orthogonal to the outflows, and has a well-ordered data cube indicative of rotation (\deleted{see spectral profiles in Appendix~\ref{sec:spectra} and the moment 1 map in }Figure~\ref{fig:irs3abc17omoment1}). \cso\space is a less abundant molecule \citep[ISM $\lbrack$\co$\rbrack/\lbrack$\cso$\rbrack\approx$1700:1; e.g.][]{1994ARAA..32..191W} isotopologue of \co\space\citep[ISM $\lbrack H_{2}\rbrack/\lbrack$\co$\rbrack\approx$10$^{4}$:1; e.g.][]{2009AA...503..323V}, and thus traces gas closer to the disk midplane. Towards IRS3B, the emission extends out to \ab1\farcs8~(\ab530~au), further than the continuum disk (\ab500~au) and has a velocity gradient indicative of Keplerian rotation. Towards IRS3A, the emission is much fainter, however, from the moment 0 maps, \cso\space still appears to trace the same region as the continuum disk.

\subsubsection{\htcop\space Line Emission}\label{sec:htcopemission}
The \htcop\space emission (Figure~\ref{fig:h13copmomentc17o}~and~\ref{fig:irs3ah13copmoment}) detected within these observations probe large scale structures ($>$5\arcsec), much larger than the size of the continuum disk of IRS3B and scales \ab1\farcs5 towards IRS3A. For IRS3B, the emission structure is fairly complicated with multiple emission peaks near line center and emission deficits near the sources IRS3B-ab$+$c, while appearing faint towards IRS3A. The data cube appears kinematically well ordered, indicating possible rotating structures. Previous studies suggested HCO$+$\space observations are less sensitive to the outer envelope structure, probing densities $\ge10^{5}$~cm$^{-3}$\space and temperatures $>25$~K\space \citep{1999ARAA..37..311E}. However, follow up surveys \citep{2009AA...507..861J} found this molecule to primarily trace the outer-circumstellar disk and inner envelope kinematics, and were unable to observe the disks of Class 0 protostars from these observations alone. \citet{2009AA...507..861J}\space postulated that in order to disentangle dynamical structures on $<$100~au scales, a less abundant or more optically thin tracer (like that of \htcop) would be required with high resolutions. However, this molecular line, as shown in the integrated intensity map of \htcop\space (Figure~\ref{fig:h13copmomentc17o}~and~\ref{fig:irs3ah13copmoment}) traces scales much larger than the continuum or gaseous disk of IRS3B and IRS3A and thus is likely tracing the inner envelope.

\subsubsection{\htcn\space Line Emission}\label{sec:htcnemission}
The \htcn/\sot\space emission (Figures~\ref{fig:irs3bh13cnmoment}~and~\ref{fig:irs3ah13cnmoment}) is a blended molecular line, with a separation of \deleted{1\kms}\added{1~\kms}\space(Table~\ref{table:obssummary2}). The integrated intensity maps towards IRS3B appear to trace an apparent outflow launch location from the IRS3B-c protostar (Figure~\ref{fig:irs3bh13cnmoment}) based on the spatial location and parallel orientation to the outflows. The \htcn/\sot\space emission towards IRS3B is nearly orthogonal to the disk continuum major axis position angle and indicates that the emission towards IRS3B is tracing predominantly \sot\space and not \htcn.

\section{Keplerian Rotation}\label{sec:keprotation}

To determine the stability of the circumstellar disks around IRS3B and IRS3A, the gravitational potentials of the central sources must be constrained. The protostars are completely obscured at $\lambda\space<\space3$~\micron, rendering spectral typing impossible and kinematic measurements of the protostar masses from disk rotation are required to characterize the protostars themselves. Assuming the gravitational potential is dominated by the central protostellar source(s), one would expect the disk to follow a Keplerian rotation pattern if the rotation velocities are large enough to support the disk against the protostellar gravity. These Keplerian motions will be observed as Doppler shifts in the emission lines of molecules due to their relative motion within the disk. Well-resolved disks with Keplerian rotation are observed as the characteristic ``butterfly'' pattern around the central gravitational potential: high velocity emission at small radii to low velocity emission at larger radii, and back to high velocity emission at small radii on opposite sides of the disk \citep[e.g., ][]{2013ApJ...774...16R, 2018AA...609A..47P}.

\subsection{PV Diagrams}
To analyze the kinematics of these sources, we first examine the moment 0 (integrated intensity) maps of the red- and blue-Doppler shifted \cso\space emission to determine if the emission appears well ordered (Figure~\ref{fig:irs3bc17omoment}) and consistent with \htcop\space(Figure~\ref{fig:h13copmomentc17o}). We then examine the sources using a position-velocity (PV) diagram which collapses the 3-D nature of these data cubes (RA, DEC, velocity) into a 2-D spectral image. \added{We specify the number of integrated pixels across the minor axis to limit bias from the large scale structure of the envelope and select emission originating from the disk.} This allows for an estimation of several parameters via examining the respective Doppler shifted components. 
\explain{Moved footnote to main text}

\subsubsection{IRS3B}
The PV diagrams for IRS3B are generated over a 105 pixel (2\farcs1) width strip at a position angle 28\deg. The PV diagram velocity axis is centered on the system velocity of 4.8~km~s$^{-1}$ \citep[][]{2016Natur.538..483T} and spans $\pm$5~km~s$^{-1}$ on either side, while the position axis is centered just off of the inner binary, determined to be the kinematic center, and spans 5\arcsec\space(\ab1500~au) on either side.

\cso\space appears to trace the gas within the disk of IRS3B on the scale of the continuum disk (Figure~\ref{fig:irs3bc17omoment}). It is less abundant and therefore less affected by outflow emission. We use it as a tracer for the kinematics of the disk (PV-diagram indicating Keplerian rotation; Figure~\ref{fig:l1448irs3b_c17o_pv}). The \cso\space emission extends to radii beyond the continuum disk, likely extending into the inner envelope of the protostar, while the \htcop\space emission~(Figure~\ref{fig:h13copmomentc17o}) appears to trace larger scale emission surrounding the disk of IRS3B and emission within the spatial scales of the disk has lower intensity. This is indicative of emission from the inner envelope as shown by the larger angular scales the emission extends to with respect to \cso\space (\htcop\space PV-diagram; Figure~\ref{fig:l1448irs3b_h13cop_pv}). Finally, the blended molecular line, \htcn/\sot\space appears to trace shocks in the outflows and not the disk kinematics for IRS3B. For these reasons, we do not plot the PV diagram of \htcn/\sot.

\subsubsection{IRS3A}
The PV diagrams for IRS3A are generated with a 31 pixel (0\farcs62) width strip at a position angle 133\deg. \cso\space is faint and diffuse towards the IRS3A disk (Figure~\ref{fig:l1448irs3a_cso_pv}) but still traces a velocity gradient consistent with rotation (Figure~\ref{fig:irs3ac17omoment}) and has a well ordered PV diagram (Figure~\ref{fig:l1448irs3a_cso_pv}). \htcn/\sot, (Figure~\ref{fig:irs3ah13cnmoment}), appears to trace the kinematics of the inner disk due to the compactness of the emission near the protostar and the appearance within the disk plane (Figure~\ref{fig:l1448irs3a_h13cn_pv}). The velocity cut is centered on the system velocity of 5.4~km~s$^{-1}$ and spans 6.2~km~s$^{-1}$ on either side. The emission from the blended \htcn/\sot\space is likely dominated by \htcn\space instead of \sot, due to the similar system velocity that is observed. \sot\space would have \ab1.05~km~s$^{-1}$\space offset which is not observed in IRS3A.

Similar\deleted{ly} to IRS3B, the \htcop\space emission likely traces the inner envelope, indicated in Figure~\ref{fig:irs3ah13copmoment}, as it extends well beyond the continuum emission but still traces a velocity gradient consistent with rotation (Figure~\ref{fig:l1448irs3a_h13cop_pv}). The circumstellar disk emission is less resolved, however, due to the compact nature of the source and has lower sensitivity to emission because it is located \ab8~arcsec (beam efficiency\ab60\%) from the primary beam center. 

\subsection{Protostar Masses: Modeling Keplerian Rotation}
The kinematic structure, as evidenced by the blue- and red-shifted integrated intensity maps (e.g., Figures~\ref{fig:irs3bc17omoment}~and~\ref{fig:irs3ac17omoment}) indicate rotation on the scale of the continuum disk.\added{ The disk red- and blue-emission emission are oriented along the disk major axis and and not along the disk minor axis, which would be expected if the emission was contaminated by outflow kinematics.} We first determined the protostellar mass by analyzing the PV diagram to determine regions indicative of Keplerian rotation. We summarize the results of our PV mass fitting in Table~\ref{table:pvtable}. PV diagram fitting provides a reasonable measurement of protostellar masses in the absence of a more rigorous modeling approach. The Keplerian rotation-velocity formula, $V(R) = (GM/R)^{0.5}$ allows several system parameters to be constrained: system velocity, kinematic center position, and protostellar mass (There is a degeneracy between mass determination and the inclination angle of the Keplerian disk. \added{We account for inclination in fitting the mass using the constraint from the major and minor axis ratio of the continuum emission.}

\subsubsection{IRS3B-\lowercase{ab}}
When calculating the gravitational potential using kinematic line tracers, one must first define the position of the center of mass.  For circum-multiple systems, the center of mass is non trivial to measure, because it is defined by the combined mass of each object and the distribution can be asymmetric. Figure~\ref{fig:kincenter} compares various ``kinematic centers'' for the circumstellar disk of IRS3B depending on the methodology used. First, by fitting the midpoint between highest velocity \cso\space emission channels, where both red and blue emission is present, for IRS3B-ab using the respective red and blue-shifted emission, the recovered center is 03$^{h}$25$^{m}$36.32$^{s}$\space 30\deg45\arcmin14\farcs92 which is very near IRS3B-a. The second method, fitting symmetry in the PV-diagram, however, requires a different center in order to reflect the best symmetry of the emission arising from the disk, at 03$^{h}$25$^{m}$36.33$^{s}$\space 30\deg45\arcmin15\farcs04 which corresponds to a position north-east of the binary pair, which is close to a region of reduced continuum emission (``deficit'' in Figure~\ref{fig:zoomincont}). The first method of fitting the highest velocity emission assumes these highest velocity channels correspond to regions that are closest to the center of mass and the emission is symmetric at a given position angle. \added{We chose the \cso\space molecule, which is not affected by the strong outflows, appears to trace the continuum disk the best, and has no outflow contamination, for fitting.} The second method of fitting the PV-diagram center assumes the source is symmetric and well described by a simple Keplerian disk across the position angle of the PV cut, ignoring the asymmetry along the minor axis. Finally, we include two other positions corresponding to the peak emission in the highest velocity blue- and red- Doppler shifted channels, respectively.  Unsurprisingly, these positions are on either side of the peak fit. The difference in the position of the kinematic centers is within \ab2 resolution elements of the \cso\space map and does not significantly affect our mass determination, as demonstrated in our following analysis.

We use a method of numerically fitting the \cso\space PV diagrams employed by \citet[][]{2018ApJ...860..119G}\space and \citet[][]{2016MNRAS.459.1892S}, by fitting the emission that is still coupled to the disk and not a part of the envelope emission. This helps to provide better constraints on the kinematic center for the Keplerian circum-multiple disk. This was achieved by extracting points in the PV-diagram that have emission \deleted{\ab}10~$\sigma$\space along the position axis for a given velocity channel and fitting these positions against the standard Keplerian rotation-velocity formula. The Keplerian velocity is the max velocity at a given radius but each position within a disk will include a superposition of \deleted{several}\added{lower} velocity components due to projection effects.

The fitting procedure was achieved using a Markov Chain Monte Carlo (MCMC) employed by the Python MCMC program  \textit{emcee} \citep{2013PASP..125..306F}. Initial prior sampling limits of the mass were set to 0.1-2~\solm. Outside of these regimes would be highly inconsistent with prior and current observations of the system. Uncertainty in the distance (22~pc) from the \textit{Gaia} survey \citep[][]{2018arXiv180803499O} and an estimate of the inclination error (10\deg) were included while the parameters \deleted{M}\added{(M$_{\*}$} and V$_{sys}$) were allowed to explore phase space. These place approximate limits to the geometry of the disk. The cyan lines in Figure~\ref{fig:l1448irs3b_c17o_pv}\space trace the Keplerian rotation curve with M$_{\*}$=1.15~\solm~with 3-$\sigma$~uncertainty$=0.09$~\solm, which fits the edge of the \cso\space emission from the source. This mass estimate describes the total combined mass of the gravitating source(s). Thus if the two clumps (IRS3B-a and -b) are each forming protostars, this mass would be divided between them. However with the current observations, we cannot constrain the mass ratio of the clumps. Thus, we can consider two scenarios (Section~\ref{sec:massacc}), an equal mass binary and a single, dominate central potential.

The \htcop\space PV-diagram (Figure~\ref{fig:l1448irs3b_h13cop_pv}) shows high asymmetry emission towards the source. However, the \htcop\space emission is still consistent with the central protostellar mass measured using \cso\space emission of 1.15~\solm\space(indicated by the white dashed line). This added asymmetry is most likely due to \htcop\space emission being dominated by envelope emission, in contrast to the \cso\space being dominated by the disk. There is considerably more spatially extended and low velocity emission that extends beyond the Keplerian curve and cannot be reasonably fitted with any Keplerian curve. Additionally, there is a significant amount of \htcop\space emission that is resolved out near line-center, appearing as negative emission, whereas, the \cso\space emission did not have as much spatial filtering as the \htcop\space emission.

\subsubsection{IRS3B-\lowercase{c}}
We also analyzed the \cso\space kinematics near the tertiary, IRS3B-c, to search for indications of the tertiary mass influencing the disk kinematics. In Figure~\ref{fig:l1448irs3b_c17o_pv_tert}, we show the PV diagram of \cso\space within a 2\farcs0 region centered on the tertiary and plot velocities corresponding to Keplerian rotation at the location of IRS3B-c within the disk, to provide an upper bound on the possible protostellar mass within IRS3B-c. Emission in excess of the red-dashed lines could be attributed to the tertiary altering the gas kinematics. The velocity profile at IRS3B-c shows no evidence of any excess beyond the Keplerian profile from the main disk, indicating that it has very low mass. Based on the non-detection, we can place upper limits on the mass of \added{the} IRS3B-c source of $<$0.2~\solm\space as shown by the white dotted lines in Figure~\ref{fig:l1448irs3b_c17o_pv_tert}. A protostellar mass much in excess of this would be inconsistent with the range of velocities observed.

\subsubsection{IRS3A}
For the IRS3A circumstellar disk, the dense gas tracers \htcn\space and \cso\space were used to analyze disk characteristics and are shown in Figures~\ref{fig:l1448irs3a_cso_pv}~and~\ref{fig:l1448irs3a_h13cn_pv}. The position cut is centered on the continuum source (coincides with kinematic center), and spans 2\arcsec (\ab576~au) on either side. This provides a large enough window to collect all of the emission from the source. The dotted white lines show the Keplerian velocity corresponding to a M$_{\*}$=1.4~\solm\space central protostar which is consistent with the PV diagram.

The spatial compactness of IRS3A limits the utility of the \htcn\space PV-diagram with the previous MCMC fitting routine. We found evidence of rotation in this line tracer from the velocity selected moment 0 map series and PV-diagram. However, from the PV diagram alone, strong constraints cannot be determined due to the compactness of the \htcn\space emission and the low S/N of \cso.

\section{Application of Radiative Transfer Models}\label{sec:kmodelresults}
\explain{Explanations about MCMC and the specifics of the code was moved to appendix.}
To further analyze the disk kinematics, we utilize the methods described in \citet{2019ApJ...874..136S} and further described in Appendix~\ref{sec:apppdspy} for modeling the molecular line emission presented thus far. The modeling framework uses RADMC-3D \citep{2012ascl.soft02015D} to calculate the synthetic channel maps using 2D axisymmetric radiative transfer models in  the \added{limit of local} local thermodynamic equilibrium (LTE)  and GALARIO \citep{2018MNRAS.476.4527T} to generate the model visibilities from those synthetic channel maps. We sample the posterior distributions of the parameters to provide fits to the visibilities by utilizing a MCMC approach 
\citep[\pdspy\deleted{: \url{https://github.com/psheehan/pdspy.git}}; ][]{2019ApJ...874..136S}. \added{\pdspy\space uses the full velocity range given by the frequency limit of the input visibilities in modeling.}

Some of the parameters are less constrained than others due to asymmetry of the disks and discussion of these parameters fall outside the scope of the kinematic models sought in this paper. Our focus for the kinematic models are: position angle (p.a.), inclination (inc.), stellar mass (M$_{*}$), disk radius (R$_D$), and system velocity (V$_{sys}$). We provide a summary of our model results in Table~\ref{table:pdspykinematic}. 

The combined fitting of the models is computationally expensive (fitting 200 models simultaneously per ``walker integration time-step''), requiring on average $1-2\times10^{4}$\space core-hours per source to reach convergence. We run these models across 5 nodes with 24~cores/nodes each for \ab150~hours on the OU (University of Oklahoma) Supercomputing Center for Education and Research supercomputers (OSCER) to reach sufficient convergence in the parameters. The convergence state is determined when the \textit{emcee} ``walkers'' reach a steady state solution where the ensemble of walkers is not changing by an appreciable amount, simply oscillating around some median value with a statistical variance.

\subsection{IRS3B}
The \pdspy\space kinematic flared disk model results for IRS3B are shown in Figure~\ref{fig:c17o_res}\space with the Keplerian disk fit compared to the data. The system velocity fitted is in agreement with the PV-diagram analysis. There is some uncertainty in the kinematic center, due to the diffuse, extended emission near the system velocity ($<$1 km s$^{-1}$) which yields degeneracy when fitting. The models yielded similar stellar masses as compared to the PV/Gaussian fitting (3-$\sigma$~uncertainties~listed, \pdspy\space 1.19$^{+0.13}_{-0.07}$~\solm; PV: 1.15$^{+0.09}_{-0.09}$~\solm), similar position angles (\pdspy:\space 27\deg$^{+  1.8}_{-  2.9}$; PV:\ab28\deg), and while the inclinations are not similar (\pdspy:\space66\deg$^{+ 3.0}_{- 4.6}$; Gaussian:\ab45\deg), this discrepancy in inclination is most likely due to a difference in asymmetric gas and dust emission. \deleted{The tertiary subtraction method (Appendix~\ref{sec:tertsub}) to Gaussian fit the dust continuum attempts to preserve the underlying disk structure and thus would fit a lower inclination than the \pdspy\space fit, which attempts to reconcile the absorption near the tertiary source with the model Keplerian disk. However, upon}\added{With the tertiary subtraction method (Appendix~\ref{sec:tertsub}), we gaussian fit the dust continuum of IRS3B-c to preserve the underlying disk structure, then fit the IRS3B-ab disk with a single gaussian. Using the PV-diagram fitting, we attempt to fit symmetric Keplerian curves to the PV-diagram. \pdspy\space attempts to also fit the asymmetric southeast side of the disk, which is an asymmetric feature, with the model symmetric Keplerian disk. Upon} further inspection of the residual map, there is significant residual emission on the south-eastern side of the disk \deleted{, mostly due to the asymmetry of the disk and the fact that the model has a smooth structure, while IRS3B has resolved spiral structure.}\added{ which is likely a second order effect in the fit, however it is confined spatially and spectrally and should not have a major effect on the overall fit.}

\subsection{IRS3A}
The \pdspy\space kinematic flared disk model results for IRS3A are shown in Figure~\ref{fig:h13cn_res}, primarily fitting the inner disk. The models demonstrate the gas disk is well represented by a truncated disk with a maximum radius of the disk of \ab40~au (most likely due to the compact nature of the emission). This disk size of 40~au is smaller than the continuum disk and results from the compact emission of \htcn. The models find a system velocity near 5.3~km~s$^{-1}$\space in agreement with the PV-diagram. The system velocity of numerous molecules (\htcop, \cso, and \htcn) are in agreement and thus likely tracing the same structure in the system. The models yielded a similar stellar mass ($1.51^{+0.06}_{-0.07}$~\solm, 3-$\sigma$~uncertainties~listed) to the estimate from the PV-diagram. Also the disk orientation of inclination (69\deg) and position angle (\ab122\deg) agree with the estimate from the continuum Gaussian fit.

\section{Discussion}\label{sec:discussion}
\explain{Determining ideal kinematic line tracer for disk was moved to appendix. Also the mass accretion was combined with discussion of FUOrs.}

\subsection{Origin of Triple System and Wide Companion}\label{sec:origin}
Protomultiple systems like that of IRS3B and IRS3A can form via several possible pathways pathways: thermal fragmentation (on scales \ab1000s of au), turbulent fragmentation (on scales \ab1000s of au), gravitational instabilities within disks (on scales \ab100s of au), and/or loose dynamical capture of cores (on scales \ab10$^{4-5}$~au). To constrain the main pathways for forming multiple systems, we must first constrain the protostellar geometrical parameters and then the (in)stability of the circum-multiple disk. Previous studies towards L1448 IRS3B \citep[see ][]{2016Natur.538..483T} achieved \ab0\farcs4\space molecular line resolution, roughly constraining the protostellar mass. The high resolution and high sensitivity data we present allows constraints on the stability of the circumstellar disk of IRS3B and sheds light on the formation pathways of the compact triple system and the wide companion. The circumstellar disk around the wide companion, IRS3A, has an orthogonal major axis orientation to the circumstellar disk of IRS3B, favoring formation mechanisms that result in wider companions forming with independent angular momentum vectors. The circumstellar disk around IRS3B is massive, has an embedded companion (IRS3B-c), and has spiral arms, which are indicative of gravitational instability, and we will more quantitatively examine the (in)stability of the disk in Section~\ref{sec:struct}.

\explain{Moved to a subsection instead of subsubsection since the content holds its own independent of the origin of the triple\&wide companion}
\subsection{Signatures of an Embedded Companion in Disk Kinematics}\label{sec:embeddedkinematics}
\added{
    Hydrodynamic simulations show that massive embedded companions within viscous disks should impact the Keplerian velocity pattern in a detectable manner \citep{2015ApJ...811L...5P}. \citet{2018MNRAS.480L..12P} showed the signatures of a massive companion embedded within a viscous, non-self gravitating disk.}\added{Their model observations are higher (\ab2$\times$) spectral and angular resolution, and more sensitive (\ab5$\times$) than the presented observations. They show a 10~M$_{J}$~mass source should be easily detectable with about 1000 orbits of evolution by analyzing the moment 1 maps. More recently, several studies of protoplanetary disks have confirmed these predictions of localized Keplerian velocity deviations for moderately massive planets \citep[][]{2018ApJ...860L..13P,2019NatAs...3.1109P}. However, these systems are much more evolved ($>3$~Myr), with quiescent, non-self gravitating disks, and likely experienced thousands of stable orbits compared to IRS3B, a self-gravitating and actively accreting Class 0 source, with a companion that likely has completed only a few dynamically changing orbits.}
    
    \added{\citet{2020arXiv200715686H}\space performed simulations of a viscous, self-gravitating disk (0.3~\msun) around a 0.6~\msun\space source much more similar in physical parameters or IRS3B than the types of systems discussed in the preceding paragraph. Their results showed that the effects of self gravity will provide ``kinks'' at high resolution and sensitivity. Additionally, \citet{2011IAUS..276..463V}\space showed due to exchange of momentum with the disk, or dispersal due to tidal torques, the fragment radius would be drastically changing up to an order-of-magnitude over the evolution of the disk. All of these work to mask definitive observable kinematic deviations of embedded companions in the disk.}
    
    \deleted{The disks presented in this work are not in quiescent states, like that of \citet{2018ApJ...860L..13P}, where the disk self-gravity is negligible. We would expect similar signatures to be readily apparent given a longer evolution or a more massive companion source.}

\subsection{Disk Structure}\label{sec:struct}
With the high resolutions observations, we can construct a radial profile of the continuum emission to analyze disk structure. The circumstellar disk of IRS3B has prominent spiral arms but the radial profile will azimuthally average this emission. In order to construct the radial profile, we have to define: an image center to begin the extraction, the geometry (position angle and inclination) of the source, and the size of each annuli. The system geometry and image center were all adapted from the PV diagram fit parameters and the radius of the annuli is defined as half the average synthesized beamsize \citep[Nyquist Sampling;][]{1928TAIEE..47..617N}. We then convert from flux density to mass via Equation~\ref{eq:dustmasseq}\space and further construct a disk mass surface density profile. To convert from flux density into dust mass, we adopt a radial temperature power law with a slope of -0.5, assuming the disk at 100~au can be described with a temperature of (30~K)$\times(L_{*}$/\lsun)$^{0.25}$\deleted{, similar to the temperature law used in other YSO regions \citep{2020ApJ...890..130T}}. The temperature profile has a minimum value of 20~K, based on models of disks embedded within envelopes \citep[][]{2003ApJ...591.1049W}. While we adopt a temperature law profile \deleted{that is similar to protostellar surveys}, protostellar multiples are expected to complicate simple radial temperature profiles.

Towards IRS3B, in order to mitigate the effects of the tertiary source in the surface density calculations, we use the tertiary subtracted images, described in the Appendix~\ref{sec:tertsub}. The system geometric parameters used for the annuli correspond to an inclination of 45\deg\space and a position angle of 28\deg\deleted{$^3$}\added{. The PV/Gaussian fits were used here for ease of reproducibility and utilizing the \pdspy\space results would still be consistent}. The largest annulus extends out to 5\arcsec, corresponding to the largest angular scale on which we can recover most emission. The temperature at 100~au for IRS3B-ab is taken to be $\approx$40.1~K. We show both the extracted flux radial profile and radial surface density profile for IRS3B-ab in Figure~\ref{fig:surfacedensity}. The radial surface density profile shows a flat surface density profile out to \ab400~au.

Towards IRS3A, the system geometry parameters used for the annuli correspond to an inclination of 69\deg\space and a position angle of 133\deg. With this method, we construct a radial surface density profile to analyze the stability of the disk (Figure~\ref{fig:irs3asurfacedensity}). The temperature at 100~au for IRS3A is taken to be $\approx$50.9~K. The circumstellar disk of IRS3A is much more compact than the circumstellar disk of IRS3B, with the IRS3A disk radius \ab150~au, and thus the assumed temperature at 100~au is a good approximation for the median disk temperature.

\subsubsection{Disk Stability}\label{sec:stability}
The radial surface density profiles allow us to characterize the stability of the disk to its self-gravity as a function of radius. The Toomre~Q parameter (herein Q) can be used as a metric for analyzing the stability of a disk. It is defined as the ratio of the rotational shear and thermal pressure of the disk versus the self-gravity of the disk, susceptible to fragmentation. When the Q parameter is $<$1, it indicates a gravitationally unstable region of the disk. 

Q is defined as: 
\begin{equation}
Q = \frac{c_{s} \kappa}{\pi G \Sigma}
\end{equation}
where the sound speed is c$_s$, the epicyclic frequency is $\kappa$ corresponding to the orbital frequency ($\kappa = \Omega$ in the case of a Keplerian disk), and the surface density is $\Sigma$, and G is the gravitational constant.

We further assume the disk is thermalized and the disk sound speed radial profile is given by the kinetic theory of gases: 
\begin{equation}\label{eq:cs}
c_{s}\left(T\right) = \left(\frac{k_{b}T}{m_{H} \mu}\right) ^{0.5}
\end{equation}
where T is the gas temperature and $\mu$ is the mean molecular weight (2.37). \added{We then evaluate the angular frequency as a function of radius,} \deleted{We can then construct the angular frequency radial profile given by the equation:}
\begin{equation}
\Omega\left(R\right) = \left(\frac{GM_{*}}{R^{3}}\right)^{0.5}
\end{equation}
 where M$_{*}=1.15$~\solm.

Simulations have shown that values of Q$<$1.7 (calculated in 1D) can be sufficient for self-gravity to drive spiral arm formation within massive disks while Q $\approx$1\space is required for fragmentation to occur in the disks \citep[][]{2010ApJ...710.1375K}. Figure~\ref{fig:irs3btoomreq} shows the Q radial profile for the circumstellar disk of L1448 IRS3B, which varies by an order of magnitude across the plotted range (0.4-4). The disk has Q$<1$\space and therefore is  gravitationally unstable starting near \ab120~au, interior to the location of the embedded tertiary within the disk and extending out to the outer parts of the disk (\ab500~au) as indicated by the IRS3B Toomre~Q radial profile. The prominent spiral features present in the circumstellar disk span a large range of radii (10s-500~au).

Figure~\ref{fig:irs3atoomreq} shows the Toomre~Q radial profile for the circumstellar disk of L1448 IRS3A. The IRS3A dust continuum emission, while having possible spiral arm detection (Figure~\ref{fig:contimage}), is more indicative of a gravitationally stable disk through the analysis of the Toomre~Q radial profile (Q$>$5 for the entire disk). This is due to the higher mass central protostar and lower disk surface density, as compared to the circumstellar disk of IRS3B. Thus substructures in IRS3A may not be gravitationally driven spiral arms and could reflect other substructure. The circumstellar disk around IRS3A has a mass of 0.04~\solm\space and the protostar has a mass of 1.4~\solm.

\subsection{Interpretation of the Formation Pathway}
The formation mechanism for the IRS3A source, IRS3B system as a whole, \added{and the more widely separated L1448 NW source \cite{2016ApJ...818...73T}}, is most likely turbulent fragmentation, which works on the 100s-1000s~au scales \citep[][]{2010ApJ...725.1485O, 2019ApJ...887..232L}. Companions formed via turbulent fragmentation are not expected to have similar orbital configurations and thus are expected to have different V$_{sys}$, position angle, inclination, and outflow orientations. For the wide companion, IRS3A, the disk and outflows are nearly orthogonal to IRS3B and have different system velocities (e.g., 5.3~km~s$^{-1}$\space and 4.9~km~s$^{-1}$, respectively; Tables~\ref{table:pvtable}~and~\ref{table:pdspykinematic}). \citet{2019ApJ...884....6M} has shown protostellar systems dynamically ejected from multi-body interactions are less likely to be disk bearing. Considering the low systemic velocity offset (IRS3A: 5.3~km~s$^{-1}$, IRS3B: 4.9~km~s$^{-1}$), the well ordered Keplerian disk of IRS3A, and relative alignment along the long axis of the natal core \citep{2017MNRAS.469.3881S}, the systems would not likely have formed via the dynamical ejection scenario from the IRS3B system \citep[][]{2012Natur.492..221R}.

\added{In contrast, the triple system IRS3B appears to have originated via disk fragmentation. The well organized \cso\space emission, which traces the disk continuum emission, indicates that the circum-multiple disk in IRS3B is in Keplerian rotation at both compact and extended spatial scales (0\farcs2 to $>$2\farcs0; 50~au to $>$600~au) (see Figure~\ref{fig:irs3bc17omoment}). The derived disk mass (M$_{d}$/M$_{s}\sim25$\%) is high, such that the effects of self-gravity are important \citep[][]{1990ApJ...358..515L}. The low-m (azimuthal wavenumber) spiral arms observed in the disk are consistent with the high mass \citep{2016ARAA..54..271K}. The protostellar disk is provided stability on scales near the central potential due to the shear effects of Keplerian rotation and higher temperatures, while, at larger radii, the rotation velocity falls off and the local temperature is lower, allowing for local gravitational instability.  Moreover, as seen in Figure~\ref{fig:irs3btoomreq}, Toomre's~Q falls below unity at radii $>120$~au, coincident with the spatial location  of the tertiary IRS3B-c, as expected if recently formed via gravitational instability in the disk. Additionally the inner binary, IRS3B-ab, could have formed via disk fragmentation prior to the IRS3B-c, resulting in the well-ordered kinematics surrounding IRS3B-ab.

The PV analysis of IRS3B-c also suggests that the central mass of the tertiary continuum source is low enough ($\sim0.02$~\solm) to not significantly alter the kinematics of the disk (Figures~\ref{fig:irs3bc17omoment}~and~\ref{fig:irs3abc17omoment1})

The apparent co-planarity of IRS3B-abc and the well-organized kinematics of both the disk and envelope tracers, \cso\space and \htcop, argue against the turbulent fragmentation pathway within the subsystem. The PV-diagram of IRS3B-ab is well structured in various disk tracing molecules and the outflow orientation of IRS3B-c is aligned with the angular momentum vector of IRS3B-ab, making dynamical capture unlikely.
}

\deleted{Massive disks (when M$_{d}$/M$_{s}>$ 10\%), such as IRS3B, are subject to the effects of self-gravity \citep{1990ApJ...358..515L}.}
\deleted{The circum-multiple disk is estimated to be 0.29~\solm\space and the central protostar(s) are estimated to be 1.15~\solm, which yields a mass ratio M$_{d}$/M$_{*}\approx0.25$\space indicating that self-gravity should be important. Furthermore, we show the Toomre~Q radial profile for IRS3B (Figure~\ref{fig:irs3btoomreq}) and we note the Q parameter is $<1$, at radii $>$120~au, encompassing IRS3B-c, further consistent with the disk having undergone gravitational fragmentation. The molecule \cso\space traces closely with the circum-multiple disk of IRS3B and is tracing the same area as the continuum disk.}
\deleted{The well-organized \cso\space emission indicates that the circum-multiple disk is well described by Keplerian rotation at both compact and extended spatial scales (0\farcs2 to $>$2\farcs0; 50~au to \ab600~au), as shown in Figure~\ref{fig:l1448irs3b_c17o_pv}, and encompasses a large amount of mass.  Combining this result with the presence of spiral arms and an embedded tertiary companion, demonstrates that the tertiary companion could have formed in situ and possibly explains the IRS3B system asymmetry. Additionally, the inner clumps (IRS3B-ab) could have formed via disk fragmentation prior to the formation of IRS3B-c, resulting in the ordered kinematics towards IRS3B-ab.}

\deleted{The organized kinematics of both the disk and envelope tracers, \cso\space and \htcop, are indicative that the formation of IRS3B-abc via turbulent fragmentation and migration is not likely. The PV analysis of IRS3B-c also suggests that the central mass of the tertiary continuum source is low enough (M$_{\*}$\space upper limit $\sim0.02$~\solm) to not significantly alter the kinematics of the disk (Figures~\ref{fig:irs3bc17omoment}~and~\ref{fig:irs3abc17omoment1}), meaning that a protostar like IRS3A could not have merged with IRS3B-ab to form the tertiary. The PV-diagram of IRS3B-ab is well structured in various disk tracing molecules and the outflow orientation of IRS3B-c is aligned with the angular momentum vector of IRS3B-ab and thus unlikely that IRS3B-c was captured. 
}
\explain{Moved paragraph explaining various kinematic centers to section 6.2}

\subsection{Protostar Masses}
Comparing the masses of IRS3A (1.51~\solm) and IRS3B (1.15~\solm) to the initial mass function (IMF) \citep[young cluster IMF towards binaries;][]{2005ASSL..327...41C} shows these protostars will probably enter the main sequence as typical, stars once mass accretion from the infalling envelope and massive disks completes. IRS3B-a and -b are likely to continue accreting matter from the disk and envelope and grow substantially in size. 

\explain{Moved from formation pathways since this section analyzes central protostellar masses.}
In addition to the symmetry in the inner clumps, further analysis towards IRS3B-ab of the spatial location of the kinematic centers \deleted {could} indicate that the kinematic center is consistent with being centered \deleted{central protostar(s) are not centered on the continuum sources, -a and -b, but could be centered} on the deficit (``deficit''; Figure~\ref{fig:zoomincont}) with a surrounding inner disk, where IRS3B-a is a bright clump moving into the inner disk. The continuum source IRS3B-b would be just outside the possible inner disk radius\deleted{ and it is unclear exactly how much mass is enclosed from these observations alone}. These various kinematic centers are within one resolving element of the \cso\space beam, and thus we are unable to break the degeneracy of the results from these observations alone\deleted{ because the three noted kinematic centers are all consistent with the \cso\space observations}. Higher resolution kinematics and continuum observations are required to understand the architecture of the inner disk and whether each dust clump corresponds to a protostar, or if the clumps are components of the inner disk \deleted{and the central mass is obscured in our observations.}\added{ and the central protostar is not apparent from dust emission in our observations.}

If we assume the IRS3B-ab clumps surround a single central source, this source would most likely form an A-type (M$_{*}\approx1.6-2.4~$~\solm) star, depending on the efficiency of accretion \citep[10-15\%;][]{2007ApJ...656..293J}. Similarly, IRS3A is likely to form an A-type star. If the IRS3B-ab clumps each represent a forming protostar, then each source would most likely form a F or G-type (M$_{*}\approx0.8-1.4~$~\solm) star depending on the ratio of the masses between the IRS3B-a and IRS3B-b components. IRS3B-c, while currently estimated to have a mass $<0.2$~\solm, it could still accrete a substantial amount mass of the disk and limit the accretion onto the central IRS3B-ab sources. This mechanism can operate without the need to open a gap \citep[][]{1996LNP...465..115A}, which remains unobserved in these systems.

More recently, \citet{2020AA...635A..15M}\space targeted several (7) Class 0 protostars in Perseus with marginal resolution and sensitivities, to fit the molecular lines emission against Keplerian curves to derive protostellar masses. Their fitting method is similar to our own PV diagram fitting and has an average protostellar mass of \ab0.5~\msun. If IRS3B-ab is a single source protostar, then this source would be significantly higher mass (\mstar\ab1.2~\msun) than the average mass of the sample, similar for IRS3A (\mstar\ab1.4~\msun). However, if IRS3B-ab is a multiple protostellar source of two equally mass protostars (\mstar\ab0.56~\msun), then these sources would be consistent with the survey's average protostellar mass.
\explain{For readability, moved to new paragraph.}
\deleted{Additionally,}\citet{2020AA...635A..15M}\space \deleted{performed a survey (\textit{CALYPSO}) with the Plateau de Bure Interferometer of well known Class 0 protostars in Perseus, which } included IRS3B (labeled L1448-NB), using the molecules \tco, \ceo, and SO, and the protostellar parameters are consistent with the results we derived here (\mstar\ab1.4~\msun, \textit{PA}\ab29.5\deg, and \textit{i}\ab45\deg)\added{, despite lower sensitivities and resolutions compared to our observations}.

\citet[][]{2017ApJ...834..178Y} targeted several well known Class 0 protostars \deleted{in Ophiuchus }and compared their stellar properties against other well known sources (see reference Table 5 and Figure 10), to determine the star/disk evolution. They derived an empirical power-law relation for Class 0 towards their observations R$_{d}=(44\pm8)\times\big(\frac{M_{*}}{0.1~M_{\odot}}\big)^{0.8\pm0.14}$~au and a Class 0$+$I relation of R$_{d}=(161\pm16)\times\big(\frac{M_{*}}{1.0~M_{\odot}}\big)^{0.24\pm0.12}$~au. The L1448 IRS3B system, with a combined mass \ab1.15~\solm, disk mass of \ab0.29~\solm, and a FWHM Keplerian gaseous disk radius of \ab300~au, positions the target well into the Class 0 stage (\ab245-500~au for the \citet[][]{2017ApJ...834..178Y} relation) and \ab2-3$\times$\space the average stellar mass and radius of these other well known targets. The protostellar mass of IRS3B is larger relative to the sample of protostars observed in \citet[][]{2017ApJ...834..178Y}, which had typical central masses 0.2 to 0.5~\solm. However, this is the combined mass of the inner binary and each component could have a lower mass. L1448 IRS3A, which has a much more compact disk (FWHM Keplerian disk radius of \ab158~au) and a higher central mass than IRS3B (\ab1.4~\solm), is more indicative of a Class I source using these diagnostics. We note there is substantial scatter in the empirically derived relations \citep{2020ApJ...890..130T}, thus the true correspondence of disk radii to an evolutionary state of the YSOs is highly uncertain and we observe no evidence for an evolutionary trend with disk radii.

\explain{Cut section on Hierarchical Comparisons between class 0/field stars and this system as to not detract from main takeaway of the paper.}

\subsection{Gravitational Potential Energy of IRS3B-c}\label{sec:gpe}
In analyzing the gravitational stability of the IRS3B circumstellar disk, we can also analyze the stability of the clump surrounding IRS3B-c. If the clump around IRS3B-c is sub-virial (i.e., not supported by thermal gas pressure) it would be likely unstable to gravitational collapse, undergoing rapid (dynamical timescale, $\tau_{dyn}$) collapse resulting in elevated accretion rates compared to the collapse of virialized clumps Additionally, it would be unlikely to observe this short-lived state during the first orbit of the clump. Dust clumps embedded within protostellar disks are expected to quickly (t$<10^{5}-yr$) migrate from their initial position to a quasi-stable orbit much closer to the parent star \citep{2019AA...631A...1V}. Thus observing the IRS3B-c clump at the wide separation within the disk is likely due to it recently forming in-situ. The virial theorem states $2E_{kin} + E_{pot}=0$, or in other words we can define an $\mathcal{R}$\space such that $\mathcal{R} := \frac{2E_{kin}}{|E_{pot}|}$\space will be $<1$\space for a gravitationally collapsing clump and $>1$\space for a clump to undergo expansion. Assuming the ideal gas scenario of N particles, we arrive at $E_{kin} = 1.5Nk_{b}T_{clump}$\space where k is the Planck constant and T$_{clump}$ is the average temperature of the particles. The potential energy takes the classic form $E_{pot} = \frac{-3}{5}\frac{GM^{2}_{clump}}{R_{clump}}$. We can define $N=\frac{M_{clump}}{\mu m_{H}}$\space where $\mu$\space is the mean molecular weight (2.37) and m$_{H}$\space is the mass of hydrogen. Assuming the clump is thermalized to the T$_{peak}=54.6$~K, the mass of the clump is $0.07$~\solm, the upper bound for the IRS3B-c protostar is $0.2$~\solm, and the diameter is 78.5~au (Table~\ref{table:obssummary3}), we calculate $\mathcal{R} \approx1.4$\space for the dust clump alone (this $\mathcal{R}$\space is likely an upper bound since our mass estimate for the dust is likely a lower limit due to the high optical depths) and $\approx0.3$\space for the combined dust clump and protostar (This $\mathcal{R}$\space is likely an lower bound since our mass estimate for the protostar an upper limit due to be consistent with the kinematic observations.). This is indicative that the core could be virialized but could also reflect a circumstellar accretion disk around IRS3B-c, or in the upper limit of the protostellar mass, could undergo contraction.

\subsection{Mass Accretion}\label{sec:massacc}
The mass in the circumstellar disks and envelopes provide a reservoir for additional mass transfer onto the protostars. However, this mass accretion can be reduced by mass outflow due to protostellar winds, thus we need to determine the maximal mass transport rate of the system to determine if winds are needed to carry away momentum \citep{1998ApJ...502..661W}. While these observations do not place a direct constraint on \mdot, from our constraints on M$_{*}$\space and the observed total luminosity we can estimate the mass accretion rate. In a viscous, accreting disk, the total luminosity is the sum of the stellar and accretion luminosity:
\begin{equation}
L_{bol}\sim L_{*}+L_{acc}
\end{equation}
and the L$_{acc}$\space is:
\begin{equation}
L_{acc}=\frac{GM_{*}\dot{M}}{R_{*}}
\end{equation}
half of which is liberated through the accretion disk and half emitted from the stellar surface. From our observations, we can directly constrain the stellar mass and thus, using the stellar birth-line in \citet{1997ApJ...475..770H}~(adopting the models with protostellar surface cooling which provides lower-estimates), we can estimate the protostellar radius. From these calculations we can estimate the mass accretion rate of the protostars. The results are tabulated in Table~\ref{table:massacc}\space but are also summarized here. For the single protostar IRS3A this is straight-forward, but for the binary source IRS3B-ab, care must be taken. We adopt the two scenarios for the system configuration: 1.) the protostellar masses are equally divided (two 0.575~\solm\space protostars) and 2.) one protostar dominates the potential (one 1.15~\solm\space protostar). From Figure~3 in \citet{1997ApJ...475..770H}\space we estimate the stellar radius to be 2.5~\rsun, 2.5~\rsun, and 2~\rsun\space for stellar masses 0.575~\solm, 1.15~\solm, and 1.51~\solm, respectively. From Figure~3 in \citet{1997ApJ...475..770H}\space we estimate the stellar luminosity to be 1.9~\lsun, 3.6~\lsun, and 2.5~\lsun\space for stellar masses 0.575~\solm, 1.15~\solm, and 1.4~\solm, respectively \added{(see Section~\ref{sec:diskmass}).}\deleted{We adopt bolometric luminosities of 13.0~\lsun\space(14.4~\lsun) for IRS3B(A) respectively (scale to a distance of 288~pc\space from \citet{2016ApJ...818...73T}). We note here in the literature there are several values for IRS3B, differing from our adopted value by a factor of a few, thus, our derived $\dot{M}$\space is uncertain within a factor of a few. Reconciling this is outside of the scope of this paper, but the difference could arise from source confusion in the crowded field and differences in SED modeling.}

Considering the bolometric luminosities for IRS3B and IRS3A given in Section~\ref{sec:diskmass}, we find the $\dot{M}\sim4.95\times10^{-7}$~\solm~yr$^{-1}$ for IRS3A. Then for IRS3B-ab, in the first scenario (two 0.575~\solm~protostars), we find $\dot{M}\sim1.5\times10^{-6}$~\solm~yr$^{-1}$ and in the second scenario (one 1.15~\solm~protostar), we find $\dot{M}\sim6.6\times10^{-7}$~\solm~yr$^{-1}$. These accretion rates are unable to build up the observed protostellar masses within the typical lifetime of the Class 0 stage (\ab160~kyr) and thus require periods of higher accretion events to explain the observed protostellar masses. This possibly indicates the IRS3B-ab system is more consistent as an equal mass binary system. However, further, more sensitive and higher resolution observations to fully resolve out the dynamics of the inner disk are needed to fully characterize the sources.

We further compare the accretion rates derived here with a similar survey towards Class 0+I protostars \citep{2017ApJ...834..178Y}. We find IRS3A is consistent with L1489 IRS, a Class I protostar with a M$_{*}\sim1.6$~\solm\space\citep[][]{2013ApJ...770..123G}\space and a $\dot{M}\sim2.3\times10^{-7}$~\solm~yr$^{-1}$~\citep[][]{2014ApJ...793....1Y}. Furthermore, in the case IRS3B-ab is an equal mass binary, the derived accretion rates as compared with the sources in \citet{2017ApJ...834..178Y}\space are in the upper echelon of rates. However, in the case IRS3B-ab is best described as a single mass protostar, the derived accretion rates are consistent with TMC-1 and TMC-1A, other Class 0+I sources in \citet{2017ApJ...834..178Y}.

\deleted{Furthermore, these rates are lower than the rate derived from maximal mass transport rate from the gravitational collapse of an isothermal sphere. \citet{1994ApJ...431..341S}\space describes the net mass accretion rate for a collapsing cloud via $\dot{M} \approx \frac{c_{s}^{3}}{G}$. Using Equation~\ref{eq:cs}\space and the average disk temperatures of IRS3A (50.9~K) and IRS3B (40.1~K), we arrive at $\dot{M} = 9.27(6.48)\times10^{-5}$~\solm~year$^{-1}$\space for IRS3A(IRS3B-ab), respectively.}

\added{
While the currently estimated accretion rates for IRS3B and IRS3A would not be able to assemble the observed protostar masses in the lifetime of a Class 0 protostar, accretion rates are not necessarily constant through the protostellar phase. The well-known FU Orionis phenomenon are exemplary examples of non-steady accretion in protostars \citep[e.g. ][]{1996ARAA..34..207H,2014prpl.conf..387A}. Accretion bursts have also been observed in both Class I and Class 0 protostars in recent years \citep{2015ApJ...800L...5S, 2013prpl.conf1H023F}. Gravitational instability in disks has been proposed as a mechanism to drive outburts with the accretion of clumps of material from the disks \citep{2011ASPC..451..213S, 2017MNRAS.465....2M, 2014ARep...58..522V,2014MNRAS.444..887D,2020arXiv201005939S}. In this scenario, the accretion luminosity increases quickly, stabilizing the disks. After the accretion event has finished, the protostars undergo a ``quiescent'' stage while the disk can re-develop gravitational instabilities and fragment. Two possible signatures for this mechanism would be in the outflow configuration: bi-polar jets with periodically spaced knots and gravitational instabilities of the disk. IRS3B does exhibit a gravitationally unstable disk (Section~\ref{sec:stability}), but the outflow, while having many bright features Appendix~\ref{sec:outflow}, does not show periodically spaced knows like the example from \citet{2015Natur.527...70P}. Thus, it is possible that both IRS3B and IRS3A have undergone past accretion burst, helping to explain their current masses and relatively low inferred accretion rates, but they do not currently exhibit features of outbursting protostars and we cannot unequivocally state that they have undergone past outbursts.
}
\section{Summary}
We present the highest sensitivity and resolution observations tracing the disk kinematics toward L1448 IRS3B and IRS3A to date, \added{\cso/\ceo\space comparison: \ab5$\times$\space higher S/N at 4.0~\kms, \ab3$\times$\space higher resolution, and \ab2$\times$\space better velocity resolution as compared to \citet{2016Natur.538..483T})}. Our observations resolve three dust continuum sources within the circum-multiple disk with spiral structure and trace the kinematic structures using \cso, \htcn/\sot, and \htcop\space surrounding the proto-multiple sources. The central gravitating mass in IRS3B, near -a and -b, dominates the potential as shown by the organized rotation in \cso\space emission. We compare the high fidelity observations with radiative transfer models of the line emission components of the disk. The presence of the tertiary source within the circum-multiple disk, detection of dust continuum spiral arms, and the Toomre~Q analysis are indicative of the disk around IRS3B being gravitationally unstable.

We summarize our empirical and modeled results:

\begin{enumerate}
    \item We resolve the spiral arm structure of IRS3B with high fidelity and observed IRS3B-c, the tertiary, to be embedded within one of the spiral arms. Furthermore, a possible symmetric inner disk and inner depression is marginally resolved near IRS3B-ab. IRS3B-b may be a high density clump just outside of the inner disk. We also marginally resolve possible spiral substructure in the disk of IRS3A. We calculate the mass of the disk surrounding IRS3B to be \ab0.29~\solm\space with \ab0.07~\solm\space surrounding the tertiary companion, IRS3B-c. IRS3A has a disk mass of \ab0.04~\solm. \explain{merged items 1 and 2 because they both are referring to the continuum observations.}
    \item We found that the \cso\space emission is indicative of Keplerian rotation at the scale of the continuum disk, and fit a central mass of 1.15$^{+0.09}_{-0.09}$~\solm\space for IRS3B using a fit to the PV diagram. \htcop\space traces the larger structure, corresponding to the outer disk and inner envelope for IRS3B. Meanwhile, the \htcn/\sot\space blended line most likely reflects \sot\space emission, tracing outflow launch locations near IRS3B-c. The \pdspy\space modeling of IRS3B finds a mass of $1.19^{+0.13}_{-0.07}$~\solm, comparable to the PV diagram fit of 1.15$^{+0.09}_{-0.09}$~\solm.
    \item \added{We find that the tertiary companion is forming a central protostar that is less than 0.2~\msun. This upper limit is based on its lack of significant disturbance of the disk kinematics. Moreover, we find that there is a jet originating from the clump, confirming that a protostar is present.} \deleted{Additionally, for the tertiary companion IRS3B-c, we find that the central forming protostar is likely less than 0.2~\solm\space to be consistent with the observed kinematics and lack of disturbance to the kinematics of the gas disk.}
    \item For IRS3A, the \htcn/\sot\space emission likely reflects \htcn\space emission due to a consistent velocity with \cso. \htcn\space emission indicates Keplerian rotation at the scale of the continuum disk corresponding to a central mass of 1.4~\solm. The molecular line, \cso, is also detected but is much fainter in the source but consistent with a central mass results of 1.4~\solm. The \pdspy\space modeling fit for IRS3A yields mass $1.51^{+0.06}_{-0.07}$~\solm\space which is also comparable to the PV diagram estimate of 1.4~\solm.
    \item The azimuthally averaged radial surface density profiles enable us to analyze the gravitational stability as a function of radius for the disks of IRS3B and IRS3A. We find the circum-multiple disk of IRS3B is gravitationally unstable (Q $<$ 1) for radii $>$ 120~au. We find the protostellar disk of IRS3A is gravitationally stable (Q $>$5) for the entire disk. We marginally detect substructure in IRS3A, but at our resolution, we cannot definitely differentiate between spiral structure and a gap in the disk. If the substructure is spiral arms due to gravitational instabilities, then the disk mass must be underestimated by a factor of 2-4 from our Toomre~Q analysis.
\end{enumerate}

Through the presented analysis, we determine the most probable formation pathway for the IRS3B and its spiral structure, is through the self-gravity \added{and fragmentation of its} \deleted{ of the} massive disk. The larger IRS3A/B system (including the even wider companion L1448 NW) likely formed via turbulent fragmentation of the core during the early core collapse, as evidenced by the nearly orthogonal disk orientation and different system velocity for IRS3A and IRS3B.

\acknowledgments{
We thank the anonymous reviewer for helpful comments. \added{N.R. and J.T. acknowledge funding from NSF grant AST-1814762. This work is supported in part by NSF AST-1910364. Z.L. is supported in part by NSF AST-1716259 and NASA 80NSSC20K0533. KMK acknowledges support from NASA Grant 80NSSC18K0726. P.D.S acknowledges support from NSF AST-2001830.} This paper makes use of the following ALMA data: 2016.1.01520.S. ALMA is a partnership of ESO (representing its member states), NSF (USA) and NINS (Japan), together with NRC (Canada), MOST and ASIAA (Taiwan), and KASI (Republic of Korea), in cooperation with the Republic of Chile. The Joint ALMA Observatory is operated by ESO, AUI/NRAO and NAOJ. The National Radio Astronomy Observatory is a facility of the National Science Foundation operated under cooperative agreement by Associated Universities, Inc. The computing for this project was performed at the OU Supercomputing Center for Education \& Research (OSCER) at the University of Oklahoma (OU). This research has made use of NASA's Astrophysics Data System. This research made use of APLpy, an open-source plotting package for Python.}

\facilities{ALMA}

\software{
Numpy~\citep[][]{numpy},
scipy~\citep[][]{2019arXiv190710121V},
emcee~\citep[][]{2013PASP..125..306F},
Matplotlib~\citep[][]{Hunter2007},
Astropy~\citep[][]{astropy2013, astropy2018},
APLpy~\citep[][]{2012ascl.soft08017R},
pdspy~\citep[][]{2019ApJ...874..136S},
CASA~\citep[][]{2007ASPC..376..127M}
}
\clearpage
\begin{small}
\bibliographystyle{apj}

\end{small}
\clearpage

\begin{deluxetable}{cccccccc}
\tablewidth{0pt}
\tabletypesize{\scriptsize}
\tablecaption{Summary of Observations}
\tablehead{
  \colhead{Source} & \colhead{RA}      & \colhead{Dec}      &\colhead{Config.\tablenotemark{a}}  & \colhead{Resolution} &\colhead{LAS\tablenotemark{b}} & \colhead{Date} & \colhead{Calibrators}\\
                   & \colhead{(J2000)} &  \colhead{(J2000)} &                   &                      &              & \colhead{(UT)} &   \colhead{(Gain, Bandpass, Flux)} \\
}
\startdata
L1448 IRS3B        & 03:25:36.382 & 30:45:14.715 & C40-6 & 0\farcs12 & 1\farcs3  & 1 and 4 October 2016 & J0336$+$3218,J0237$+$2848,J0238$+$1636    \\
L1448 IRS3B        & 03:25:36.382 & 30:45:14.715 & C40-3 & 0\farcs59 & 5\farcs6 & 19 December 2016     &  J0336$+$3218,J0237$+$2848,J0238$+$1636 \\
\enddata
\tablenotetext{a}{C40-6 - Extended and C40-3 - Compact} 
\tablenotetext{b}{LAS- Largest Angular Scale} 
\end{deluxetable}\label{table:obssummary1}

\begin{deluxetable}{cccccccc}
\tablewidth{0pt}
\tabletypesize{\scriptsize}
\tablecaption{Continuum and Spectral Line Data}
\tablehead{
  \colhead{} & \colhead{MFS Continuum\tablenotemark{a}} & \colhead{\co} & \colhead{\sio\tablenotemark{b}} & \colhead{\htcn/\sot\tablenotemark{c}} & \colhead{\htcop} & \colhead{\cso}  & \colhead{335.5GHz Continuum} \\
}
\startdata
 Rest. Freq. (GHz)   & 341.0     & 346.0             & 347.000030579  & 345.339756     & 346.998347     & 337.061104      & 335.5       \\
 Center Freq. (GHz)  & 341.0     & 346.778059        & 347.2698586    & 345.3520738    & 347.010582     & 337.0730133     & 335.4708304 \\
 Chan. Width (km/s)  & 2747.96   & 0.212             & \replaced{0.053}{0.210}          & 0.053          & 0.053          & 0.054           & 0.873       \\
 Num. Chan.          & 1         & 1920              & 1920           & 1920           & 1920           & 3840            & 1920        \\
 RMS/chan. (mJy)     & 0.069     & 4.0               & 0.5            & 4.5            & 4.5            & 3.7             & -           \\
 Integr. (Jy) IRS3B\tablenotemark{d}  & 1.5       & 512.7,809.8\tablenotemark{e} & 7.8, 3.5\tablenotemark{e} & 0.2, 0.1\tablenotemark{e} & 2.6, 4.1\tablenotemark{e}& 2.9, 3.2\tablenotemark{e} & -           \\
 Integr. (Jy) IRS3A\tablenotemark{d}  & 0.2       & 3.16,67.6\tablenotemark{e}   & 0,0 \tablenotemark{e}     & 2.3, 2.7\tablenotemark{e} & 0.2, 0.4\tablenotemark{e} & 0.1, 0.1\tablenotemark{e} & -           \\
 Synth. Beam\tablenotemark{f}         & \contbeam & \cobeam           & \siobeam       & \htcnbeam      & 0\farcs21$\times$0\farcs13     & \csobeam        & \tttfbeam   \\
 Briggs Robust       & 0.5       & 0.5               & 0.5            & 0.5            & 0.5            & 0.5             & -           \\
 Taper (k$\lambda$)  & -         & -                 & -              & 1000           & 1500            & 1500            & -           \\
\enddata 
\tablecomments{The setup of the correlator for the observations}
\tablenotetext{a}{Multi-Frequency Synthesis (MFS) utilizing the extracted emission from line free spectral channels}
\tablenotetext{b}{\sio\space was tuned incorrectly for the C40-6 observations.}
\tablenotetext{c}{The \htcn\space line is blended with the \sot\space line (345.3385377~GHz) and have a velocity separation of \ab1.06~km~s$^{-1}$.}
\tablenotetext{d}{The integrated flux density for the source, measured by integrated the full emission whose origin is the source. In the case of continuum emission, this is given in \textit{Jy}; in the case of molecular line emission this is given in \textit{Jy~km~s$^{-1}$}.}
\tablenotetext{e}{The molecular line emission is given as the total integrated flux (Jy~km~s$^{-1}$) for the blue and red-Doppler shifted emission, denote blue, red, respectively.}
\tablenotetext{f}{The synthesized beam size is provided from the \textit{clean} task for the molecular lines (or continuum for the MFS column) using the briggs robust weighing parameter of 0.5 during image reconstruction.}
\end{deluxetable}\label{table:obssummary2}

\movetabledown=2in
\begin{splitdeluxetable}{cccccccBccccccccc}
\rotate
\tablewidth{0pt}
\tabletypesize{\scriptsize}
\tablecaption{Source Properties}
\tablehead{
 \colhead{Source} & \colhead{RA}      & \colhead{Dec}     & \colhead{Inc.\tablenotemark{a}} & \colhead{P.A.\tablenotemark{b}}  & \colhead{Outflow?} & \colhead{V$_{sys}$}      & \colhead{L$_{bol}$}  & \colhead{M$_{dust}$}  & \colhead{FWHM$_{Dust}$\tablenotemark{c} Major Axis} & \colhead{FWHM$_{Dust}$\tablenotemark{c} Minor Axis} & \colhead{FWHM$_{Gas}$\tablenotemark{c} Major Axis} & \colhead{FWHM$_{Gas}$\tablenotemark{c} Minor Axis}  & \colhead{$<$T$_{0}>$} & \colhead{$<$Optical Depth$>$} \\
                  & \colhead{(J2000)} & \colhead{(J2000)} & \colhead{(\deg)}                & \colhead{(\deg)}                 &                    & \colhead{(km~s$^{-1}$)}  & \colhead{(\lsun)}    & \colhead{(\solm)}                             & \colhead{(\arcsec, au)}                 & \colhead{(\arcsec, au)} & \colhead{(\arcsec, au)}                 & \colhead{(\arcsec, au)}   & \colhead{(K)} & & \\
}
\startdata
 IRS3B-ab & 03:25:36.317 & 30:45:15.005 & 45 & 28  & Joint & 4.75 & 13.0\tablenotemark{e}  & 0.29 & 1.73$\pm$0.05, 498$\pm$14 & 1.22$\pm$0.04, 351$\pm$12 & 2.38$\pm$0.09, 685$\pm$26 & 2.25$\pm$0.08, 648$\pm$23 & 40 & 0.34 \\
 IRS3B-c  & 03:25:36.382 & 30:45:14.715 & 27 & 21  & Yes   & 4.75 & -\tablenotemark{d}     & 0.07 & 0.28$\pm$0.05, 81$\pm$14  & 0.25$\pm$0.04, 72$\pm$12  & -                         & -                         & 55 & 2.14 \\
 IRS3A    & 03:25:36.502 & 30:45:21.859 & 69 & 133 & No    & 5.2  & 14.4\tablenotemark{e}  & 0.04 & 0.70$\pm$0.02, 202$\pm$6  & 0.25$\pm$0.01, 72$\pm$3   & 0.52$\pm$0.08, 150$\pm$23 & 0.42$\pm$0.07, 121$\pm$20 & 51 & 0.57 \\
\enddata
\tablecomments{Summary of the empirical parameters based from the observations of the system. The sizes were derived from a 2-D Gaussian fit to the continuum and moment 0 emission maps, directly to the visibilities. IRS3B-c is blended with the underlying disk continuum and estimates here are extracted from a 2-D gaussian fit with a zero-level offset to preserve the underlying disk flux and is discussed in Appendix~\ref{sec:tertsub}.}
\tablenotetext{a}{Inclination is defined such that 0\deg\space is a face-on disk.}
\tablenotetext{b}{Position angle is defined such that at 0\deg, the major axis of the disk is aligned North and the angle corresponds to East-of-North.}
\tablenotetext{c}{The circumstellar disks surround IRS3B and IRS3A are ellipsoidal in the dust continuum and molecular line emission.}
\tablenotetext{d}{The bolometric luminosity is not known at this time.}
\tablenotetext{e}{The bolometric luminosity is scaled to a distance of 288~pc\space from \citet{2016ApJ...818...73T}.}
\end{splitdeluxetable}\label{table:obssummary3}

\explain{Cut IRS33B-ab and merged with IRS3B for clarity.}
\begin{deluxetable}{ccccccc}
\tablewidth{0pt}
\tabletypesize{\scriptsize}
\tablecaption{PV Diagram Fitting}
\tablehead{
  \colhead{Source} & \colhead{Center RA} & \colhead{Center Dec} & \colhead{Inclination} & \colhead{Position Angle} & \colhead{Stellar Mass} & \colhead{Velocity} \\
  & \colhead{(\arcsec)} & \colhead{(\arcsec)} & \colhead{(\deg)} & \colhead{(\deg)} & \colhead{(\solm)} & \colhead{(km~s$^{-1}$)}\\
}
\startdata
IRS3B    & 03$^{h}$25$^{m}$36.317$^{s}$ & 30\deg45\arcmin15\farcs005 & 45 & 29 & 1.15$^{+0.09}_{-0.09}$ & 4.8\\
IRS3B-c  & 03$^{h}$25$^{m}$36.382$^{s}$ & 30\deg45\arcmin14\farcs715 & - & - & $<$0.2\tablenotemark{a} & - \\
IRS3A    & 03$^{h}$25$^{m}$36.502$^{s}$ & 30\deg45\arcmin21\farcs859 & 69 & 125 & 1.4\tablenotemark{b}& 5.4\\
\enddata
\tablecomments{Summary of PV diagram stellar parameter estimates with 3-$\sigma$\space confidence interval of the best fit walkers generated from emcee. The inclination and position angle estimates are provided by 2-D Gaussian fitting of the uv-truncated data and is further confirmed with the PV diagram analysis.}
\tablenotetext{a}{\replaced{IRS3B-c, was only marginally constrained through the PV diagram analysis. The source dust component, while optically thick, is estimated to be no more than 0.2~\solm\space to be consistent with the data.}{The upper limit for IRS3B-c of $<$0.2~\msun\space is derived from its apparent lack of significant influence on the disk kinematics within its immediate proximity. Furthermore, we estimate from the dust emission that the mass of the gas and dust clum surrounding the protostar is \ab0.07~\msun. So the combined mass of the clump and protostar must be $<$0.2~\msun.} \deleted{Through analyzing the \cso\space PV diagram emission and considering the gravitational potential of IRS3B-ab, we estimate the upper mass limit of the tertiary source. }Figure~\ref{fig:l1448irs3b_c17o_pv_tert} shows the mass limit estimates of the tertiary of the source, with emission outside of the dotted lines indicating additional mass if perturbing the disk.}
\tablenotetext{b}{IRS3A, was marginally resolved and no sufficient numeric fits could be achieved with simple PV-diagram fitting. These estimates are provided by fitting the curve by eye and are not designated to be the final results and simply provide further constraints for the priors for the more rigorous kinematic modeling.}
\end{deluxetable}\label{table:pvtable}

\begin{deluxetable}{cccccccccccc}
\rotate
\tablewidth{0pt}
\tabletypesize{\scriptsize}
\tablecaption{Kinematic \pdspy Modeling}
\tablehead{
  \colhead{Source} & \colhead{RA Offset} & \colhead{Dec Offset} & \colhead{Inc.} & \colhead{P.A.} & \colhead{M$_{*}$} & \colhead{M$_{gas}$\tablenotemark{a}} &\colhead{R$_{disk}$} & \colhead{V$_{sys}$}& \colhead{Turbulence}& \colhead{Surface Density Index $\gamma$} & \colhead{T$_{0}$}\\
  & \colhead{(\arcsec)} & \colhead{(\arcsec)} & \colhead{(\deg)} & \colhead{(\deg)} & \colhead{(\solm)} & \colhead{(\solm)}& \colhead{(au)} & \colhead{(km~s$^{-1}$)}& \colhead{(km~s$^{-1}$)}& & \colhead{(K)}  \\
}
\startdata
IRS3B  & $0.031^{+0.019}_{-0.011}$ & $0.025^{+0.020}_{-0.015}$ & $66.0^{+ 3.0}_{- 4.6} $ & $26.7^{+  1.8}_{-  2.9} $ & $1.19^{+0.13}_{-0.07}$ & $0.079^{+0.021}_{-0.016}$ & $299.0^{+ 24.9}_{- 47.6}$ & $4.880^{+0.110}_{-0.090}$ & $0.012^{+ 0.005}_{- 0.003}$ & $  1.2^{+  0.1}_{-  0.1}$&$   50^{+    3}_{-    5}$\\
IRS3A  & $0.034^{+0.003}_{-0.003}$ & $0.015^{+0.003}_{-0.003}$ & $69.5^{0.38}_{0.37} $  & $122.4^{1.4}_{1.4}$ & $1.51^{+0.06}_{-0.07}$ & $6.3^{+1.6}_{-1.3}x10^{-6}$ & $ 39.9^{+  2.4}_{-  1.4}$ & $5.288^{+0.090}_{-0.084}$ & $0.015^{+ 0.06}_{-0.009}$ & $0.4^{+0.2}_{-0.1}$ & $163^{+9}_{-8}$ \\
\enddata
\tablecomments{Summary of kinematic model parameters. The RA and DEC offsets of the \pdspy modeling are defined from the central positions given in PV Analysis, Table~\ref{table:pvtable}. The errors presented are the 3-$\sigma$\space confidence intervals of the best fit walkers generated from \textit{emcee}.}
\tablenotetext{a}{\added{The reported values of M$_{gas}$\space depend on the assumed abundance for each of the molecules. For the IRS3B source, we used the \cso\space emission, which has an assumed abundance of $2\times10^{-7}$ relative to H$_{2}$, while for the IRS3A source we used the \htcn\space emission which has an assumed abundance of $2.9\times10^{-11}$ relative to H$_{2}$.}}
\end{deluxetable}\label{table:pdspykinematic}

\movetabledown=2in
\begin{deluxetable}{cccccc}
\rotate
\tablewidth{0pt}
\tabletypesize{\scriptsize}
\tablecaption{Mass Accretion}
\tablehead{
 \colhead{Source} & \colhead{L$_{bol}$} & \colhead{M$_{*}$} & \colhead{R$_{*}$} & \colhead{L$_{*}$} & \colhead{$\dot{M}_{acc}$} \\
                  & \colhead{(\lsun)}   & \colhead{(\solm)} & \colhead{(R$_{\odot}$)} & \colhead{(L$_{\odot}$)}& \colhead{(10$^{-7}$~M$_{\odot}$~yr$^{-1}$)} \\
}
\startdata
 IRS3B-ab\tablenotemark{a} & 13.0\tablenotemark{b}  & (0.575, 1.2) & (2.5, 2.5) & (1.91, 3.57) & (15.3, 6.56) \\
 IRS3A                     & 14.4\tablenotemark{b}  & 1.5 & 2 & 2.53 & 5.43 \\
\enddata
\tablecomments{Summary of the derived parameters from \citet{1997ApJ...475..770H}\space to estimate the amount of mass accretion that is consistent with protostellar models and the observations. The methodology for estimating R$_{*}$, L$_{*}$, $\dot{M}$, and M$_{accreted}$\space are provided in Section~\ref{sec:massacc}.}
\tablenotetext{a}{When constraining R$_{*}$, L$_{*}$, $\dot{M}$, and M$_{accreted}$\space, IRS3B can be analyzed at two scenarios, 1.) equally mass binary and 2.) one protostar with most of the mass; we reference these delineations as (equal mass, single massive protostar), respectively.}
\tablenotetext{b}{The bolometric luminosity is scaled to a distance of 288~pc\space from \citet{2016ApJ...818...73T}.}
\end{deluxetable}\label{table:massacc}
 
\movetabledown=2in
\begin{deluxetable}{lcrrcc}
\rotate
\tablewidth{0pt}
\tabletypesize{\scriptsize}
\tablecaption{Self-Calibration}
\tablehead{
 \colhead{Step} & \colhead{RMS} & \colhead{IRS3B S/N} & \colhead{IRS3A S/N} & \colhead{Iterations} & \colhead{Solution Integration} \\
                  & \colhead{(mJy~beam$^{-1}$)}   & \colhead{} & \colhead{} & \colhead{}& \colhead{(s)} \\
}
\startdata
No-selfcal.  & 6.5 | 74  & 82   | 43  & 26  | 13  & 100  | 100 & \\
phase-cal. 1 & 4.2 | 25  & 140  | 140 & 48  | 40  & 100  | 110 & ``inf'' \\
phase-cal. 2 & 1.7 | 11  & 310  | 330 & 120 | 100 & 300  | 500 & 30.25 \\
phase-cal. 3 & 1.3 | 5.8 & 540  | 620 & 200 | 190 & 3000 | 1500& 12.1 \\
ampl.-cal.   & 0.7 | 4.3 & 1000 | 840 & 390 | 260 & 2500 | 2500 & ``inf''\\
\enddata
\tablecomments{Summary of the parameters required to reproduced the gain and amplitude self-calibrations. The configurations are delineated as C40-6 | C40-3, respectively in the table. ``inf'' indicates the entire scan length, dictated by the time on a single pointing, which is typically 6.05 seconds.}
\end{deluxetable}\label{table:selfcal}

\clearpage

\begin{figure}[H]
\begin{center}
 \begin{minipage}{0.49\textwidth}
  \includegraphics[width=\textwidth]{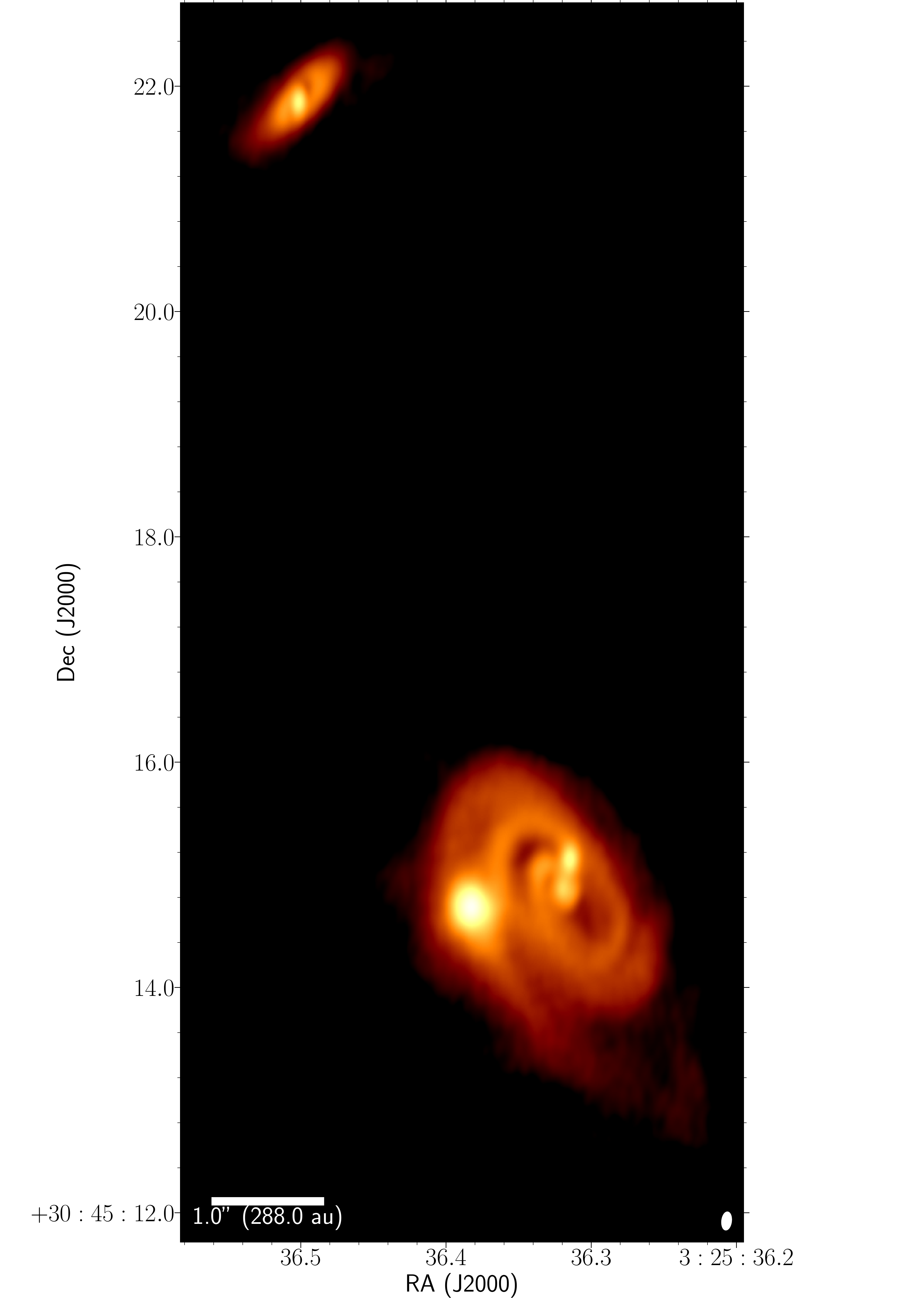}
 \end{minipage}
 \begin{minipage}{0.49\textwidth}
  \vfill
  \includegraphics[width=\textwidth]{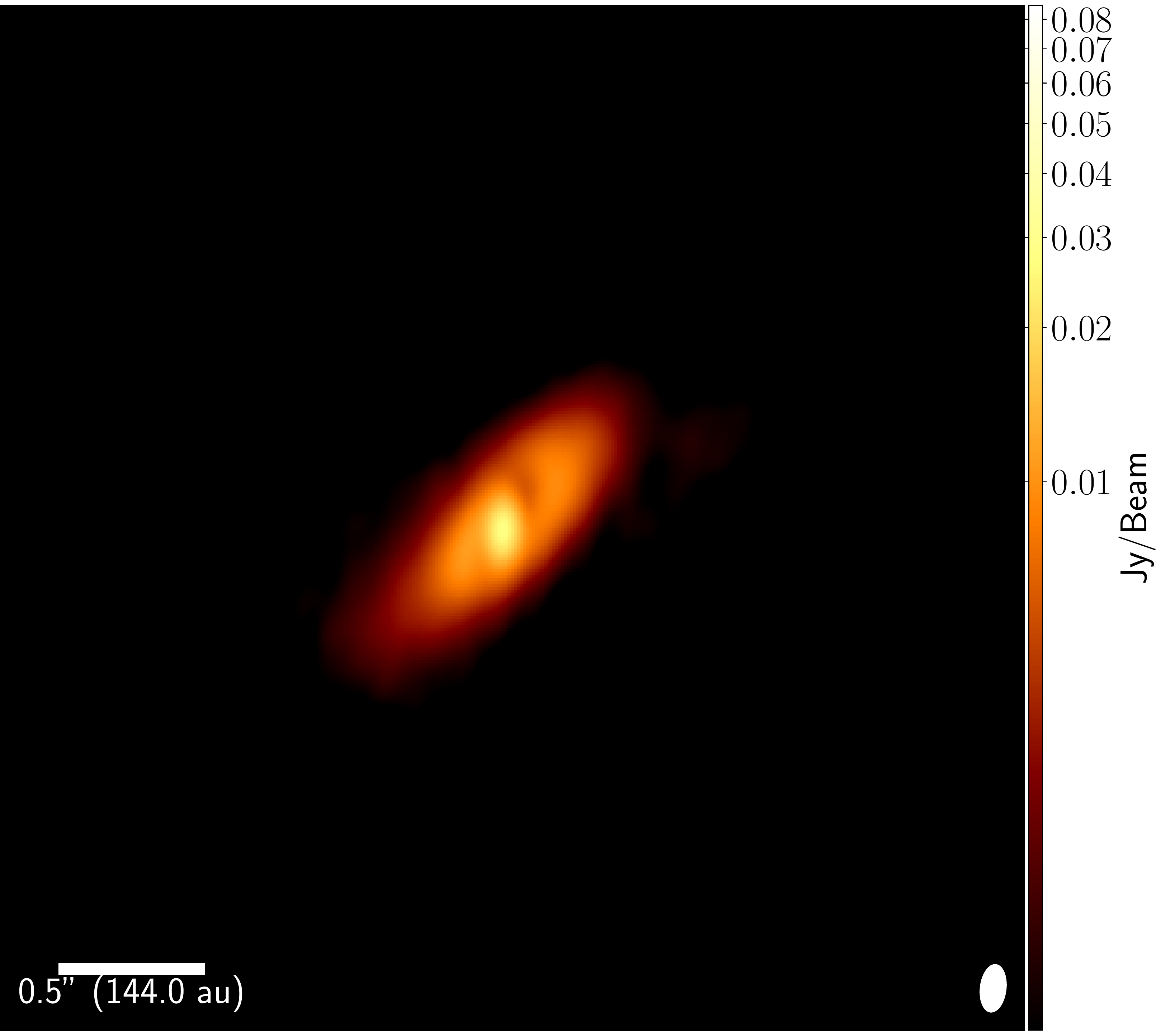}
  \includegraphics[width=\textwidth]{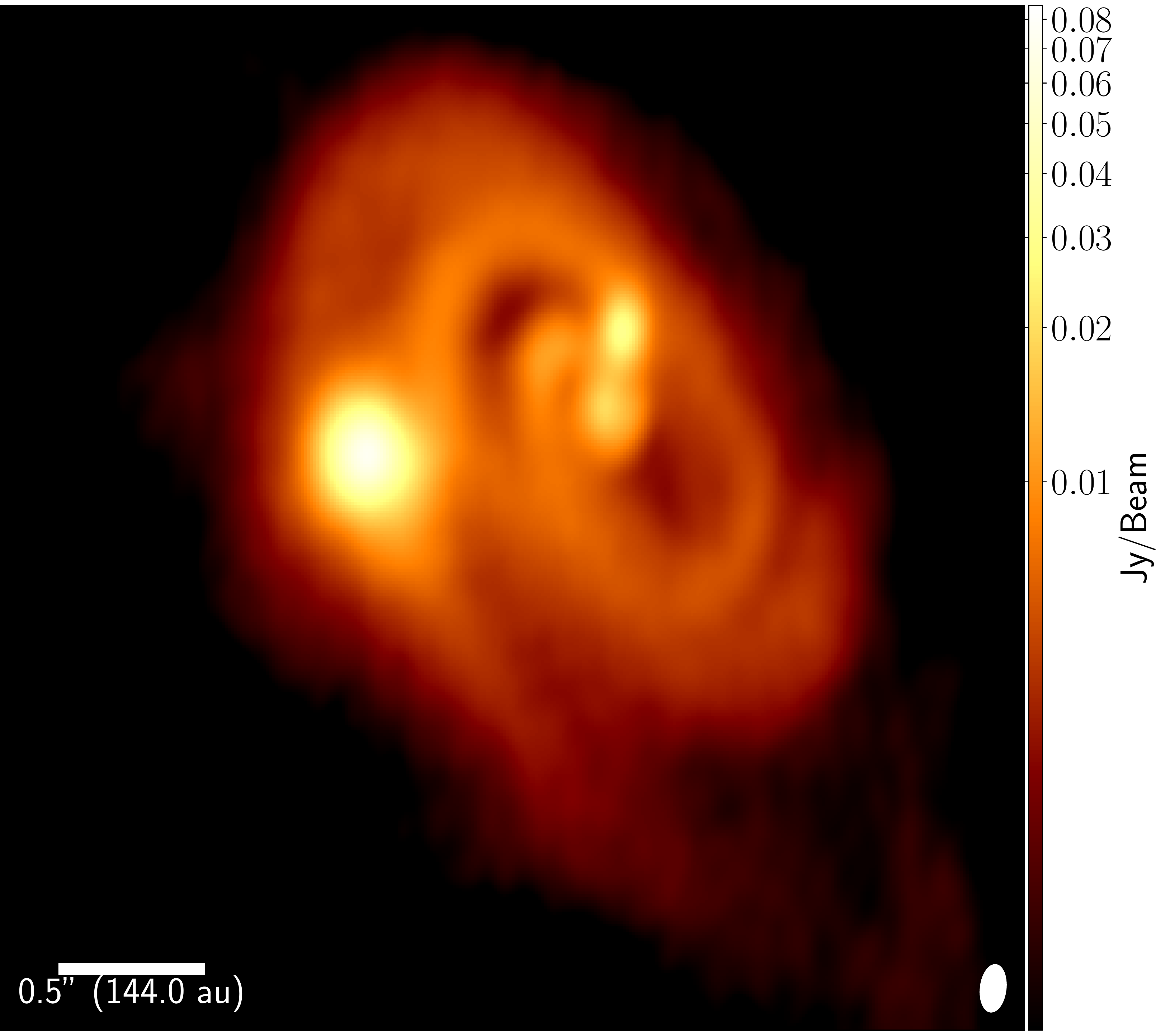}
  \vfill
 \end{minipage}
\end{center}
\caption{ALMA 879~\micron\space continuum observations of the triple protostellar system L1448 IRS3B and its wide companion IRS3A (\textbf{left}). The right panels are \ab2$\times$\space zoom-ins on IRS3B~and~IRS3A. The top right image shows the wide companion, IRS3A, (d\ab7\farcs9$\approx$2300~au), featuring possible spiral structure. The bottom right image zooms in on the proto-multiple system, IRS3B. The inner binary is separated by 0\farcs25 (75~au) and has a spiral circum-binary disk with the embedded source \ab0\farcs8 (230~au) away from the binary within one of the arms. The beam size of each panel is shown in lower right (\contbeam).}\label{fig:contimage}
\end{figure}

\begin{figure}[H]
\begin{center}
   \includegraphics[width=0.48\textwidth]{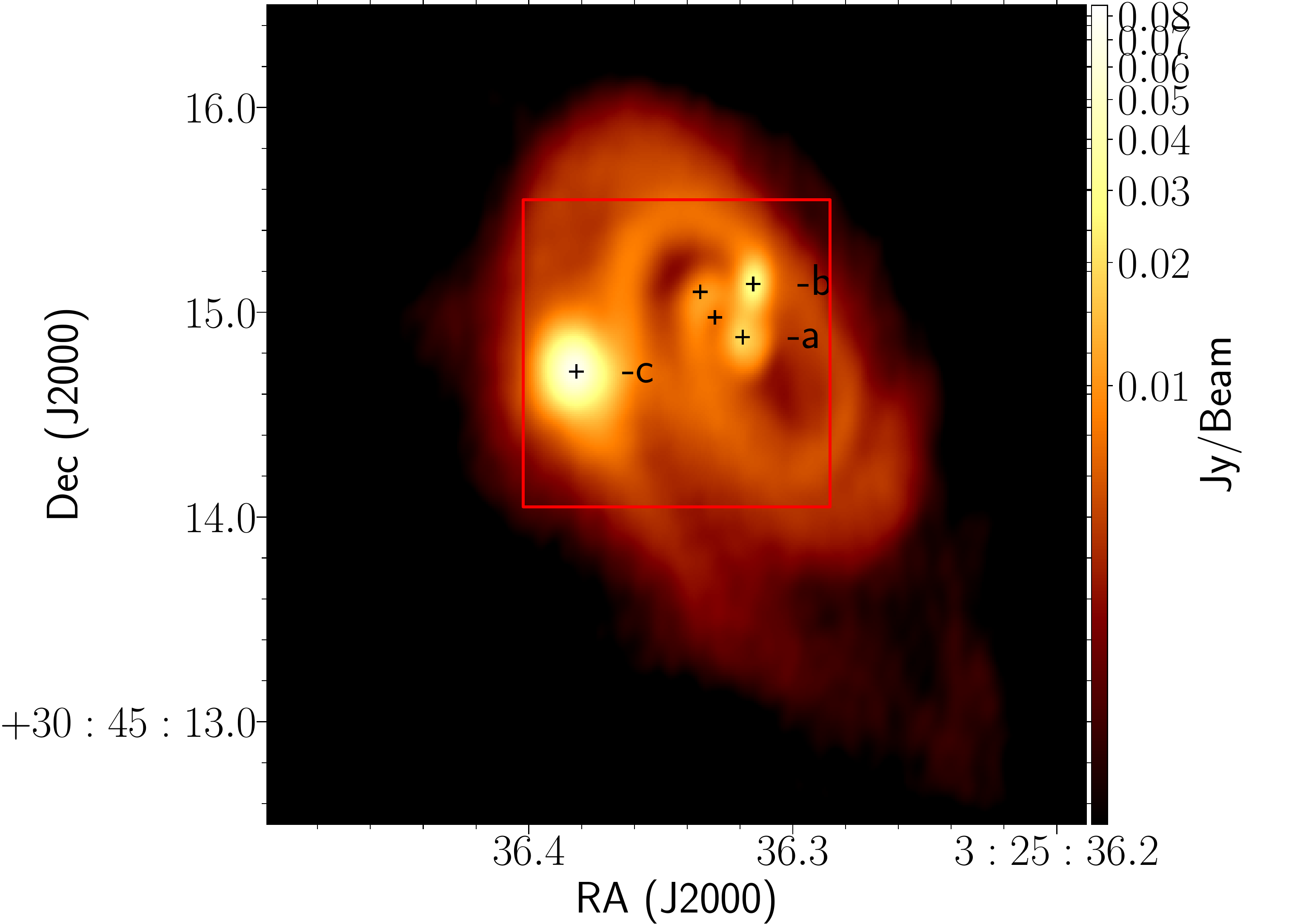}
   \includegraphics[width=0.48\textwidth]{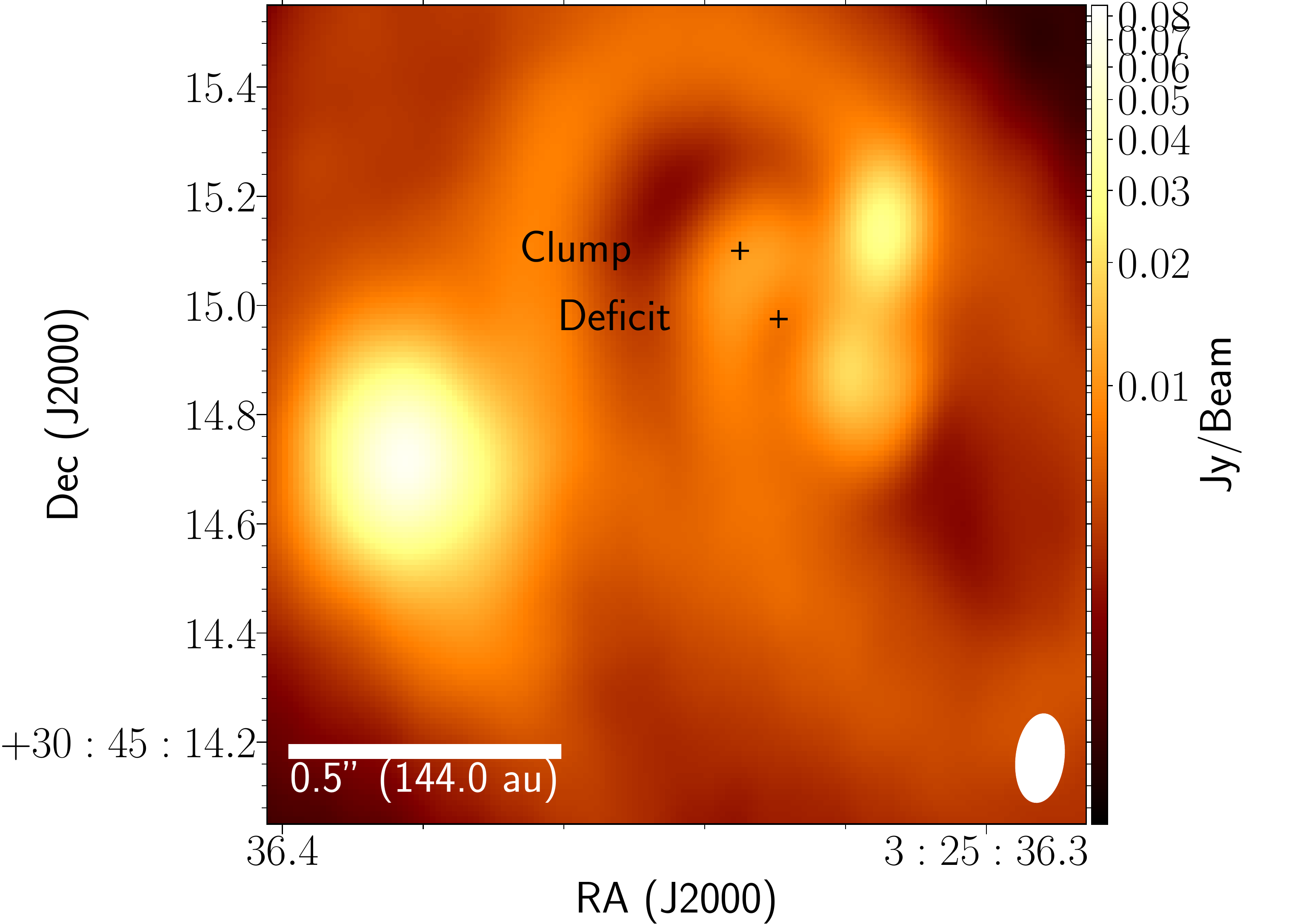}
\end{center}
   \caption{ALMA 879~\micron\space continuum observations of the triple protostellar system L1448 IRS3B\added{ with the difference continuum sources marked}. The left colored image is zoomed in on IRS3B and is plotting with a log color stretch. The inner binary is separated by 0\farcs25 (75~AU) and has a circum-binary disk with spiral structure and the tertiary is separated from the binary by \ab0\farcs8 (230~AU) within one of the arms. The ``protostars'' are the continuum positions previously discovered in \citet{2016Natur.538..483T}, while the ``clump'' is a new feature, resolved in these observations. The ``deficit'' indicates the location of depression of flux between IRS3B-a and the ``clump''. This is discussed in Sections~\ref{sec:dcont}~and~\ref{sec:discussion}. The beam size of each panel is shown in lower right (\contbeam\space using Briggs Robust parameter of 0.5).}\label{fig:zoomincont}
\end{figure}

\begin{figure}[H]
  \begin{center}
   \includegraphics[width=0.48\textwidth]{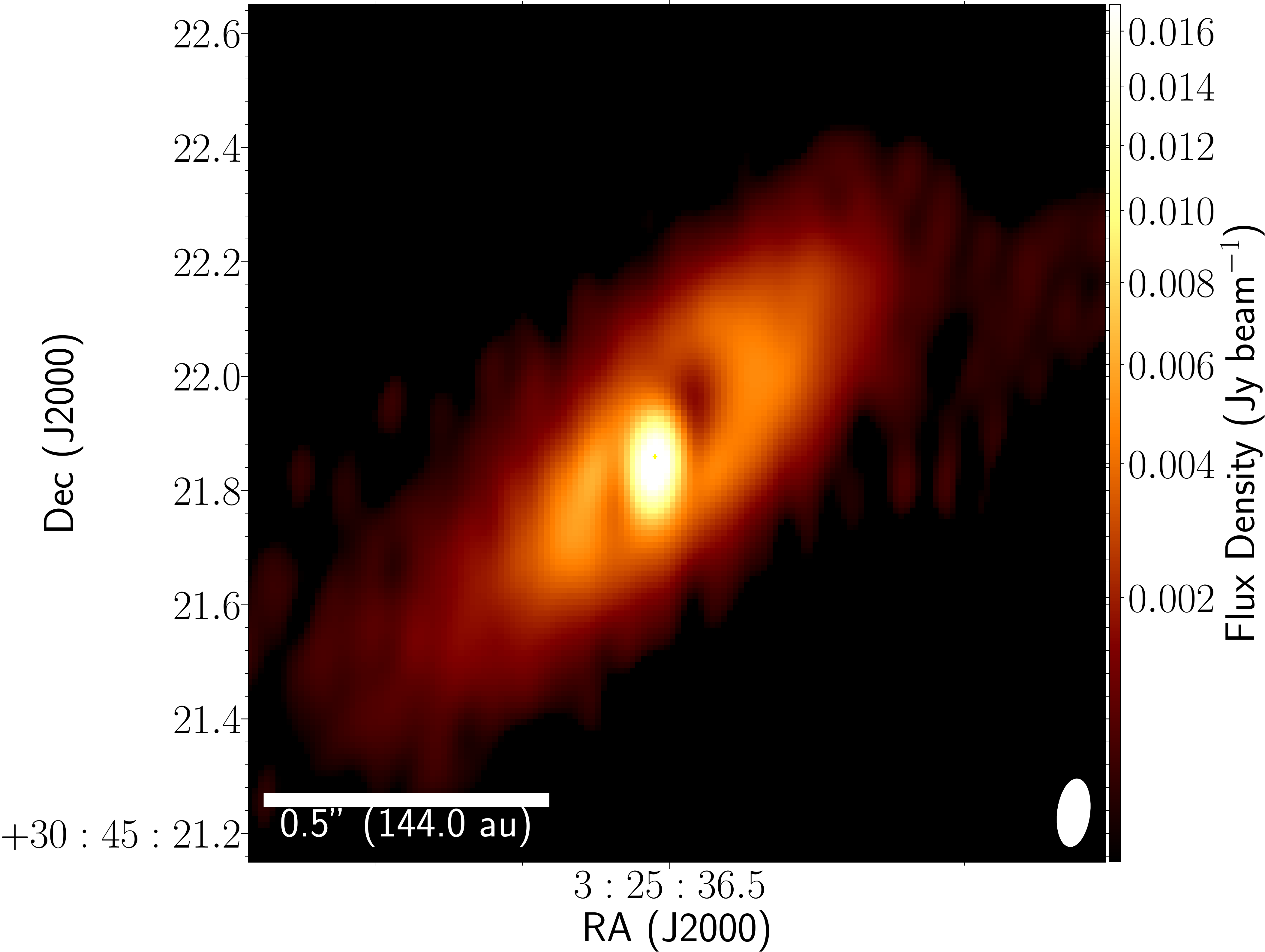}
   \end{center}
   \caption{Continuum (879~\micron) image of IRS3A, reconstructed with the \textit{superuniform} weighing scheme, half of the cell size, and zoomed 2x from the images in Figure~\ref{fig:contimage}\space to highlight the possible spiral substructure.} \label{fig:widesuperuniform}
\end{figure}

\begin{figure}[H]
\begin{center}
   \includegraphics[width=0.45\textwidth]{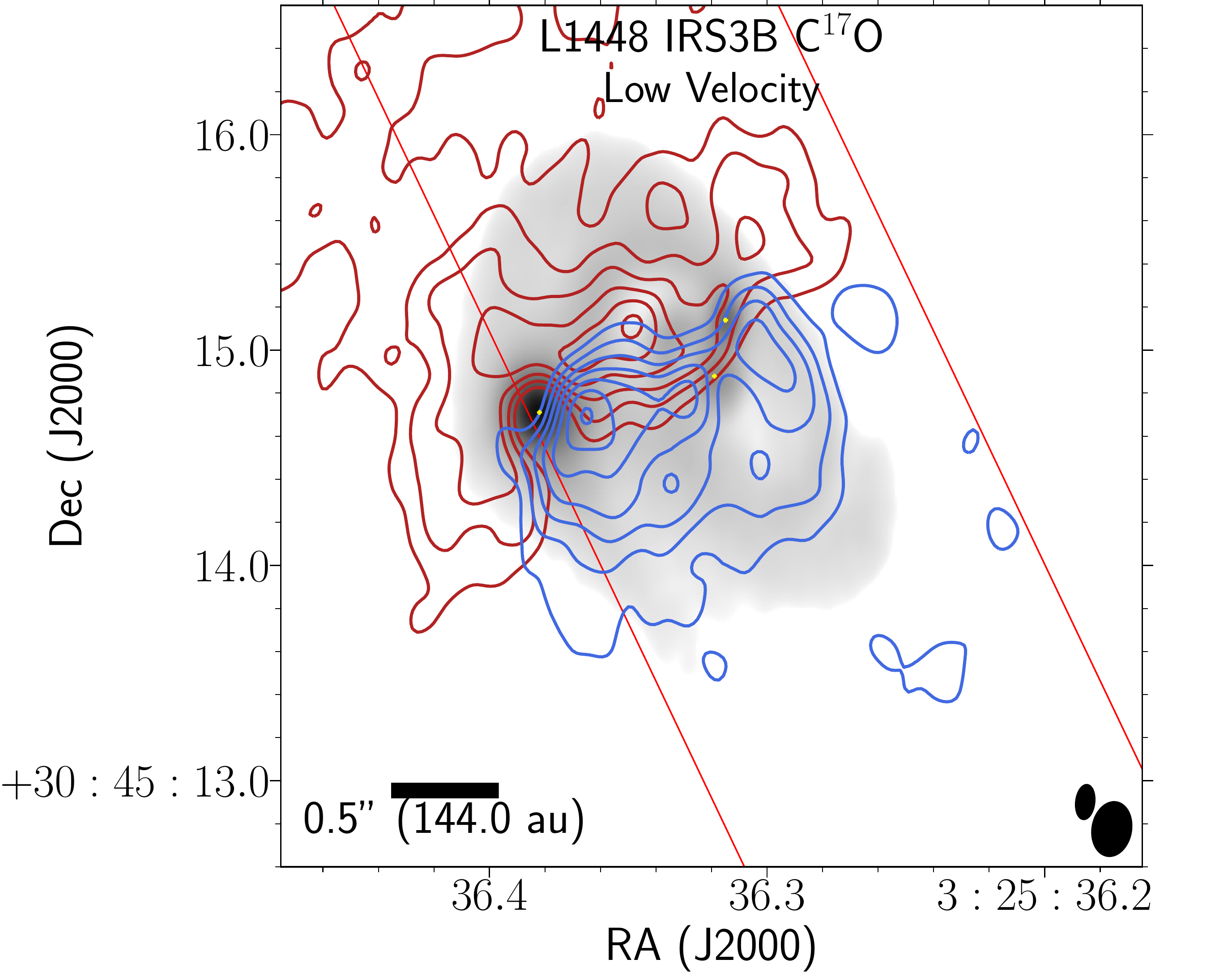} 
   \includegraphics[width=0.45\textwidth]{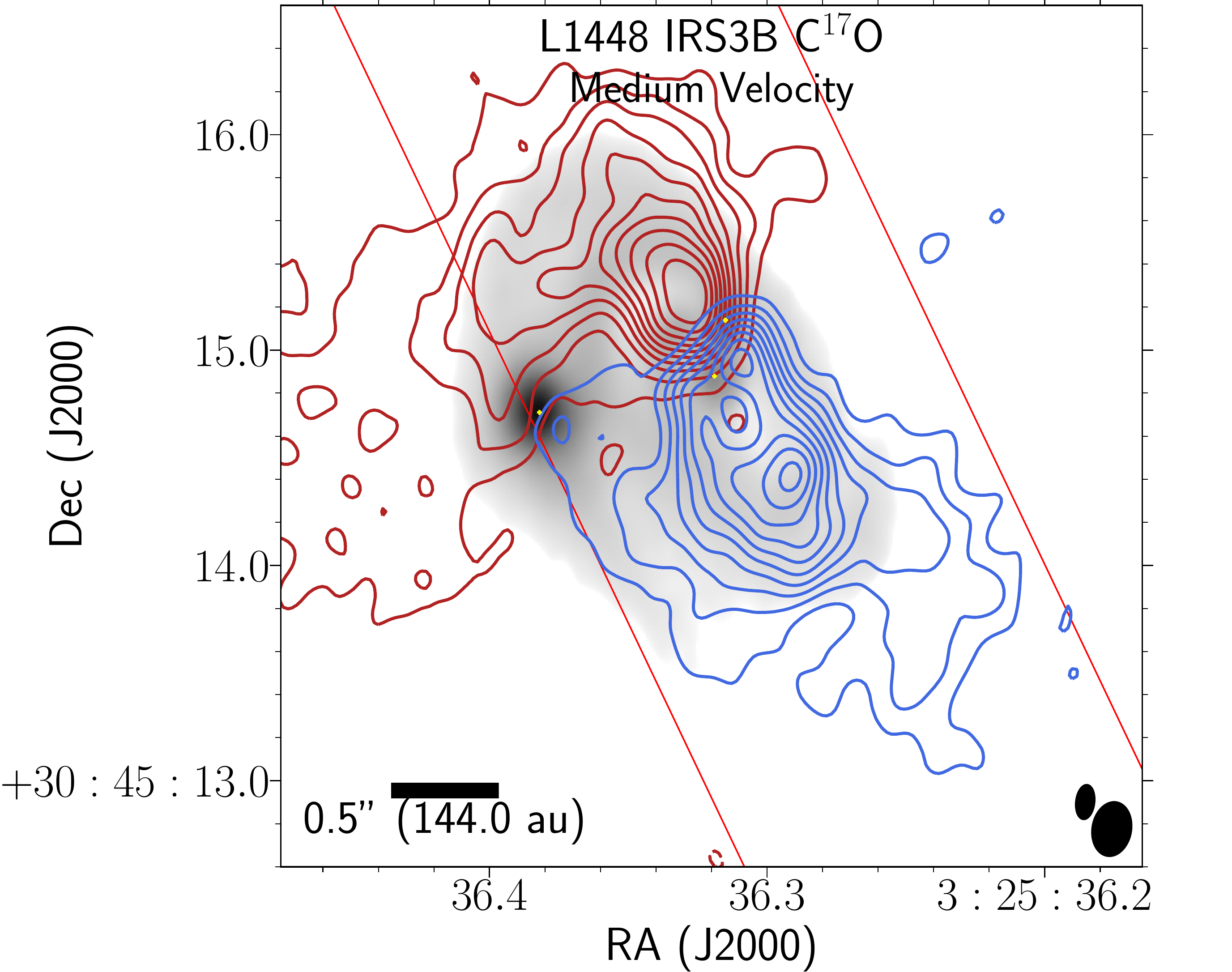} 
   \includegraphics[width=0.45\textwidth]{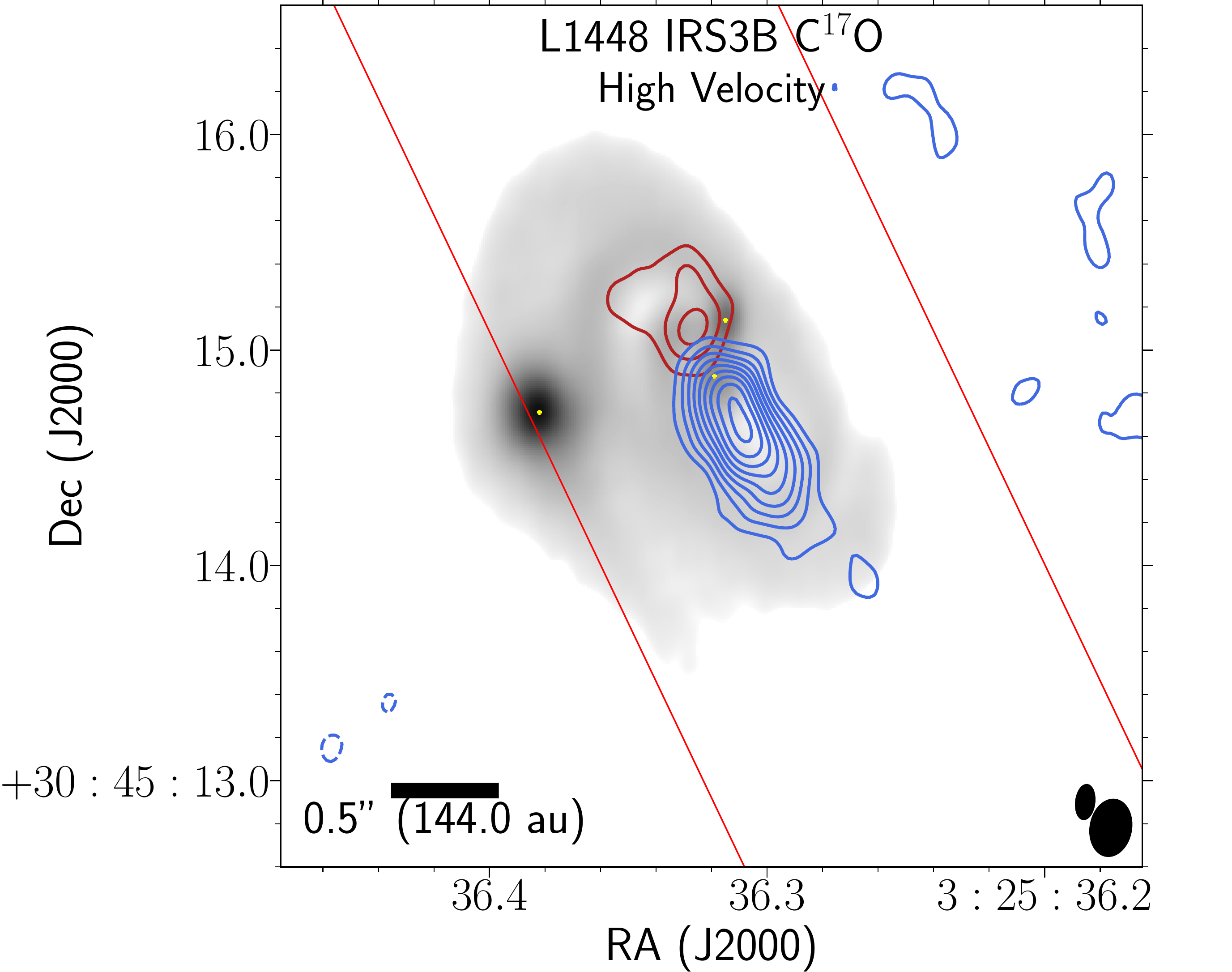} 
\end{center}
   \caption{\cso\space integrated intensity maps towards IRS3B over a selected range of velocities overlayed on continuum (grayscale). The \cso\space emission traces the rotating gas within the disk via Doppler-shifted emission. The panels correspond to low, medium, and high velocity ranges which are delineated as red(blue), respectively. Negative contours are not present in these integrated intensity maps; however, at the location of IRS3B-c, there is strong absorption that is evident in the high spectral resolution data cube, but is not represented here. The red lines indicate the region extracted for PV diagram construction, along the position angle of the major axis. \textbf{Low Velocity:} Velocity range starts at 4.68$\rightarrow$5.67~\kms (3.58$\rightarrow$4.68~\kms) and contours start at 8(8)$\sigma$ and iterate by 3(3)$\sigma$ with the 1$\sigma$~level starting at 0.0023(0.0025)~Jy~beam$^{-1}$ for the red(blue) channels respectively. \textbf{Medium Velocity:} Velocity range starts at 5.67$\rightarrow$6.66~\kms (2.48$\rightarrow$3.58~\kms) and contours start at 3(5)$\sigma$ and iterate by 3(3)$\sigma$ with the 1$\sigma$~level starting at 0.002(0.0016)~Jy~beam$^{-1}$ for the red(blue) channels respectively. \textbf{High Velocity:} Velocity range starts at 6.66$\rightarrow$7.65~\kms (1.27$\rightarrow$2.48~\kms) and contours start at 5(5)$\sigma$ and iterate by 3(3)$\sigma$ with the 1$\sigma$~level starting at 0.0018(0.0012)~Jy~beam$^{-1}$ for the red(blue) channels respectively. The \cso\space synthesized beam (\csobeam) is the bottom-right most ellipse on each of the panels and the continuum synthesized beam (\contbeam) is offset diagonally.}\label{fig:irs3bc17omoment}
\end{figure}

\begin{figure}[H]
\begin{center}
   \includegraphics[width=0.45\textwidth]{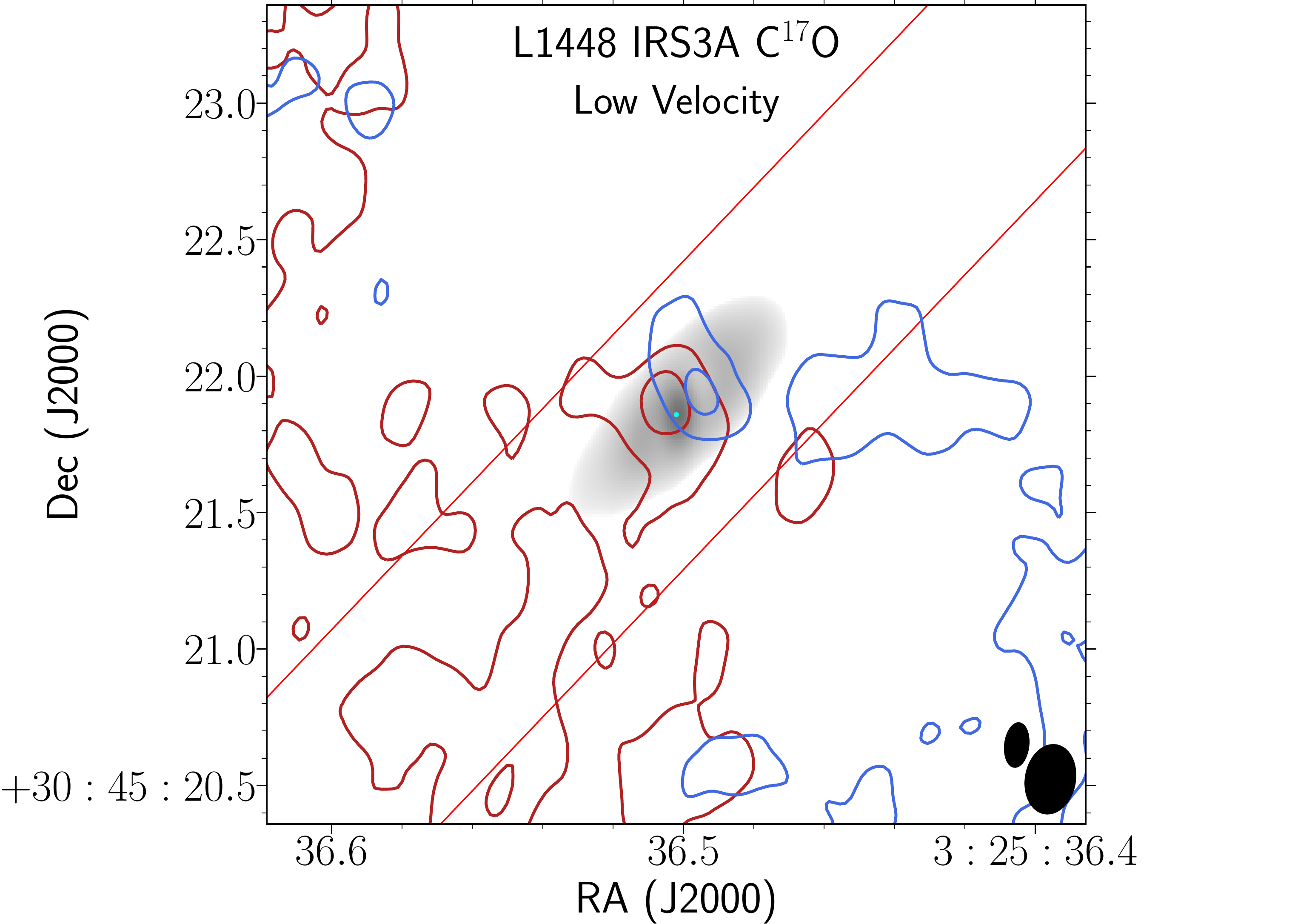}
   \includegraphics[width=0.45\textwidth]{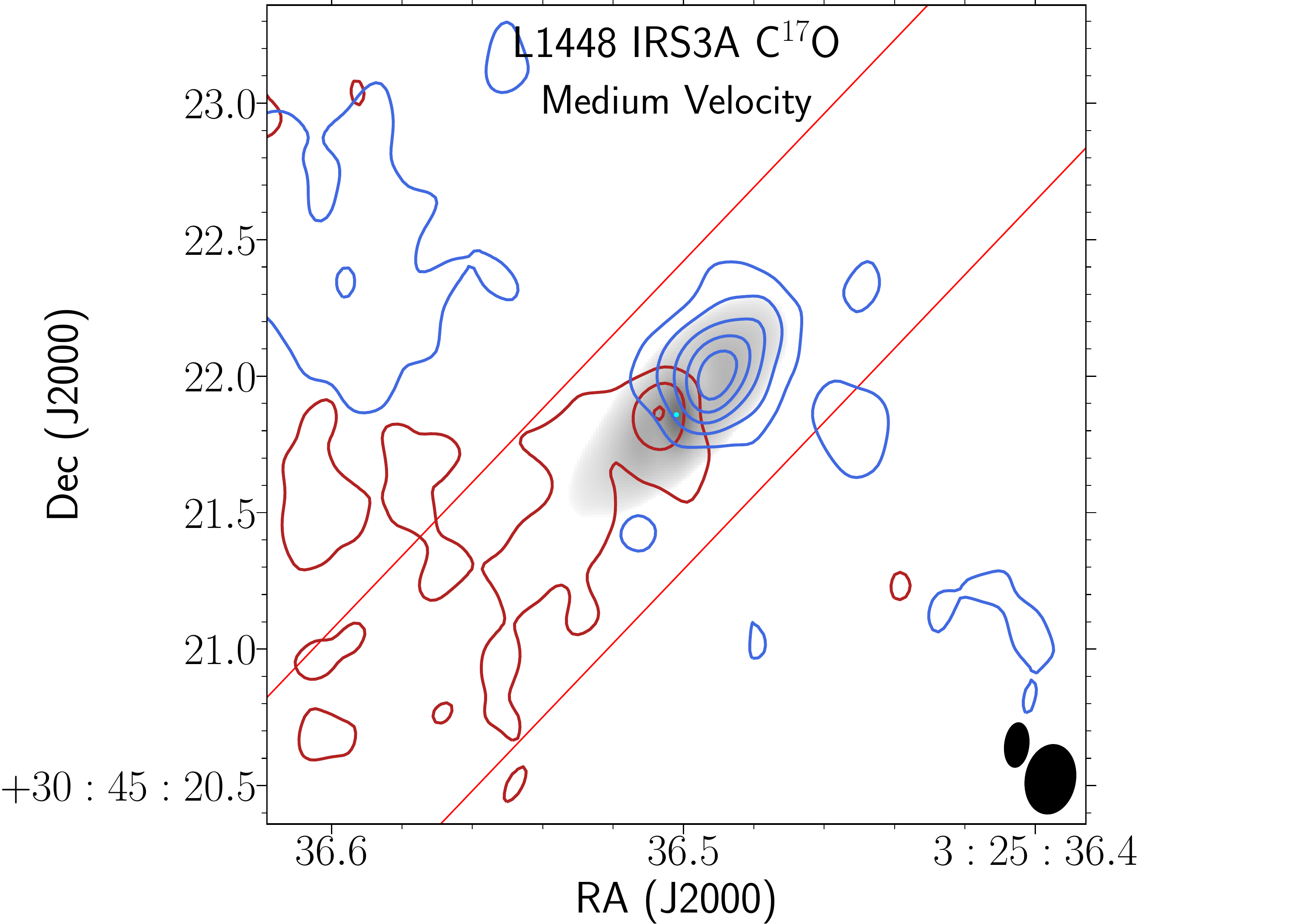}
   \includegraphics[width=0.45\textwidth]{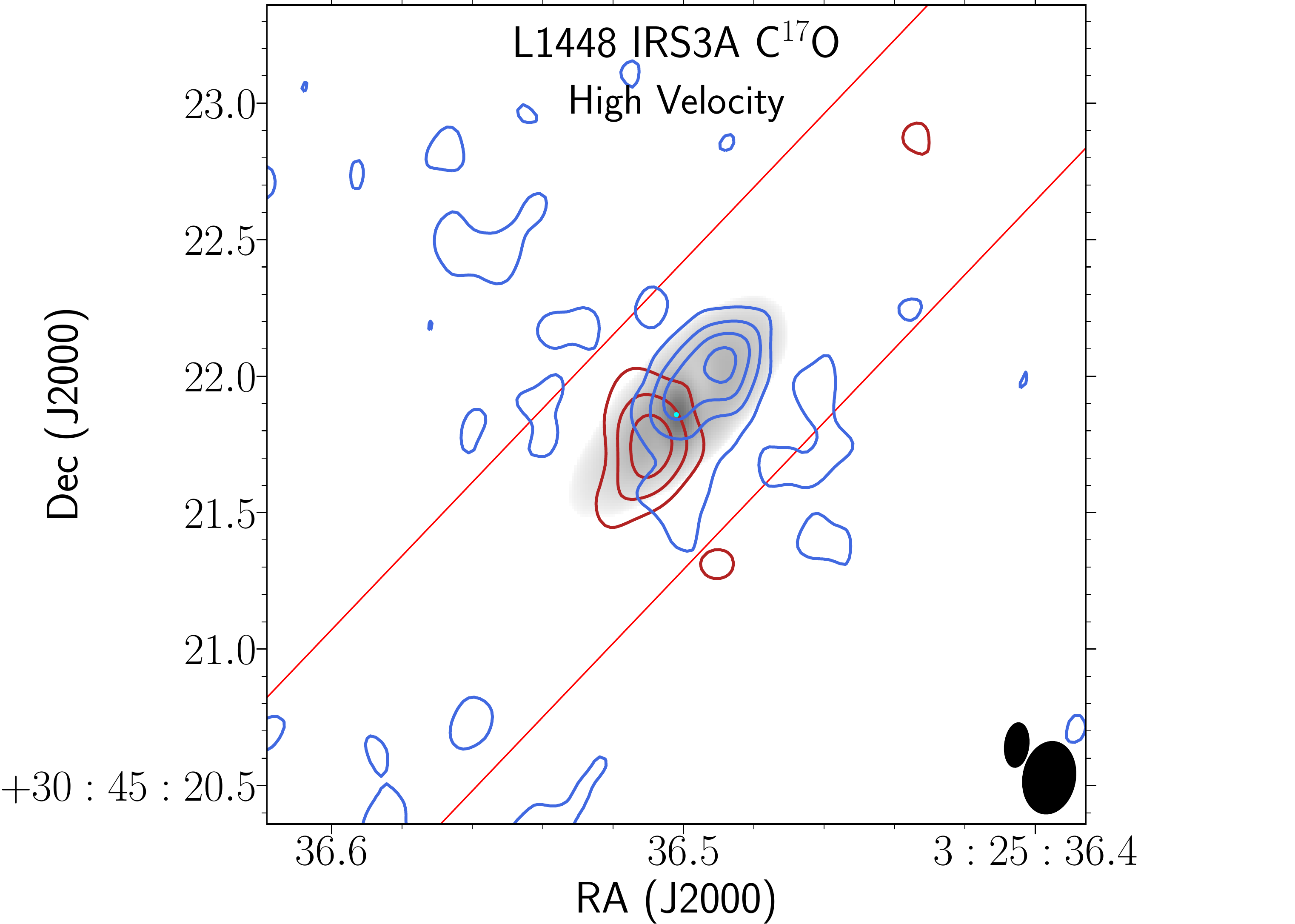} 
\end{center}
   \caption{\cso\space integrated intensity maps toward IRS3A over a selected range of velocities overlayed on continuum (grayscale). The \cso\space emission exhibits a velocity gradient across the continuum emission. However, S/N is low in comparison to IRS3B. The panels correspond to low, medium, and high velocity ranges which are delineated as red(blue), respectively. The red lines indicate the region extracted for PV diagram construction, along the position angle of the major axis. \textbf{Low Velocity:} velocity ranges 5.2$\rightarrow$6.5~\kms (4.1$\rightarrow$5.2~\kms), contours start at 3(3)$\sigma$ and iterate by 3(3)$\sigma$ with the 1$\sigma$~level starting at 0.0023(0.0025)~Jy~beam$^{-1}$ for the red(blue) channels respectively. \textbf{Medium Velocity:}  velocity ranges 6.5$\rightarrow$7.4~\kms (3.0$\rightarrow$4.1~\kms), contours start at 3(3)$\sigma$ and iterate by 3(3)$\sigma$ with the 1$\sigma$~level starting at 0.002(0.0016)~Jy~beam$^{-1}$ for the red (blue) channels respectively. \textbf{High Velocity:} velocity ranges 7.4$\rightarrow$8.6~\kms (1.8$\rightarrow$3.0~\kms), contours start at 3(3)$\sigma$ and iterate by 3(3)$\sigma$ with the 1$\sigma$~level starting at 0.0018(0.0012)~Jy~beam$^{-1}$ for the red(blue) channels respectively. The \cso\space synthesized beam (\csobeam) is the bottom-right most ellipse on each of the panels and the continuum synthesized beam (\contbeam) is offset diagonally.}\label{fig:irs3ac17omoment}
\end{figure}

\begin{figure}[H]
\begin{center}
   \includegraphics[width=1\textwidth]{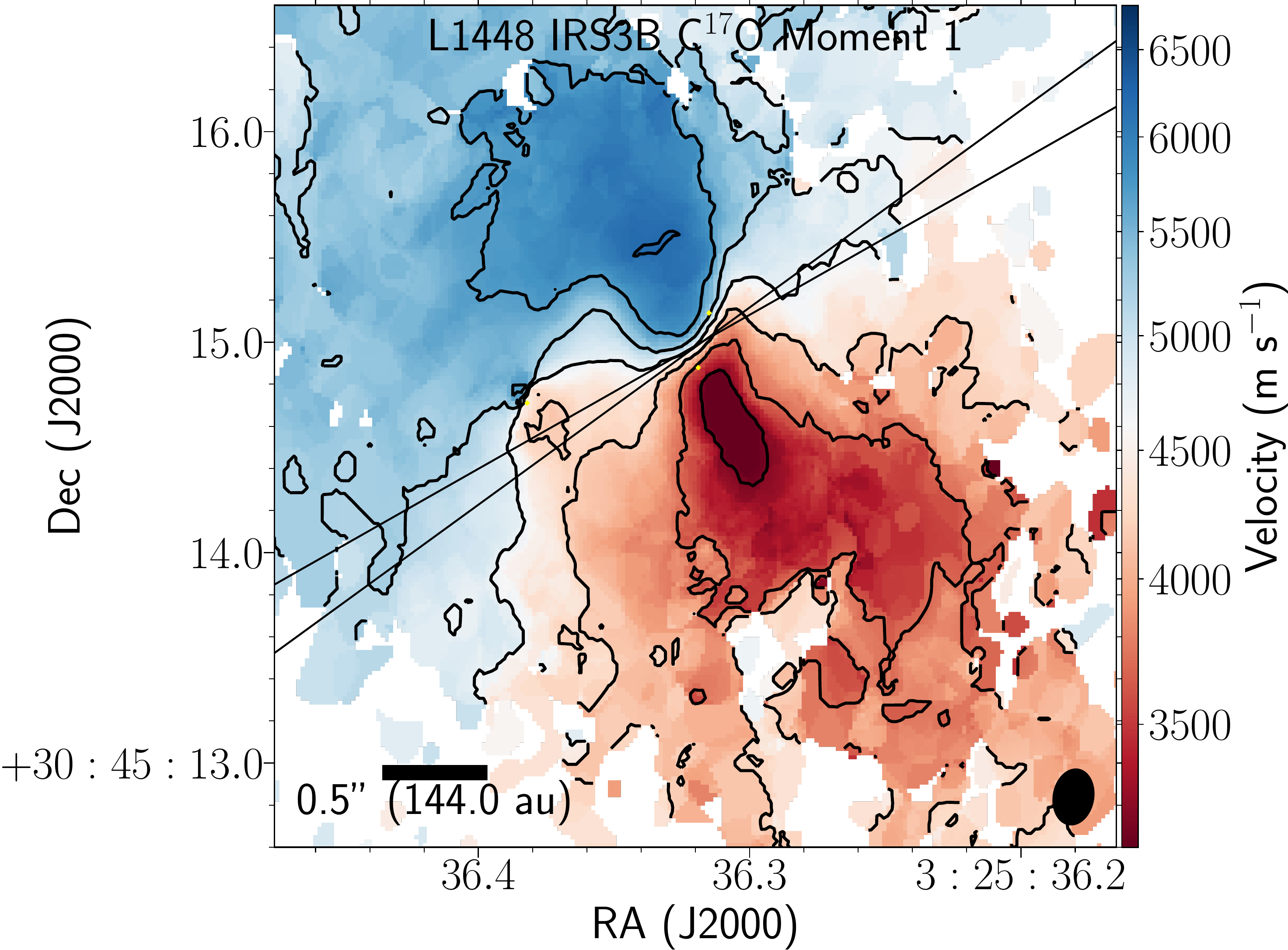} 
\end{center}
   \caption{\cso\space velocity-weighted integrated intensity maps toward IRS3B and IRS3A over a selected range of velocities (1.27$\rightarrow$7.65~\kms) The \cso\space emission appears well ordered across the semi-major axis. The contours denote the 0.5~km~s$^{-1}$ velocity offsets from system velocity of 4.8~km~s$^{-1}$. The yellow markers indicate the three continuum sources. \added{The black lines indicate the position angle of the minor disk estimates as given by the \pdspy\space fitting routine in Table~\ref{table:pvtable}, of $90+26.7^{+  1.8}_{-  2.9}$\deg.} The \cso\space synthesized beam (\csobeam) is the bottom-right most ellipse.}\label{fig:irs3abc17omoment1}
\end{figure}

\begin{figure}[H]
\begin{center}
   \includegraphics[width=0.45\textwidth]{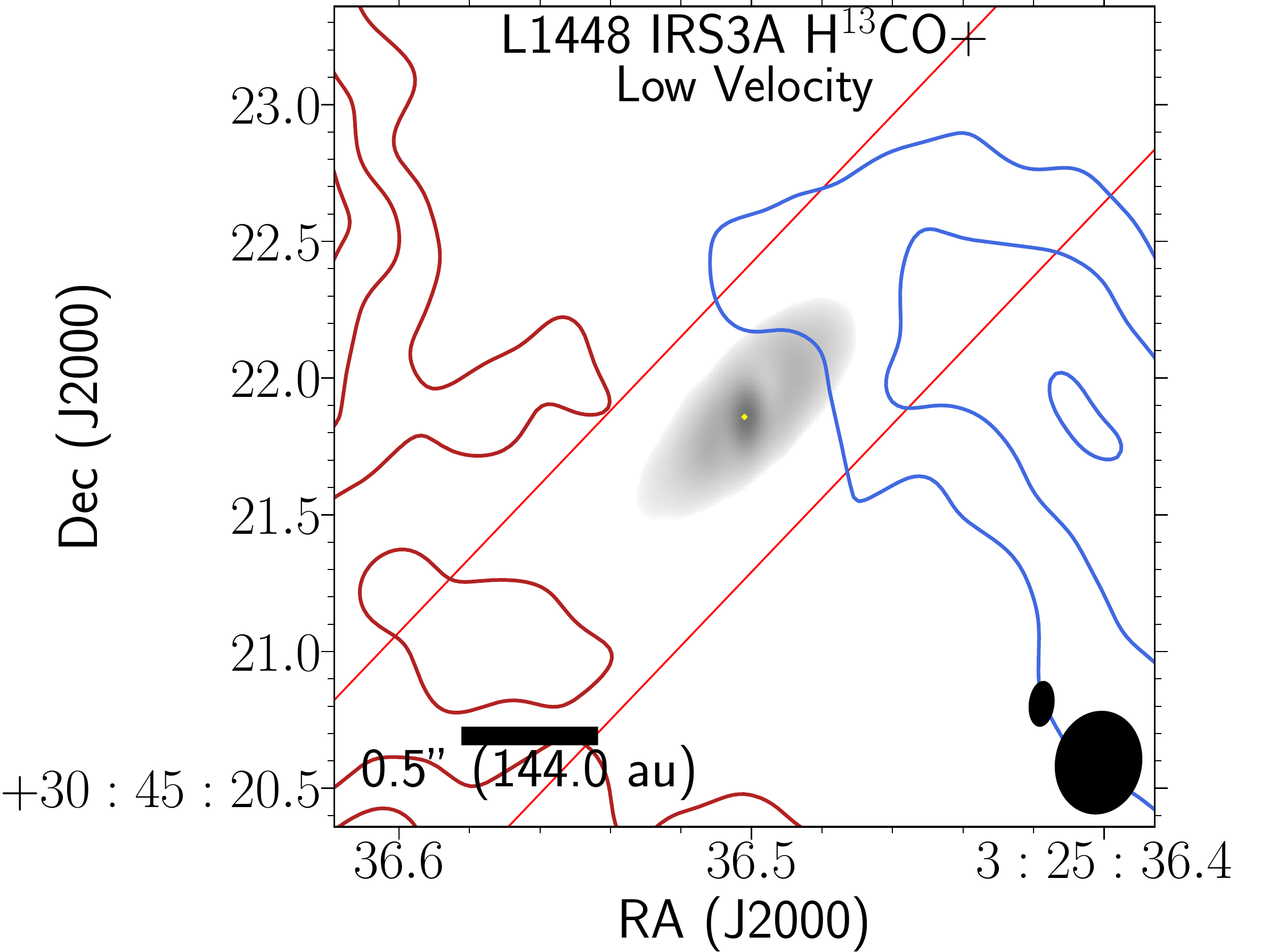}
   \includegraphics[width=0.45\textwidth]{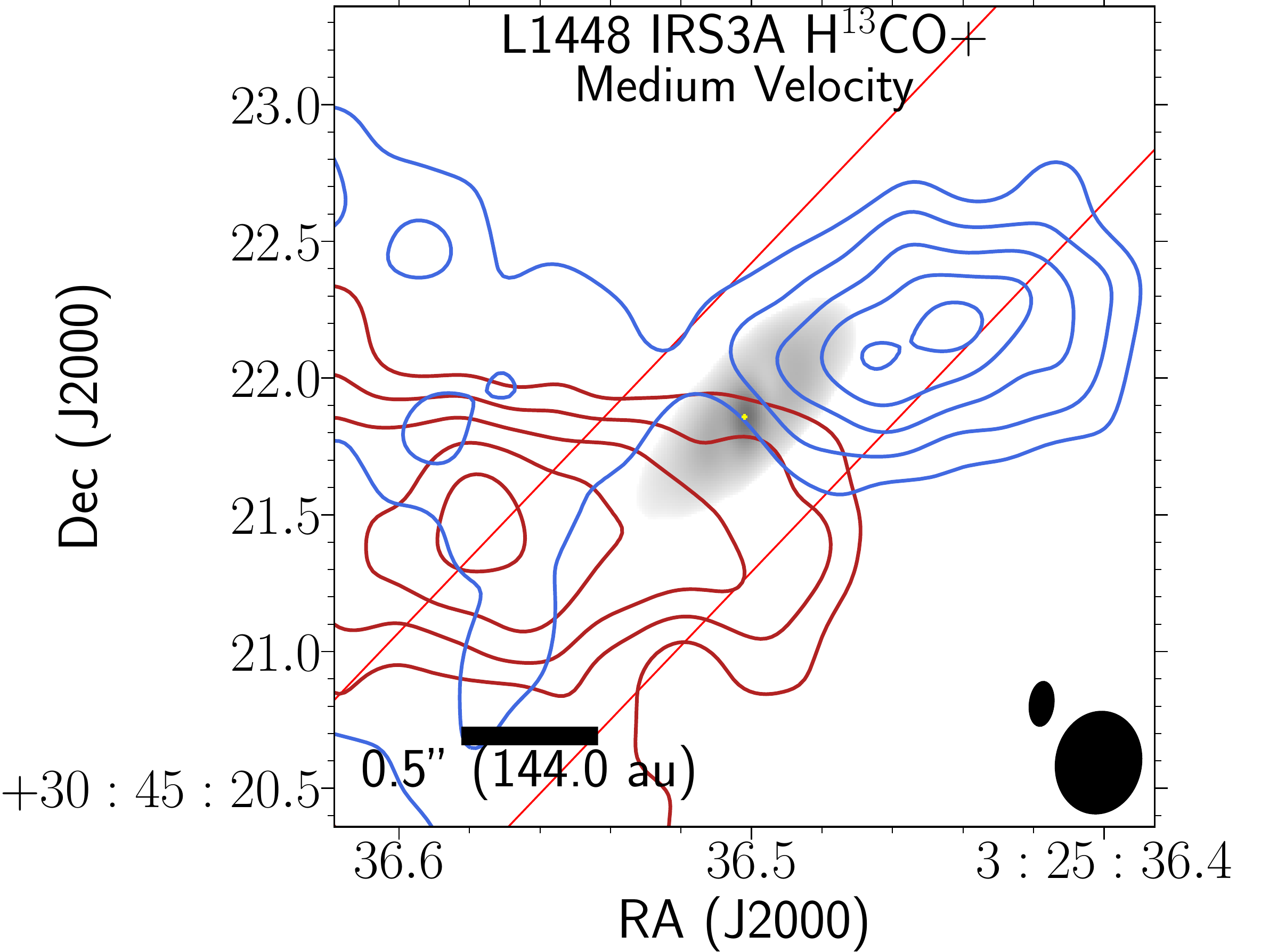}
   \includegraphics[width=0.45\textwidth]{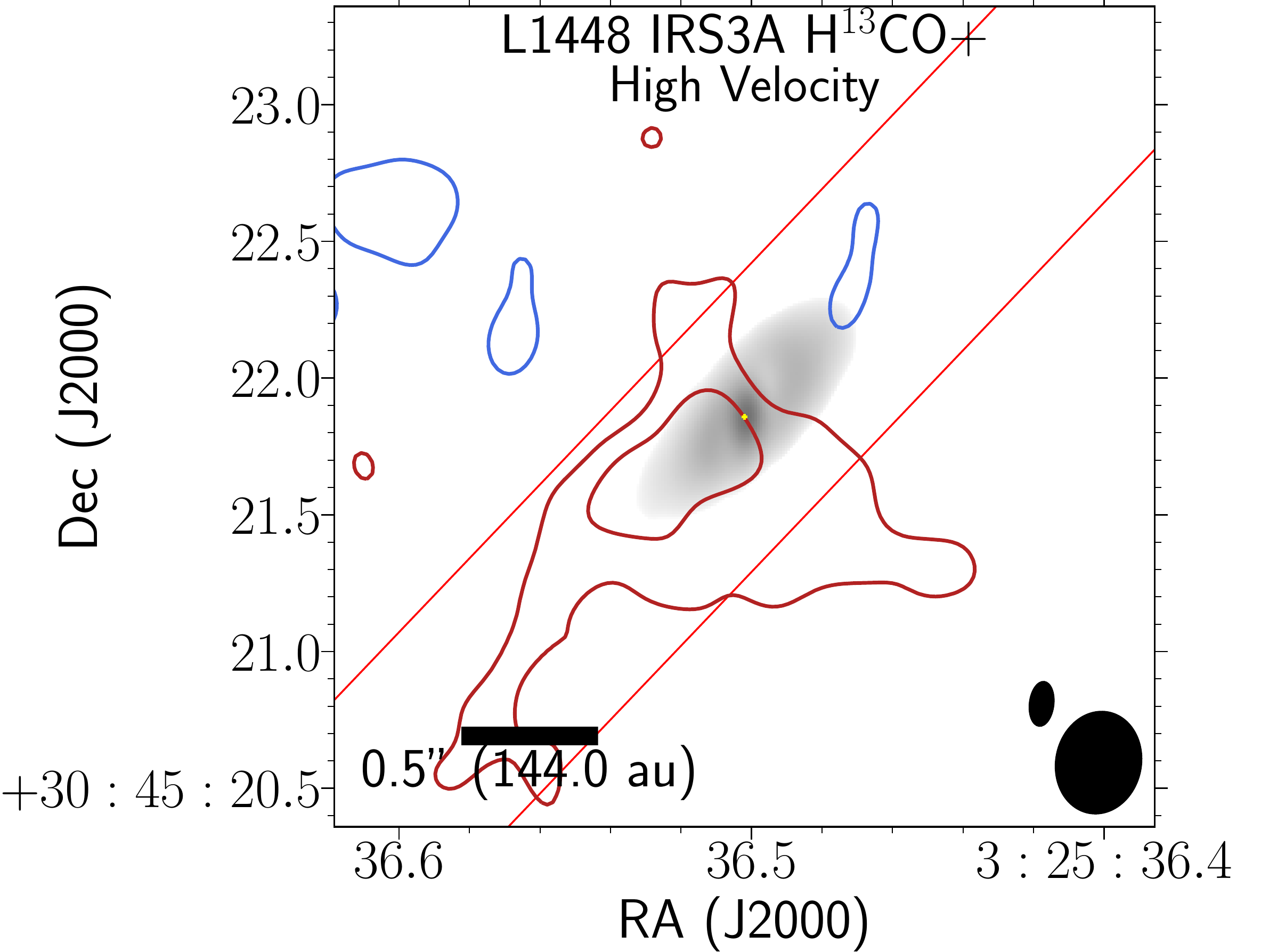}
\end{center}
   \caption{\htcop\space integrated intensity maps towards IRS3B over a selected range of velocities overlayed on continuum (grayscale). The top row spatial scale is set to match those of Figure~\ref{fig:irs3bc17omoment} and the bottom row scale is set to encapsulate the entire IRS3B system, to better demonstrate the spatial scales probed with this molecule. The top row is tapered with a 400~k$\lambda$\space Gaussian to best reduce the amount of noise and show the proper resolution to the spatial scales shown. The \htcop\space emission is primarily tracing the intermediate dense, gaseous material within the inner envelope, but the higher-velocity emission does originate near the protostars. The columns correspond to similar velocity ranges of \cso\space emission as shown in the previous figure, with low, medium, and high Doppler-shifted velocity ranges delineated as red(blue), respectively. Negative contours do not show additional structure and are suppressed for visual aid. The red lines indicate the region extracted for PV diagram construction, along the position angle of the major axis in a region much larger than the \cso\space PV diagram extraction to fully capture the emission. \textbf{Low Velocity:} velocity ranges 4.7$\rightarrow$5.7~\kms (3.6$\rightarrow$4.7~\kms)), contours start at 10(10)$\sigma$ and iterate by 2(2)$\sigma$ with the 1$\sigma$~level starting at 0.003(0.003)~Jy~beam$^{-1}$ for the top row and 0.005(0.005)~Jy~beam$^{-1}$ for the bottom row,  red(blue) channels respectively.  \textbf{Medium Velocity:} velocity ranges 5.7$\rightarrow$6.7~\kms (2.4$\rightarrow$3.5~\kms), contours start at 5(5)$\sigma$ and iterate by 5(3)$\sigma$ with the 1$\sigma$~level starting at 0.005(0.005)~Jy~beam$^{-1}$ for the top row and 0.005(0.005)~Jy~beam$^{-1}$ for the bottom row, red(blue) channels respectively. \textbf{High Velocity:} velocity ranges 6.7$\rightarrow$7.7~\kms (1.3$\rightarrow$2.4~\kms), contours start at 5(5)$\sigma$ and iterate by 2(2)$\sigma$ with the 1$\sigma$~level starting at 0.002(0.002)~Jy~beam$^{-1}$ for the top row and 0.005(0.005)~Jy~beam$^{-1}$ for the bottom row, for the red(blue) channels respectively. The \htcop\space synthesized beam (top:0\farcs374$\times$0\farcs310, bottom: \htcopbeam) is the bottom-right most ellipse on each of the panels and the continuum synthesized beam (\contbeam) is offset diagonally.}\label{fig:h13copmomentc17o}
\end{figure}

\begin{figure}[H]
\begin{center}
   \includegraphics[width=0.45\textwidth]{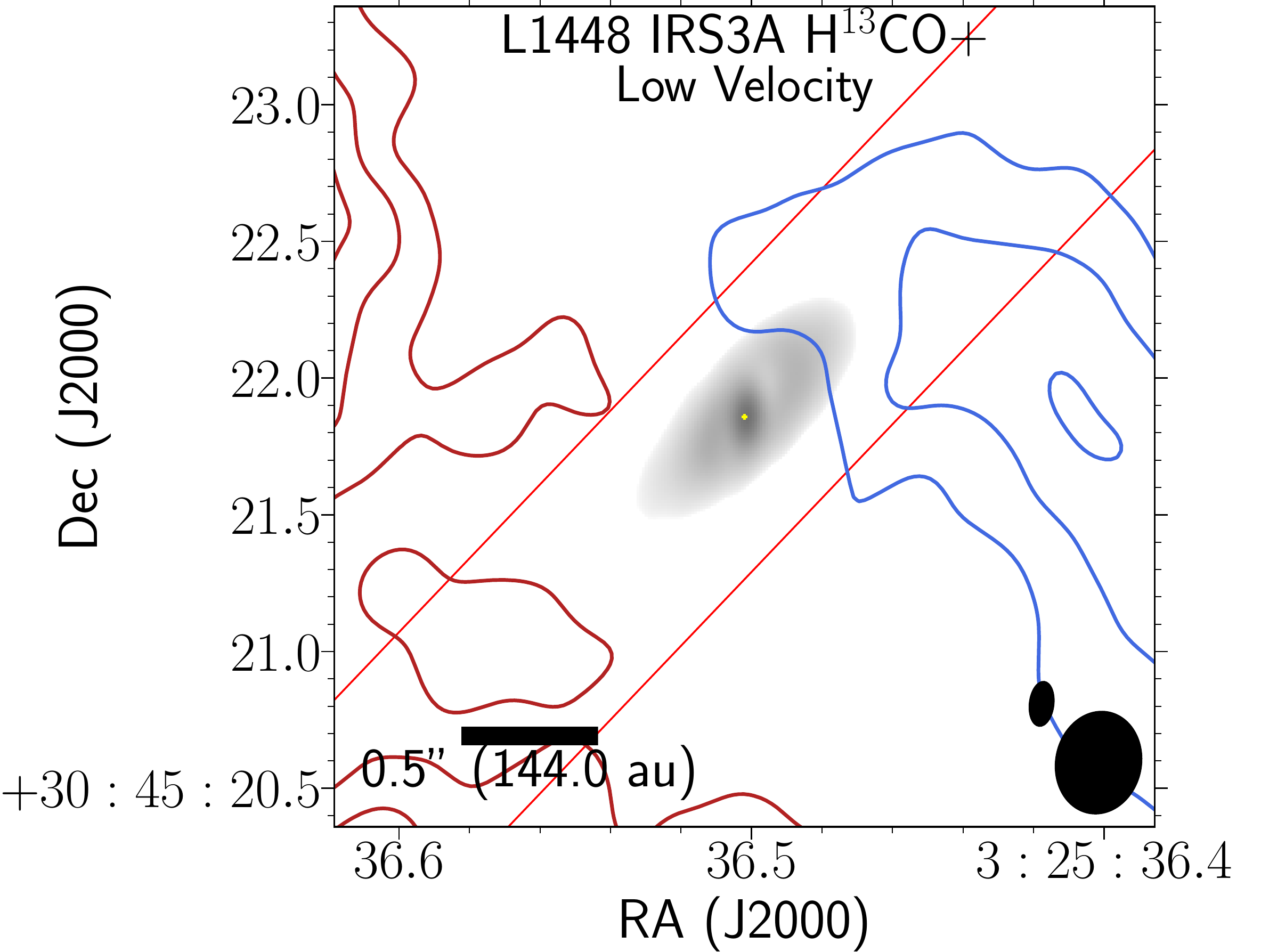}
   \includegraphics[width=0.45\textwidth]{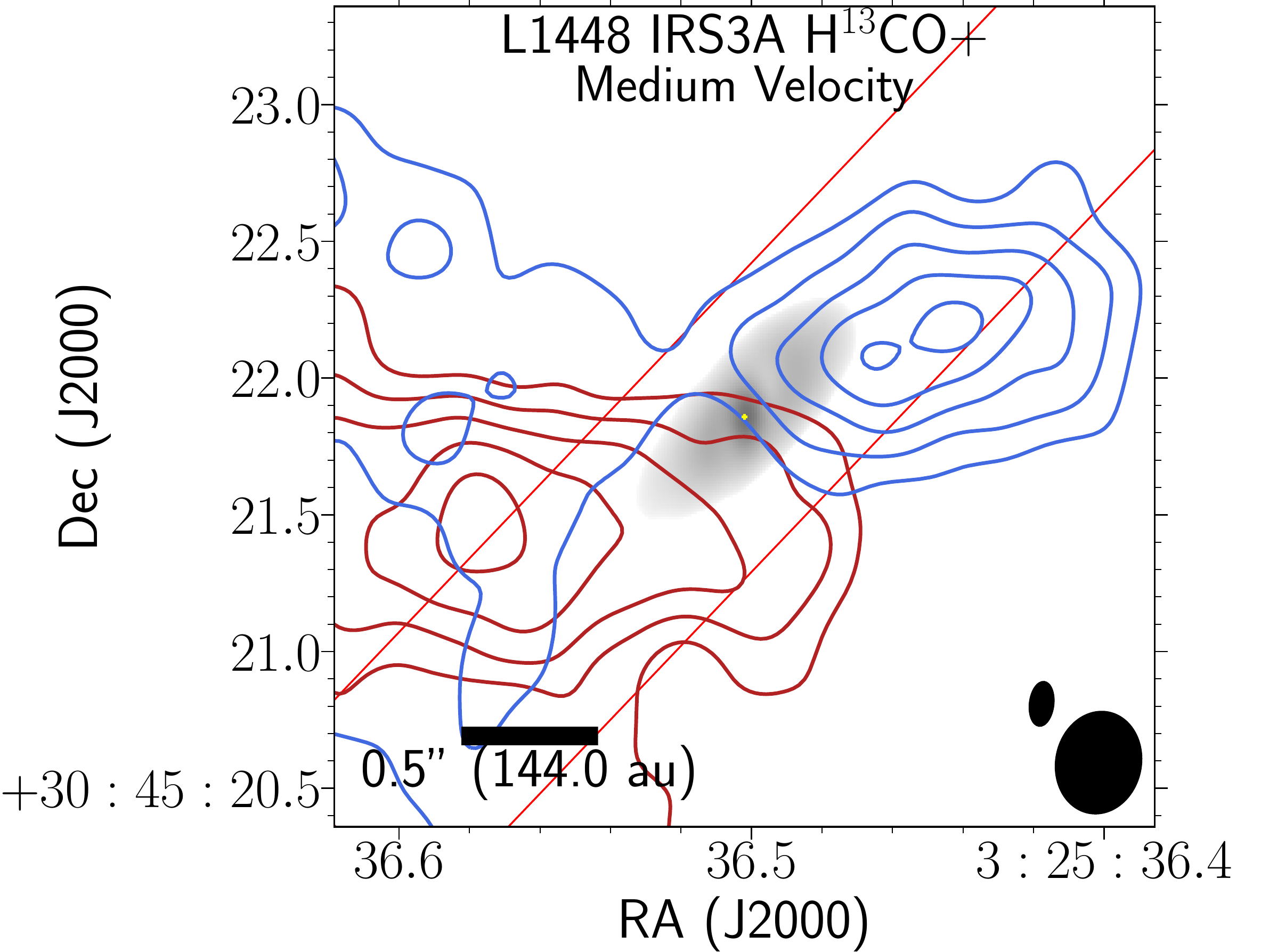}
   \includegraphics[width=0.45\textwidth]{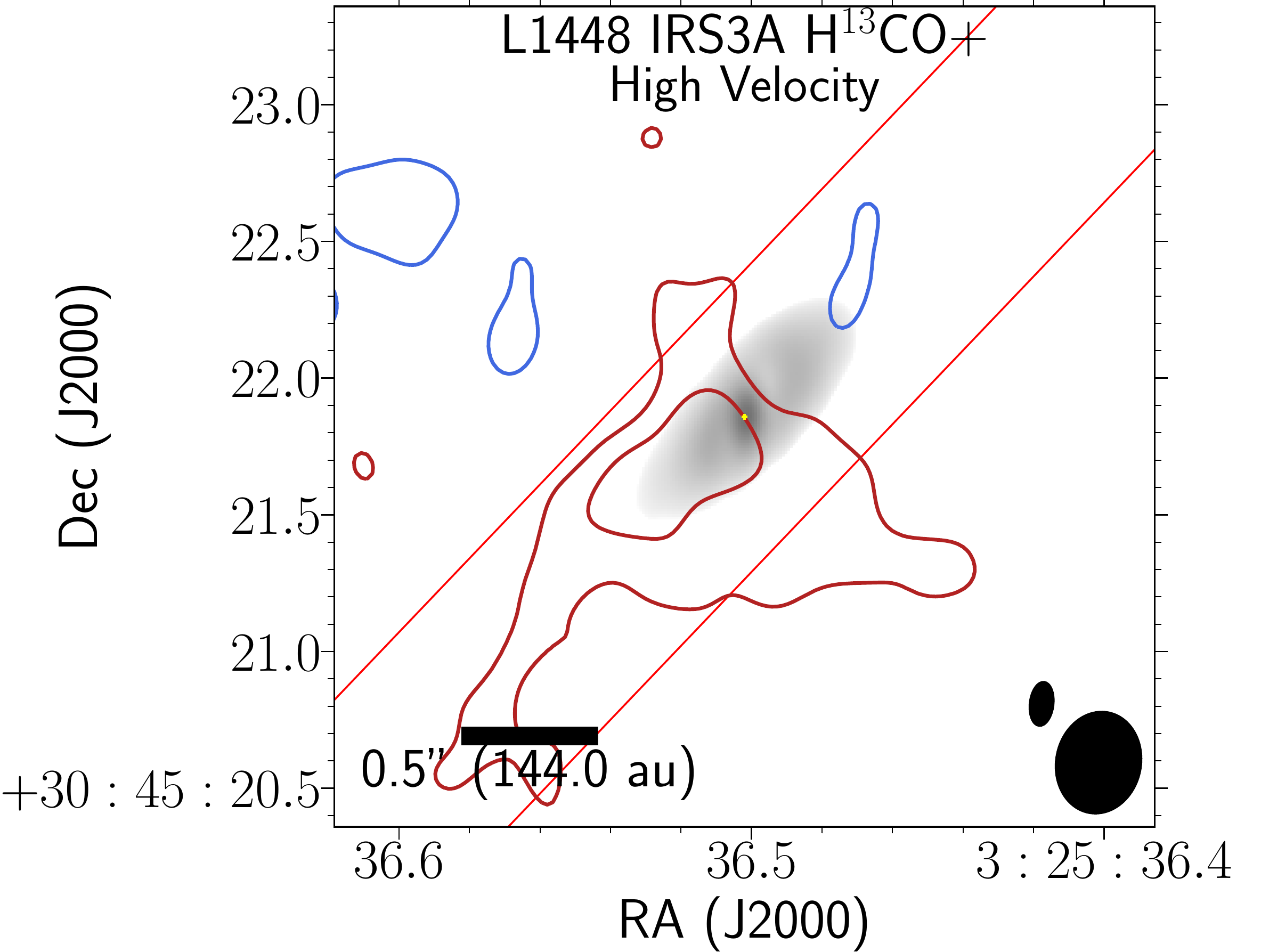}
\end{center}
   \caption{\htcop\space integrated intensity map towards IRS3A generated at a position angle of 125\deg; whose emission predominately traces the intermediate dense, gaseous material of the inner envelope. The image is tapered with a 400~k$\lambda$\space Gaussian to best reduce the amount of noise and show the proper resolution to the spatial scales shown. The \htcop\space emission might trace a velocity gradient across the source, but \replaced{the compactness of the source}{the lack of strong emission coming from the disk itself} hinders resolving the kinematics. The columns correspond to low, medium, and high velocity ranges which are delineated as red(blue), respectively. \textbf{Low Velocity:} velocity ranges 5.2$\rightarrow$6.5~\kms (4.1$\rightarrow$5.2~\kms), contours start at 5(5)$\sigma$ and iterate by 2(2)$\sigma$ with the 1$\sigma$~level starting at 0.004(0.007)~Jy~beam$^{-1}$ for the red(blue) channels respectively. \textbf{Medium Velocity:}  velocity ranges 6.5$\rightarrow$7.4~\kms (3.0$\rightarrow$4.1~\kms), contours start at 3(3)$\sigma$ and iterate by 2(2)$\sigma$ with the 1$\sigma$~level starting at 0.003(0.003)~Jy~beam$^{-1}$ for the red (blue) channels respectively. \textbf{High Velocity:} velocity ranges 7.4$\rightarrow$8.6~\kms (1.8$\rightarrow$3.0~\kms), contours start at 3(3)$\sigma$ and iterate by 2(2)$\sigma$ with the 1$\sigma$~level starting at 0.002(0.0025)~Jy~beam$^{-1}$ for the red(blue) channels respectively. The \htcop\space synthesized beam (\htcopbeam) is the bottom-right most ellipse on each of the panels and the continuum synthesized beam (\contbeam) is offset diagonally.}\label{fig:irs3ah13copmoment}
\end{figure}

\begin{figure}[H]
\begin{center}
   \includegraphics[width=0.45\textwidth]{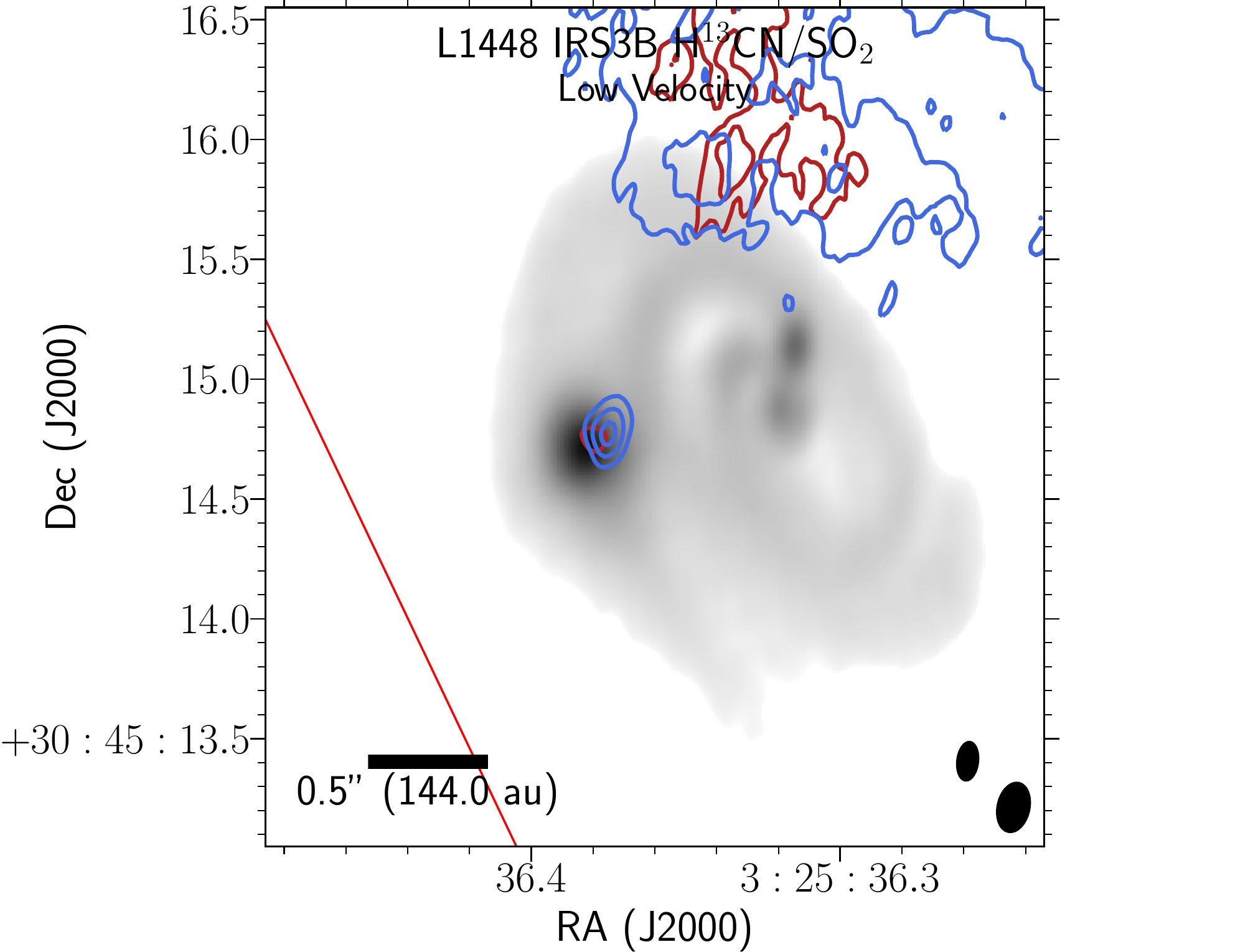}
   \includegraphics[width=0.45\textwidth]{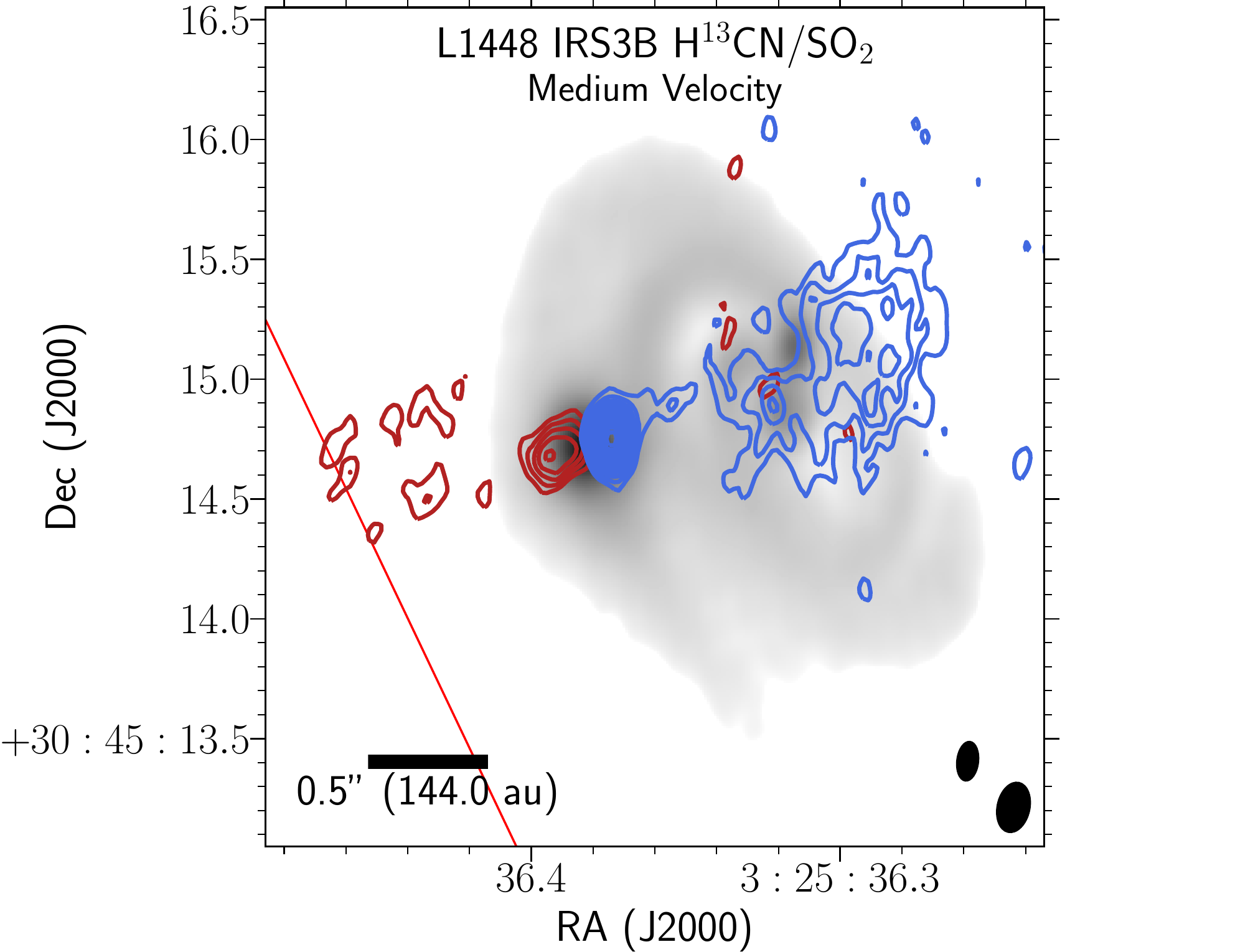}
   \includegraphics[width=0.45\textwidth]{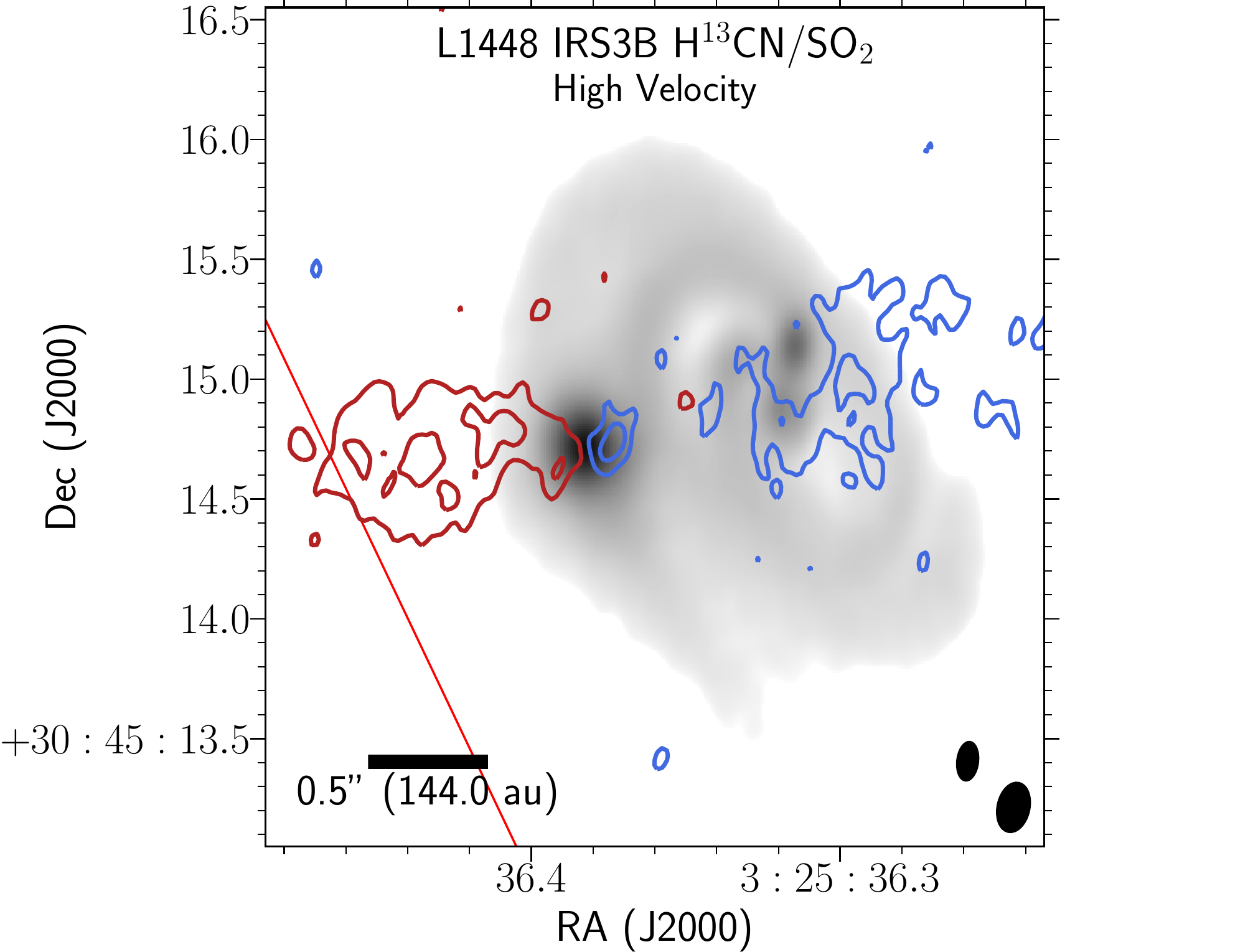}
\end{center}
   \caption{\htcn/\sot\space integrated intensity map towards IRS3B, appears to trace near the outflow launch location from the tertiary, IRS3B-c. There is pretty large asymmetry in the velocity channels covered by the red and blue-shifted emission. The panels correspond to low, medium, and high velocity ranges which are delineated as red(blue), respectively. \textbf{Low Velocity:} velocity ranges 5.2$\rightarrow$7.2~\kms (4$\rightarrow$4.8~\kms), contours start at 5(5)$\sigma$ and iterate by 2(5)$\sigma$ with the 1$\sigma$~level starting at 0.0025(0.0021)~Jy~beam$^{-1}$ for the red(blue) channels respectively. \textbf{Medium Velocity:} velocity ranges 7.2$\rightarrow$9.2~\kms (3.2$\rightarrow$4~\kms), contours start at 5(5)$\sigma$ and iterate by 2(2)$\sigma$ with the 1$\sigma$~level starting at 0.0016(0.0016)~Jy~beam$^{-1}$ for the red(blue) channels respectively. \textbf{High Velocity:} velocity ranges 9.2$\rightarrow$11.2~\kms (1.6$\rightarrow$3.2~\kms), contours start at 4(4)$\sigma$ and iterate by 3(3)$\sigma$ with the 1$\sigma$~level starting at 0.0021(0.0021)~Jy~beam$^{-1}$ for the red(blue) channels respectively. The \htcn\space synthesized beam (\htcnbeam) is the bottom-right most ellipse on each of the panels and the continuum synthesized beam (\contbeam) is offset diagonally.}\label{fig:irs3bh13cnmoment}
\end{figure}

\begin{figure}[H]
\begin{center}
   \includegraphics[width=0.45\textwidth]{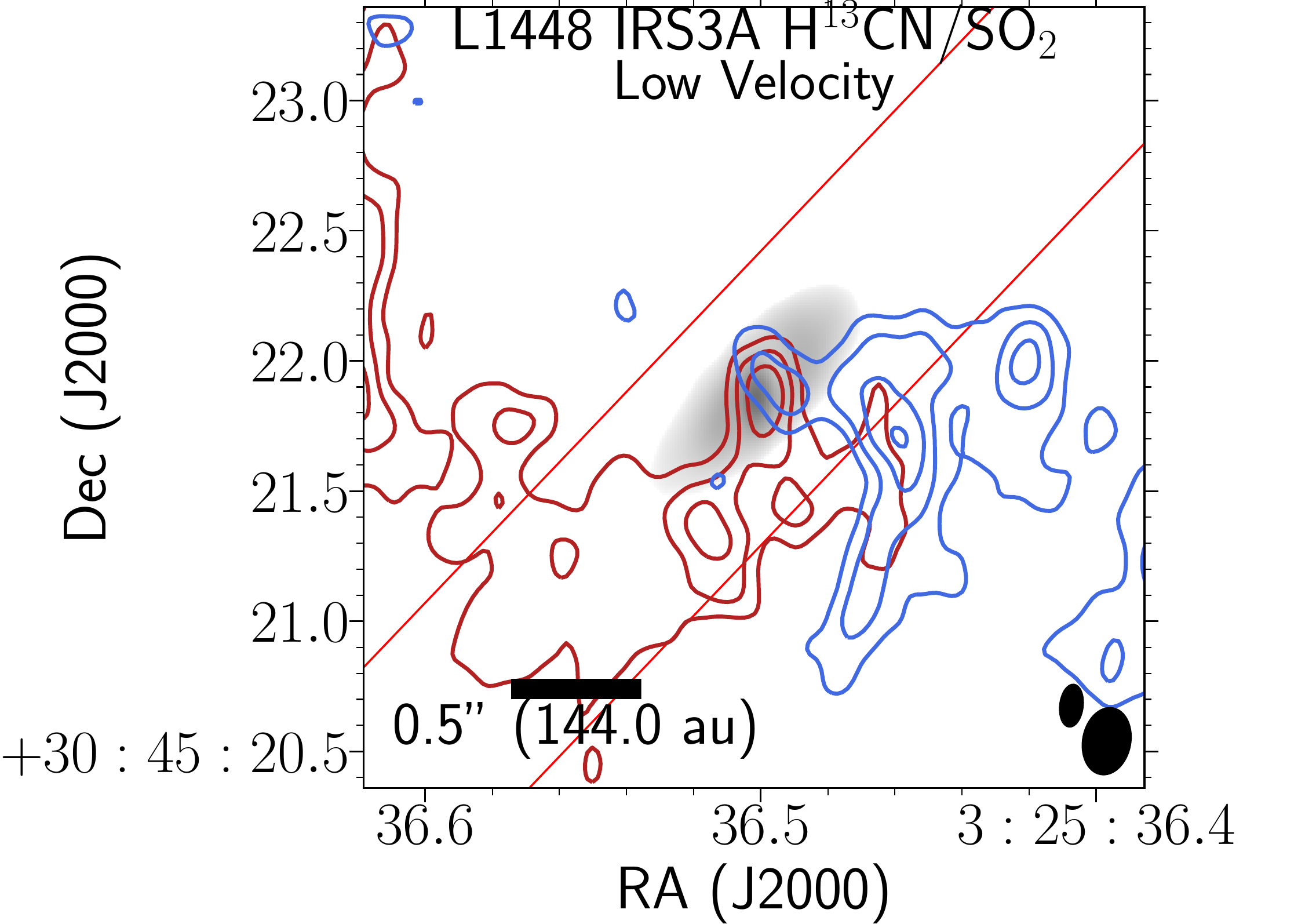}
   \includegraphics[width=0.45\textwidth]{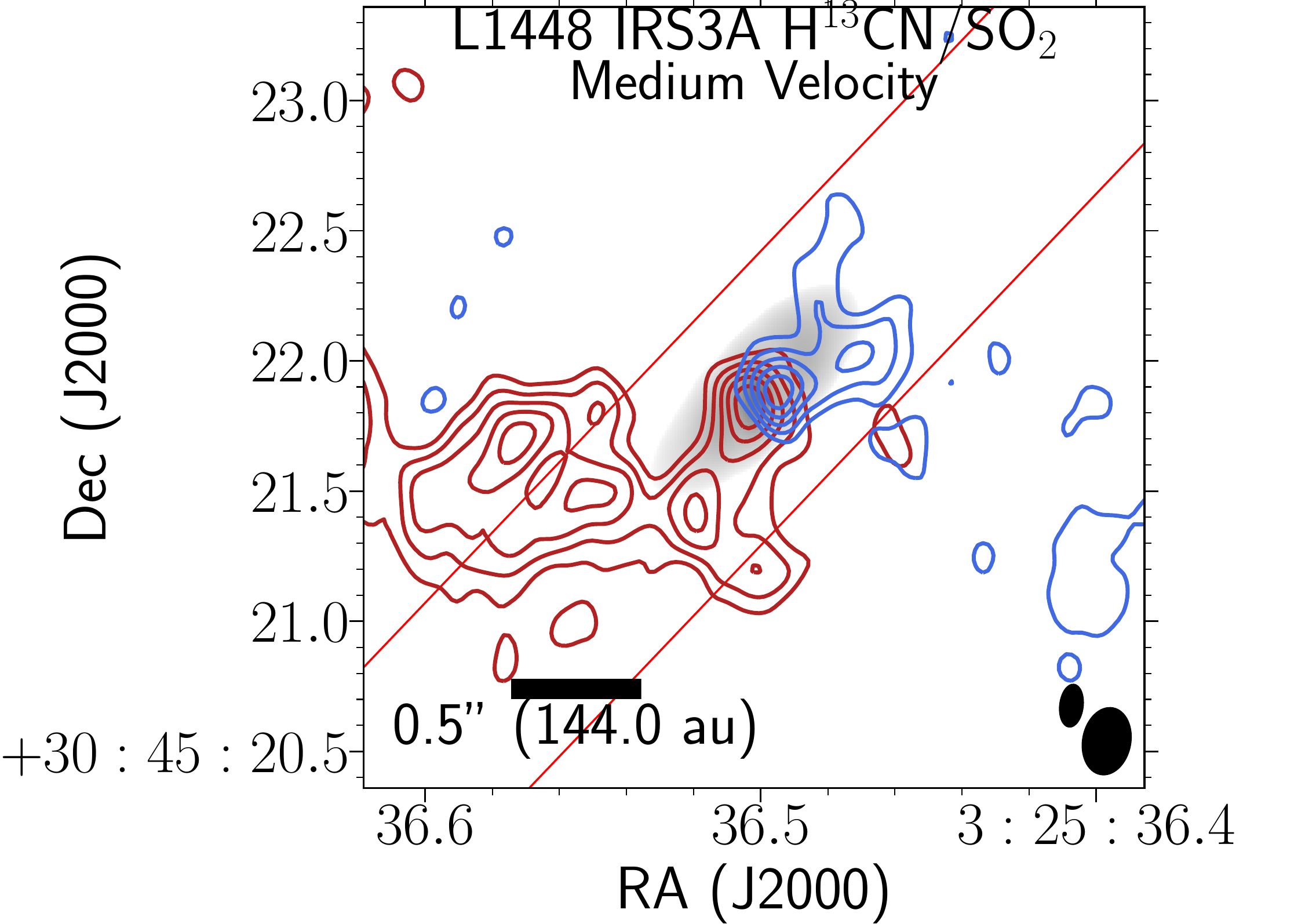}
   \includegraphics[width=0.45\textwidth]{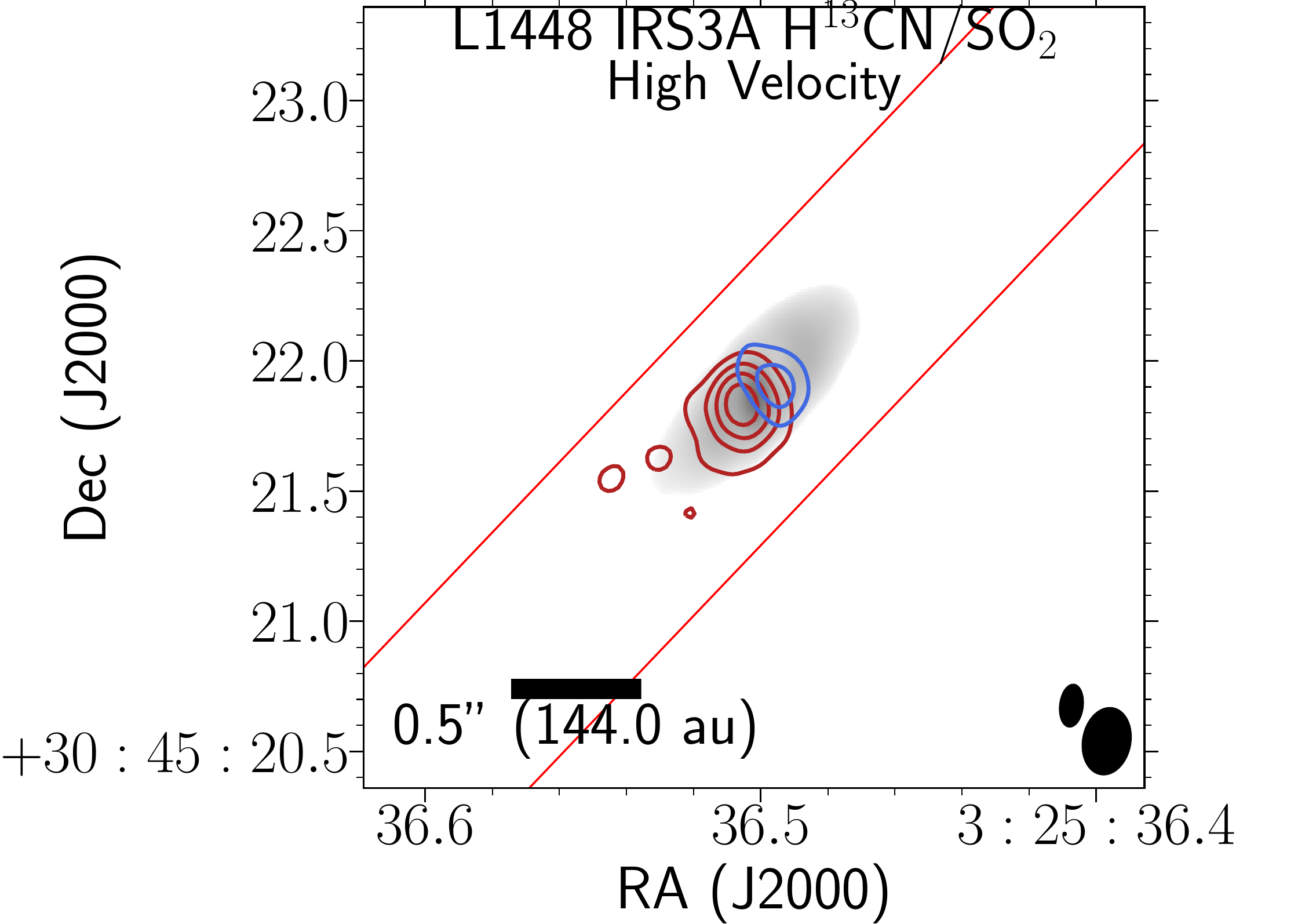}
\end{center}
   \caption{\htcn/\sot\space integrated intensity map towards IRS3A; whose emission appears to trace rotation within the inner disk. The panels correspond to low, medium, and high velocity ranges which are delineated as red(blue), respectively. The system velocity of the \htcn/\sot\space emission (\ab5.4~km~s$^{-1}$) agrees with system velocity of \cso\space, likely tracing \htcn\space emission and not \sot\space emission. \textbf{Low Velocity:} velocity ranges 5.2$\rightarrow$6.5~\kms (4.1$\rightarrow$5.2~\kms), contours start at 4(4)$\sigma$ and iterate by 2(2)$\sigma$ with the 1$\sigma$~level starting at 0.0021(0.0021)~Jy~beam$^{-1}$ for the red(blue) channels respectively. \textbf{Medium Velocity:}  velocity ranges 6.5$\rightarrow$7.4~\kms (3.0$\rightarrow$4.1~\kms), contours start at 4(4)$\sigma$ and iterate by 2(2)$\sigma$ with the 1$\sigma$~level starting at 0.0016(0.0016)~Jy~beam$^{-1}$ for the red (blue) channels respectively. \textbf{High Velocity:} velocity ranges 7.4$\rightarrow$8.6~\kms (1.8$\rightarrow$3.0~\kms), contours start at 4(4)$\sigma$ and iterate by 3(3)$\sigma$ with the 1$\sigma$~level starting at 0.0021(0.0021)~Jy~beam$^{-1}$ for the red(blue) channels respectively. The \htcn\space synthesized beam (\htcnbeam) is the bottom-right most ellipse on each of the panels and the continuum synthesized beam (\contbeam) is offset diagonally.}\label{fig:irs3ah13cnmoment}
\end{figure}

\begin{figure}[H]
\begin{center}
\includegraphics[width=0.44\textwidth]{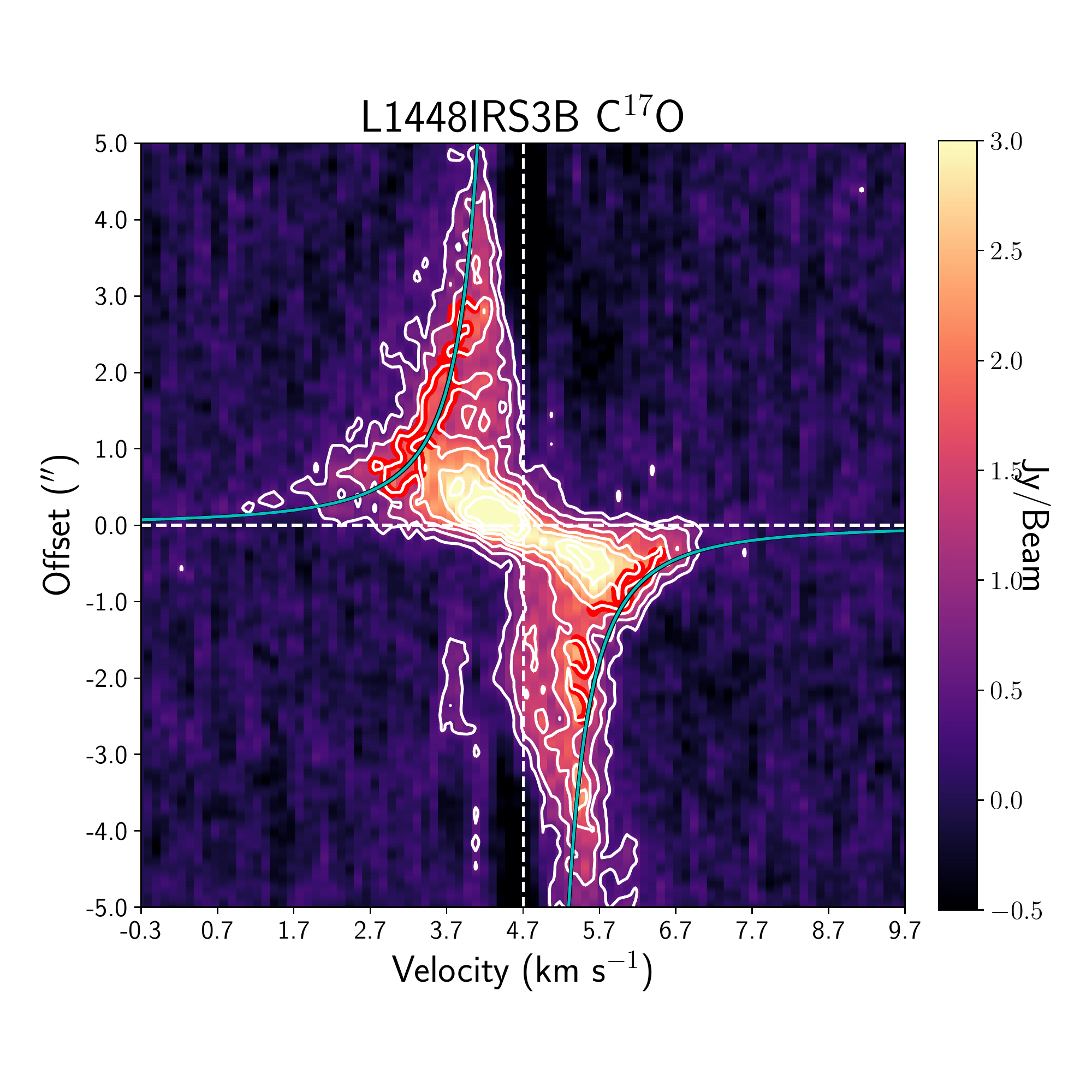}
\includegraphics[width=0.44\textwidth]{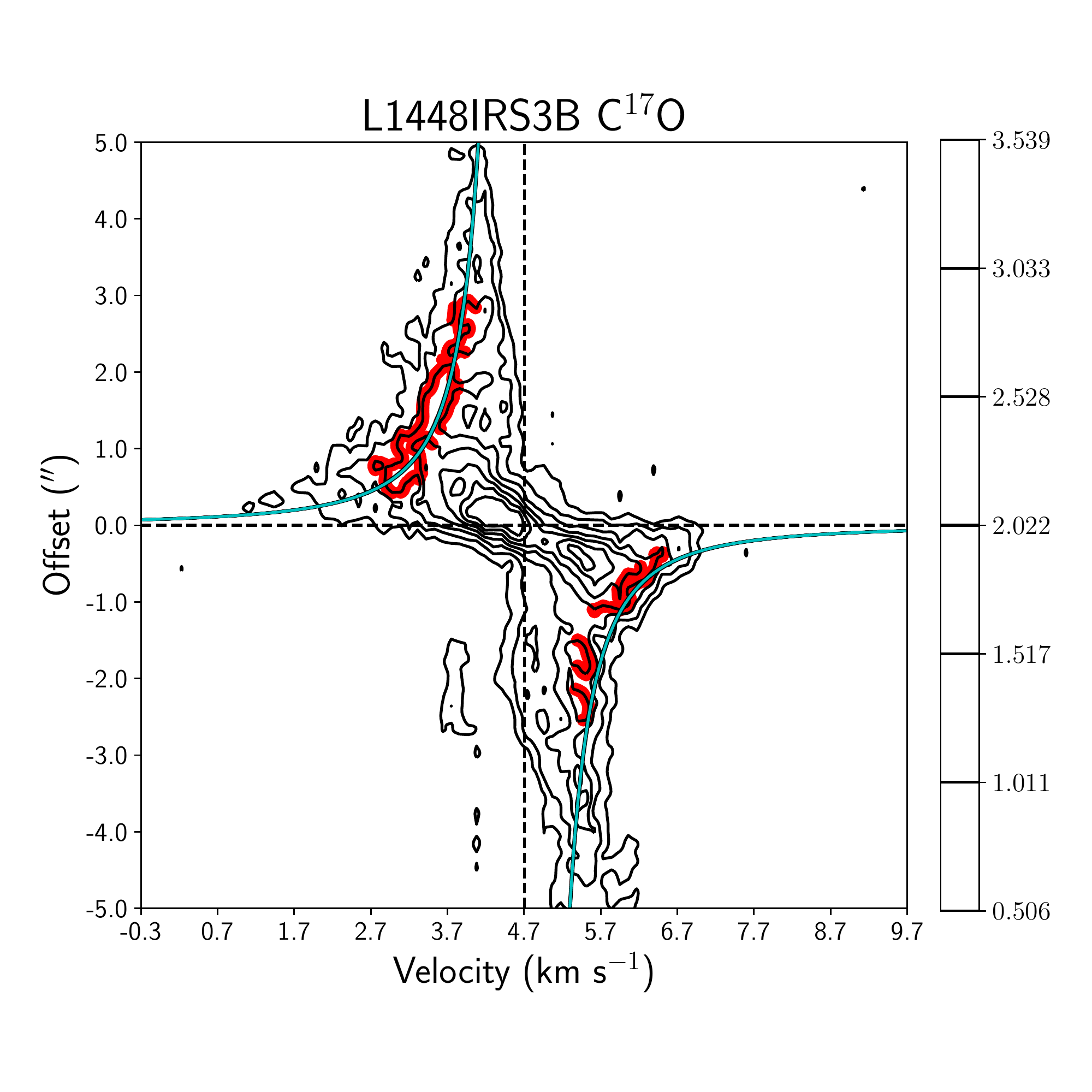}
\end{center}
\caption{PV-diagrams of IRS3B \cso\space emission generated at a position angle of 29\deg, with the cyan lines corresponding to the fit of 1.15~\solm, demonstrating the data could be reproduced reasonably well with a Keplerian disk orbiting a 1.15~\solm\space protostar. The cyan line traces the median fit for numeric Keplerian orbital fit routine while the black lines represent 100 randomly sampled MCMC fits, used to estimate errors. As evident, this methodology selectively fits the highest velocity emission that is symmetric in the protostellar system. The white/black contours trace regions starting from 3$\sigma$ at 2$\sigma$\space intervals, where $\sigma\approx$0.14~Jy~beam$^{-1}$. The red contours trace the regions selected for the MCMC fit which are defined as the 10 and 12$\sigma$\space levels as to not fit the diffuse large-scale emission.}\label{fig:l1448irs3b_c17o_pv}
\end{figure}

\begin{figure}[H]
\begin{center}
\includegraphics[width=0.5\textwidth]{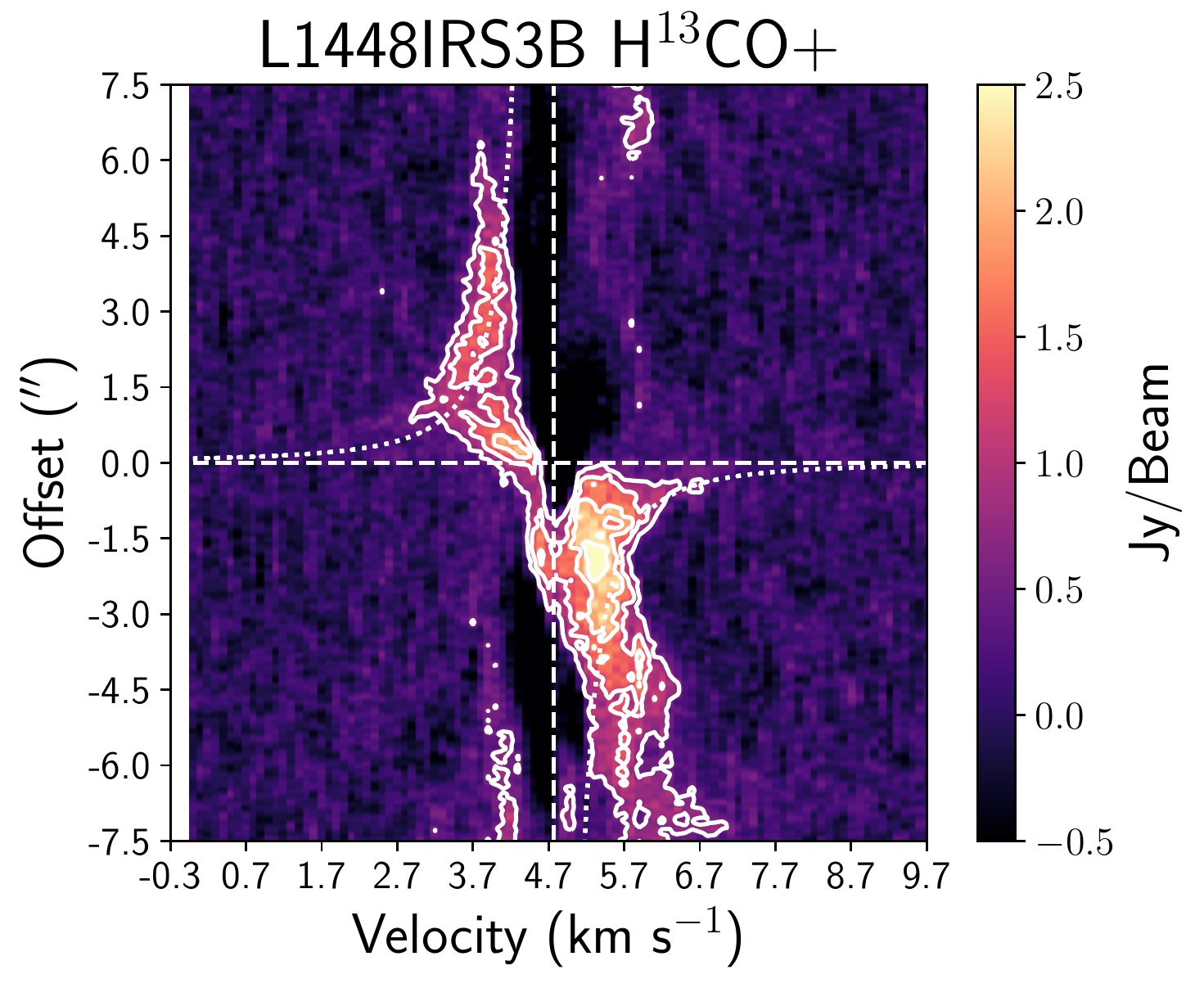}
\end{center}
\caption{\htcop\space emission towards IRS3B generated at a position angle of 29\deg, with the white dashed lines corresponding to the Keplerian fit of 1.15~\solm\space from the fit to \cso, demonstrating the data are not inconsistent with a 1.15~\solm\space protostar, similarly demonstrated from the \cso\space emission Keplerian fits. The PV diagram shows a large amount of asymmetry in the molecular line emission close to system velocity, with emission  at velocities in  excess of Keplerian particularly at the red-shifted velocities. These are possible indications of infalling material from the envelope given the spatial location this emission. The white contours trace regions starting from 3$\sigma$ at 2$\sigma$\space intervals, where $\sigma\approx$0.15~Jy}\label{fig:l1448irs3b_h13cop_pv}
\end{figure}

\begin{figure}[H]
\begin{center}
\includegraphics[width=0.44\textwidth]{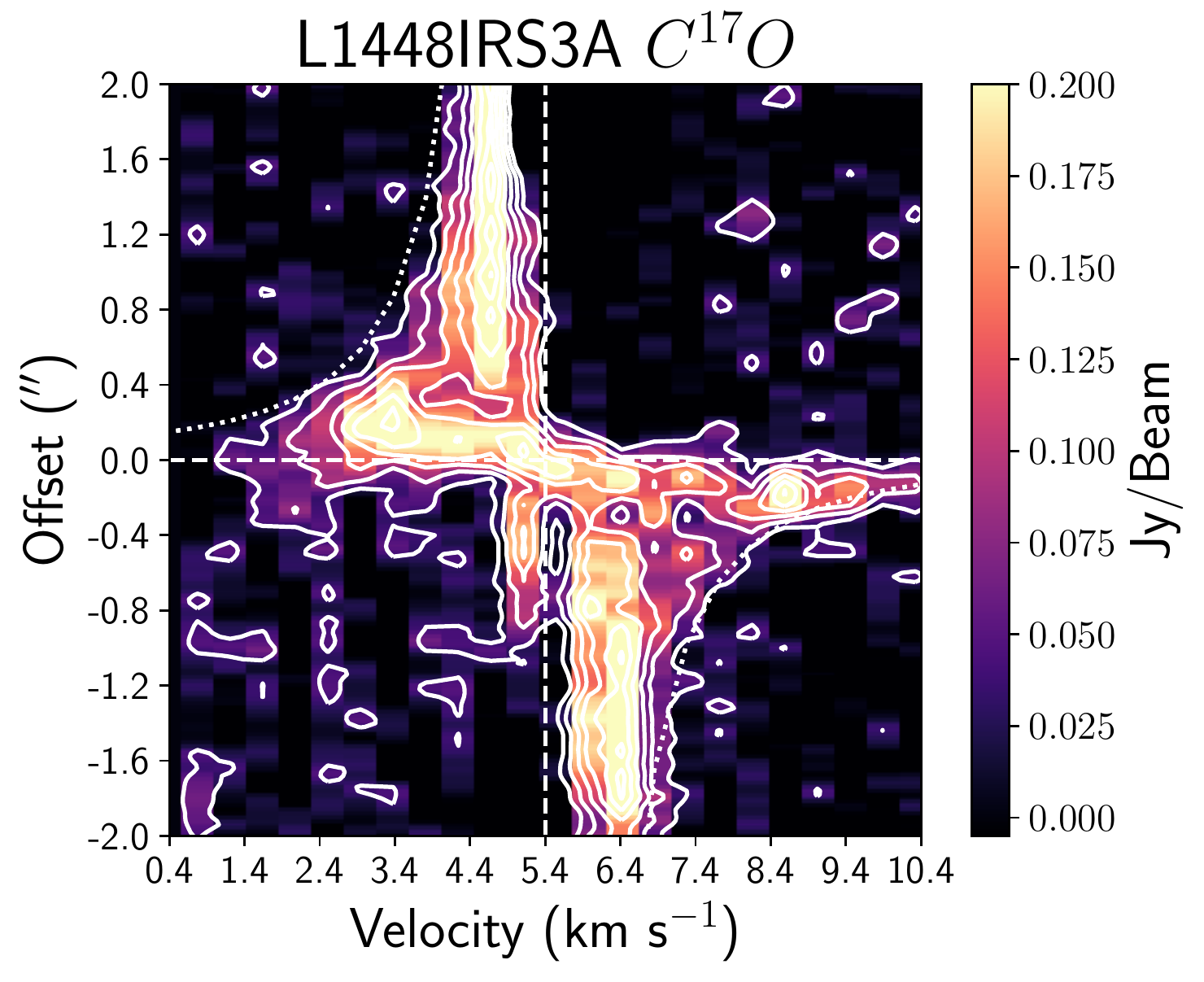}
\end{center}
\caption{Position-velocity diagrams of IRS3A \cso\space emission generated at a position angle of 125\deg, with the dotted lines corresponding to 1.4~\solm. The emission suffers from the lower spatial sampling across the source and the extended, resolved-out emission from the IRS3B$+$A envelope/core. Similarly, strong spatial integration (width of slice 0\farcs3) restrictions were placed when making the PV diagram to limit the inclusion of large-scale emission. \added{The white contours trace regions starting from 3$\sigma$ at 2$\sigma$\space intervals, where $\sigma\approx$0.15~Jy.}\deleted{Further constraints on the viewing distance (0\farcs5 displacement from the source center) were placed to help avoid the large scale emission.}}\label{fig:l1448irs3a_cso_pv}
\end{figure}

\begin{figure}[H]
\begin{center}
\includegraphics[width=0.44\textwidth]{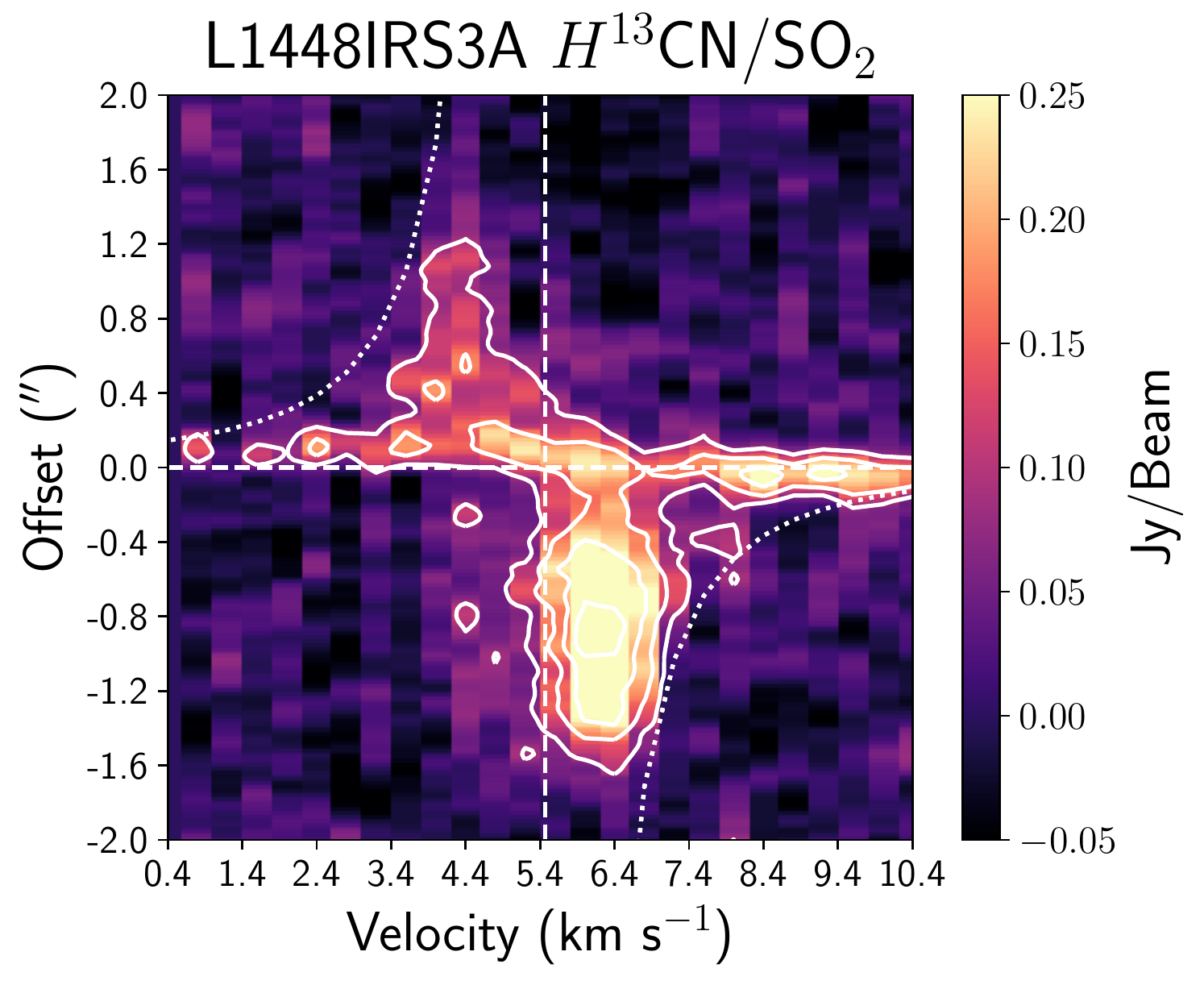}
\end{center}
\caption{Position-velocity diagram of IRS3A \htcn/\sot\space emission with the dotted lines corresponding to Keplerian velocities for a 1.4~\solm\space protostar. This PV diagram places a constraint on the possible protostellar mass parameter of \ab1.4~\solm. The IRS3A mass is less well constrained due to the compactness of the emission. Strong spatial integration (width of slice 0\farcs3) restrictions were placed when making the PV diagram to help limit the inclusion of large scale emission.\deleted{Further constraints on the viewing distance (0\farcs5 displacement from the source center) were placed to help limit contributions from the large scale emission.}}\label{fig:l1448irs3a_h13cn_pv}
\end{figure}

\begin{figure}[H]
\begin{center}
   \includegraphics[width=0.44\textwidth]{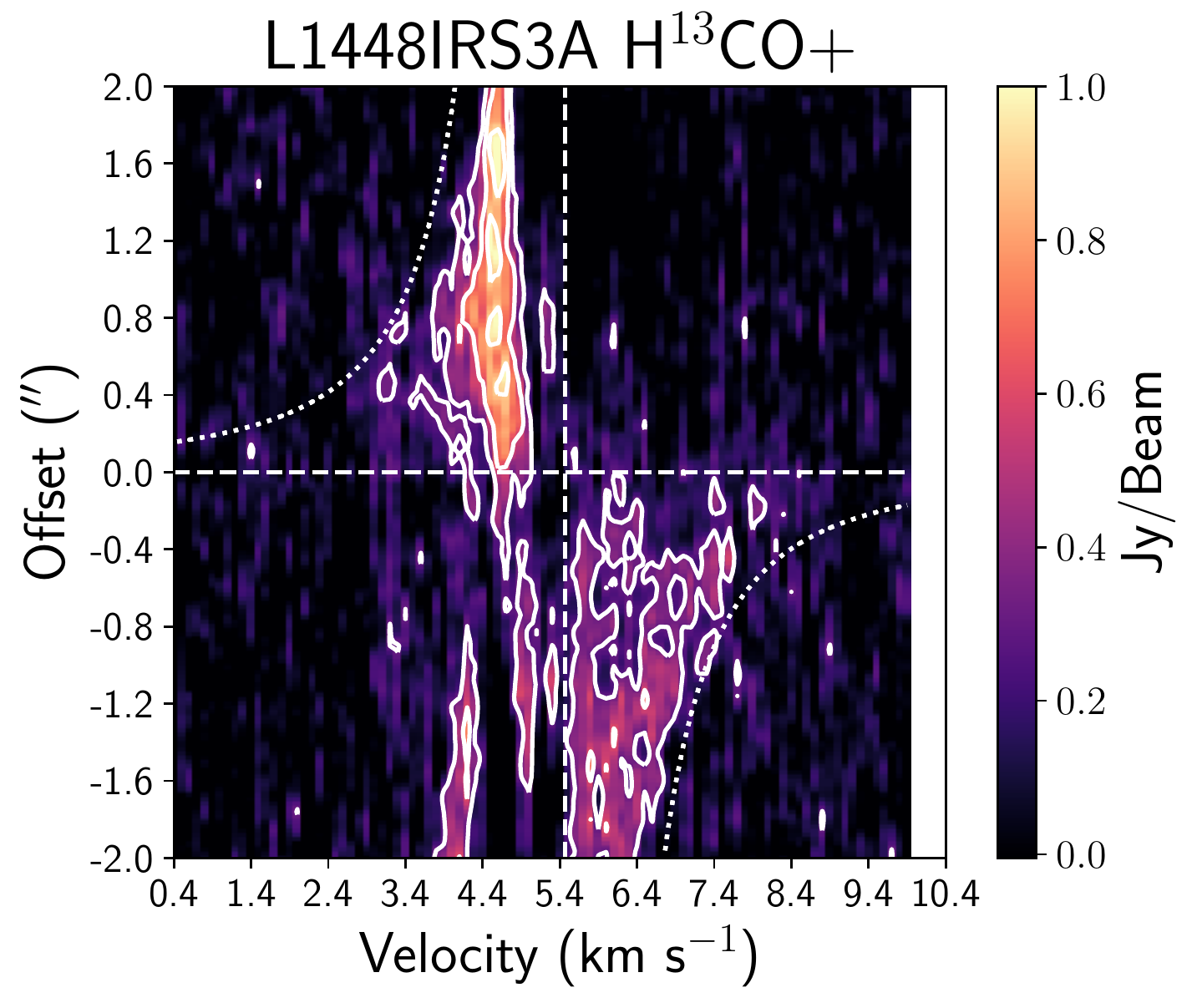}
\end{center}
   \caption{Position-velocity diagram of IRS3A \htcop\space  generated at a position angle of 125\deg; whose emission predominately traces the intermediate dense, gaseous material of the inner envelope. The emission \deleted{suffers lower spatial sampling across the source and less emission is present towards the source}\added{is fainter and is coming from the outerdisk/inner-envelope}. The dotted line corresponds to a central protostellar mass of 1.4~\solm.}\label{fig:l1448irs3a_h13cop_pv}
\end{figure}

\begin{figure}[H]
\begin{center}
   \includegraphics[width=0.48\textwidth]{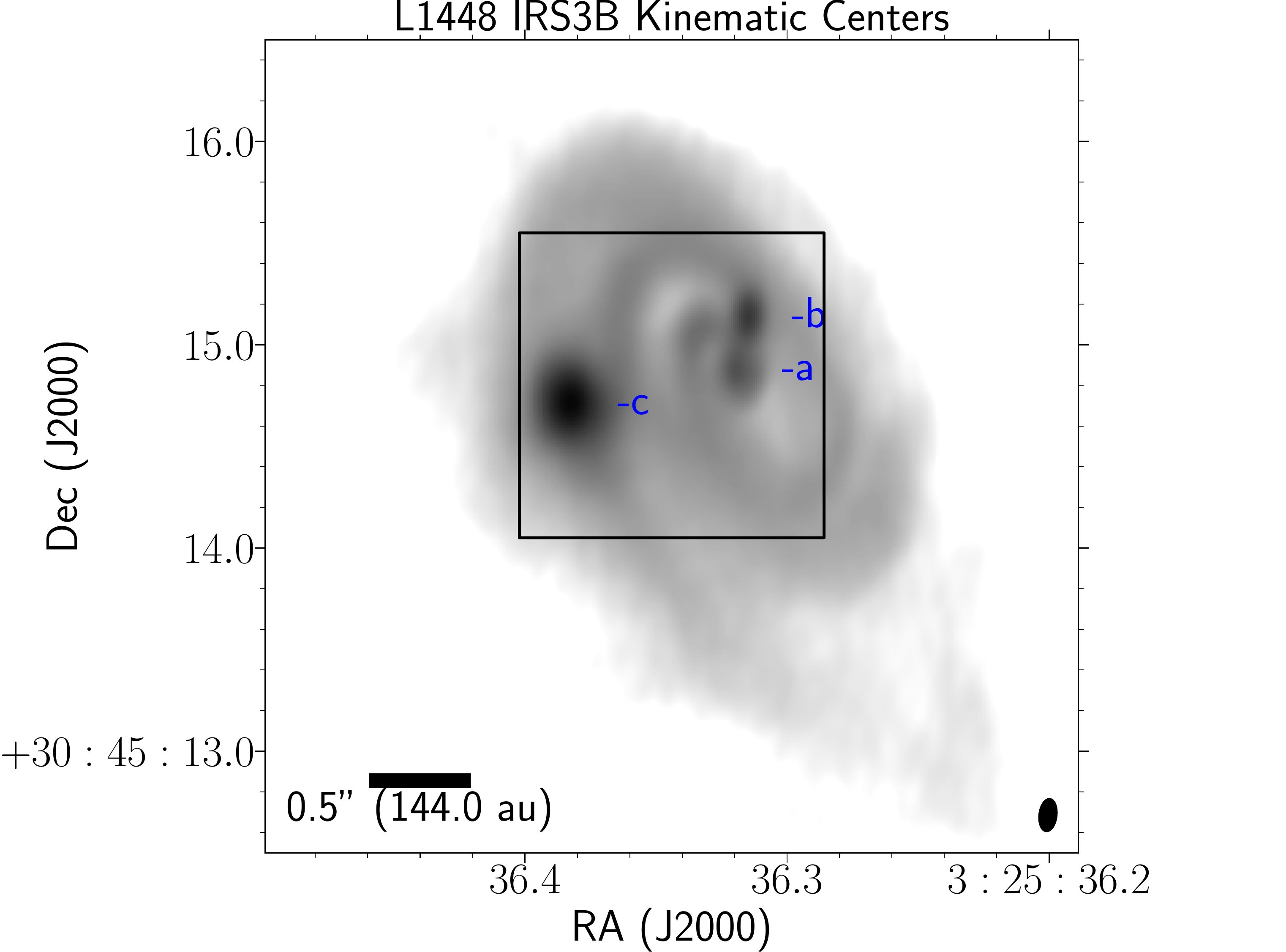}
   \includegraphics[width=0.48\textwidth]{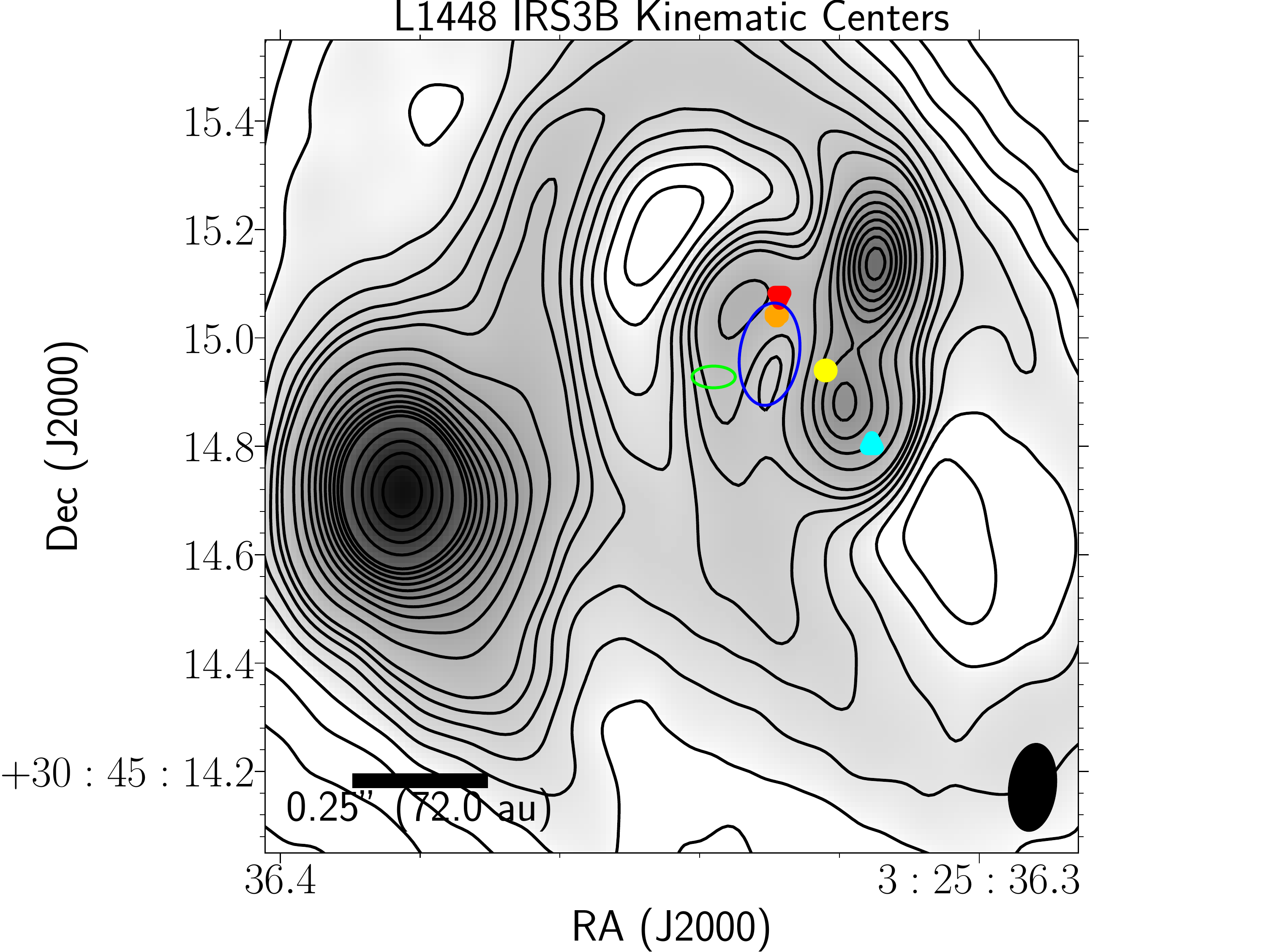}
\end{center}
   \caption{Positions of the various ``kinematic centers'' that have been fit from \cso\space emission at IRS3B in relation to continuum structure. The grayscale is the dust continuum from Figure~\ref{fig:contimage}. Left: The red colored texts detail the locations of continuum sources, presumed to be protostars. Right: A zoom in on the region indicated by the black rectangle in the left image. The red and blue triangles indicate the central Gaussian fit of the highest Doppler-shifted velocity emission with the yellow circle indicating the midpoint. The orange circle indicates the center that best constructs the PV diagram symmetrically. The green ellipse is the model Keplerian centroid fit with the respective error as indicated by the size of the ellipse (see Section~\ref{sec:kmodelresults}). The blue ellipse is the \cso\space beam (\csobeam) centered on the region of emission deficit for size comparison. The contours start at 10$\sigma$\space and iterate by 10$\sigma$\space with the 1$\sigma$~level starting at 8.5$\times10^{-5}$~Jy~beam$^{-1}$. The region of deficit, first identified in Figure~\ref{fig:zoomincont}\space is shown to be centered within the three various kinematic center fits and are marginally separated by less than a few beams.}\label{fig:kincenter}
\end{figure}

\begin{figure}[H]
\begin{center}
\includegraphics[width=0.5\textwidth]{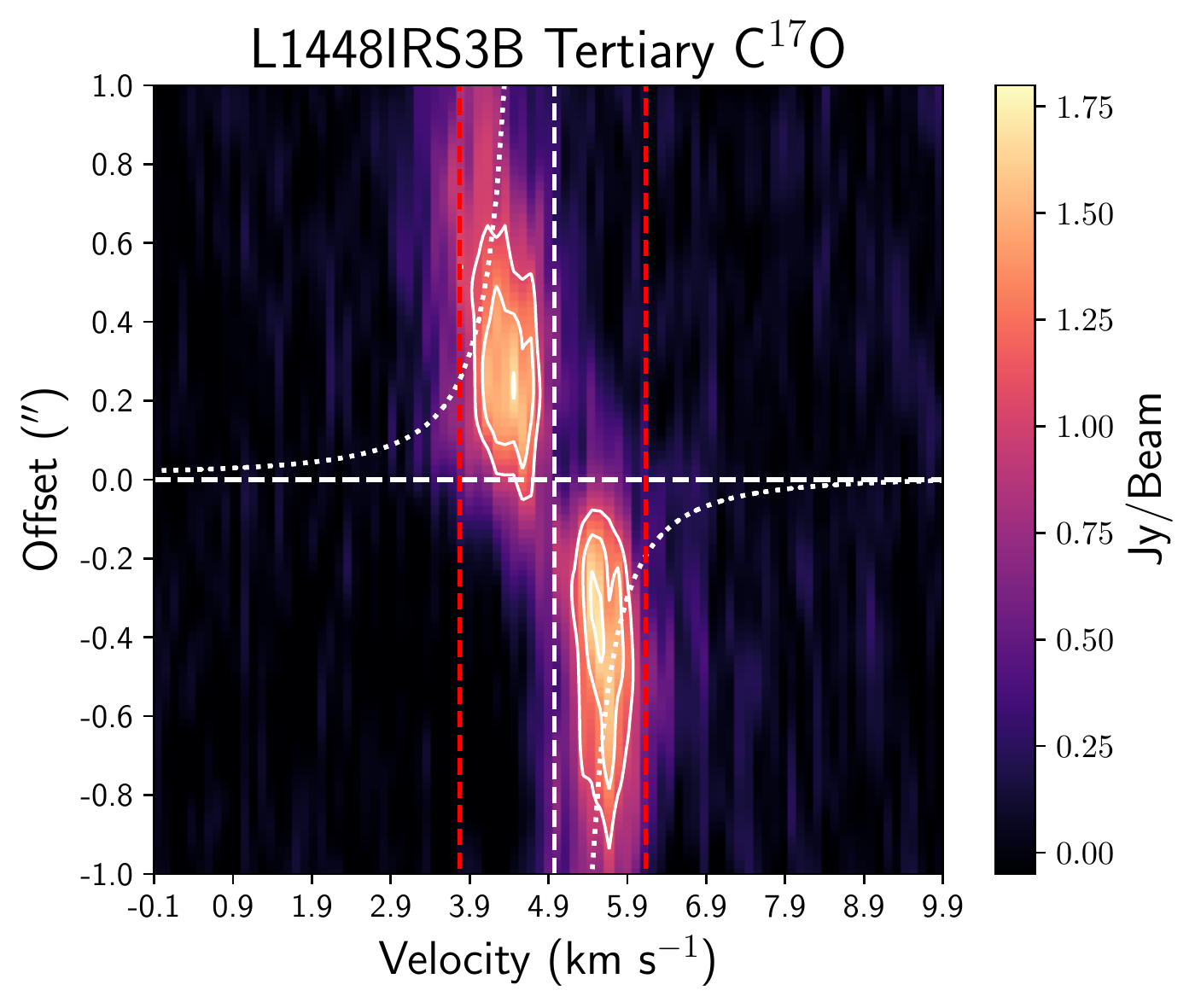}
\end{center}
\caption{PV-diagram of \cso\space toward IRS3B-c, the tertiary. The white lines corresponding to a Keplerian curve of a 0.2~\solm\space protostellar source. These fits place an upper limit to the mass of the tertiary companion to $<$0.2~\solm, since any larger mass and we would expect to see emission extending to high velocity, indicating the tertiary would be affecting disk kinematics. The red dashed lines indicate the maximum Keplerian velocities at the radius of IRS3B-c in the rotating disk corresponding to the 1.15~\solm\space mass of the central potential. Emission outside of these bounds could be due to the tertiary affecting disk kinematics, but from this analysis, we cannot detect an obvious effect of the tertiary on the disk kinematics. The white/black contours trace regions starting from 14$\sigma$ at 4$\sigma$\space intervals, where $\sigma\approx$0.1~Jy~beam$^{-1}$.}\label{fig:l1448irs3b_c17o_pv_tert}
\end{figure}
\begin{figure}[H]
\begin{center}
\includegraphics[width=0.5\textwidth]{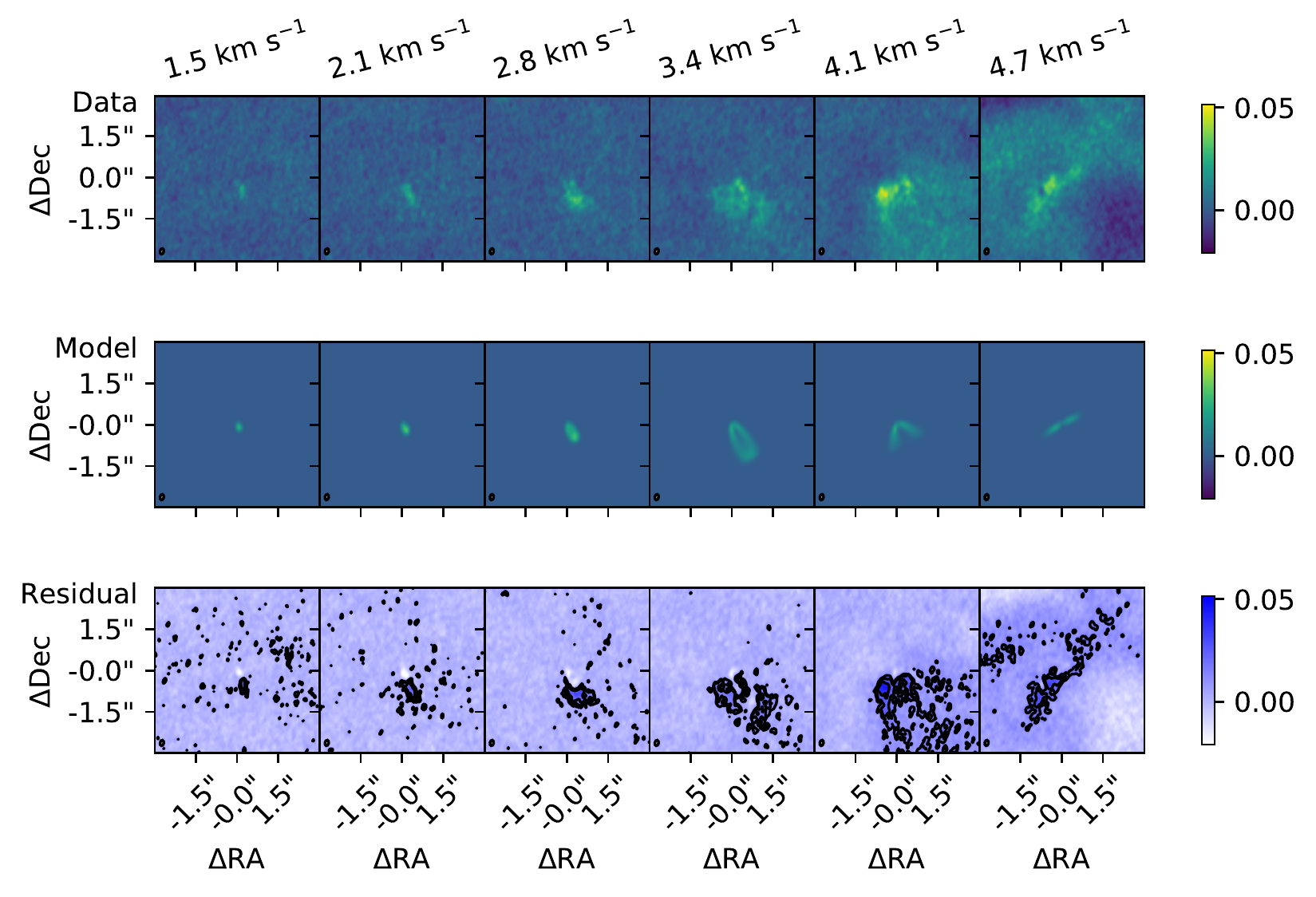}
\includegraphics[width=0.5\textwidth]{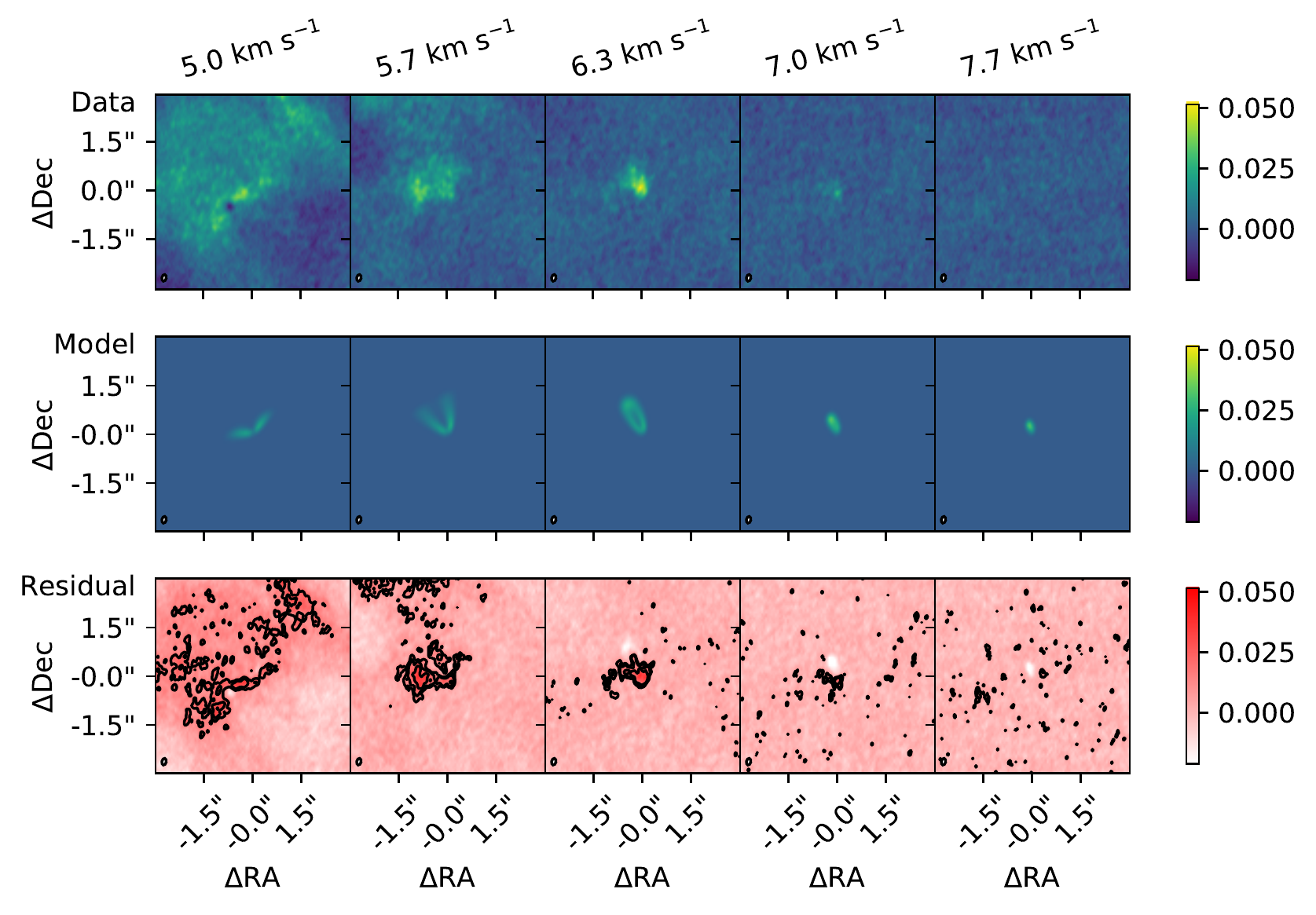}
\end{center}
\caption{IRS3B Kinematic Model comparison: A representative selection of channel maps that demonstrate the fit of the model to the data. The top figure is the blue Doppler shifted emission while the bottom figure is the red Doppler shifted emission. The first row contours are the model contours, generated at the 2, 3, 5, and 10$\sigma$ level overlaid the data channels selected at the same velocity. The second row is the residual contours (2 and 3$\sigma$) overlaid the same data channels. System velocity is \ab4.8~km~s$^{-1}$. It should be noted the highly correlated structure visible in the residuals. This reflects an imperfect fit to the data given that the circumstellar disk itself is asymmetric.}\label{fig:c17o_res}
\end{figure}

\begin{figure}[H]
\begin{center}
\includegraphics[width=0.5\textwidth]{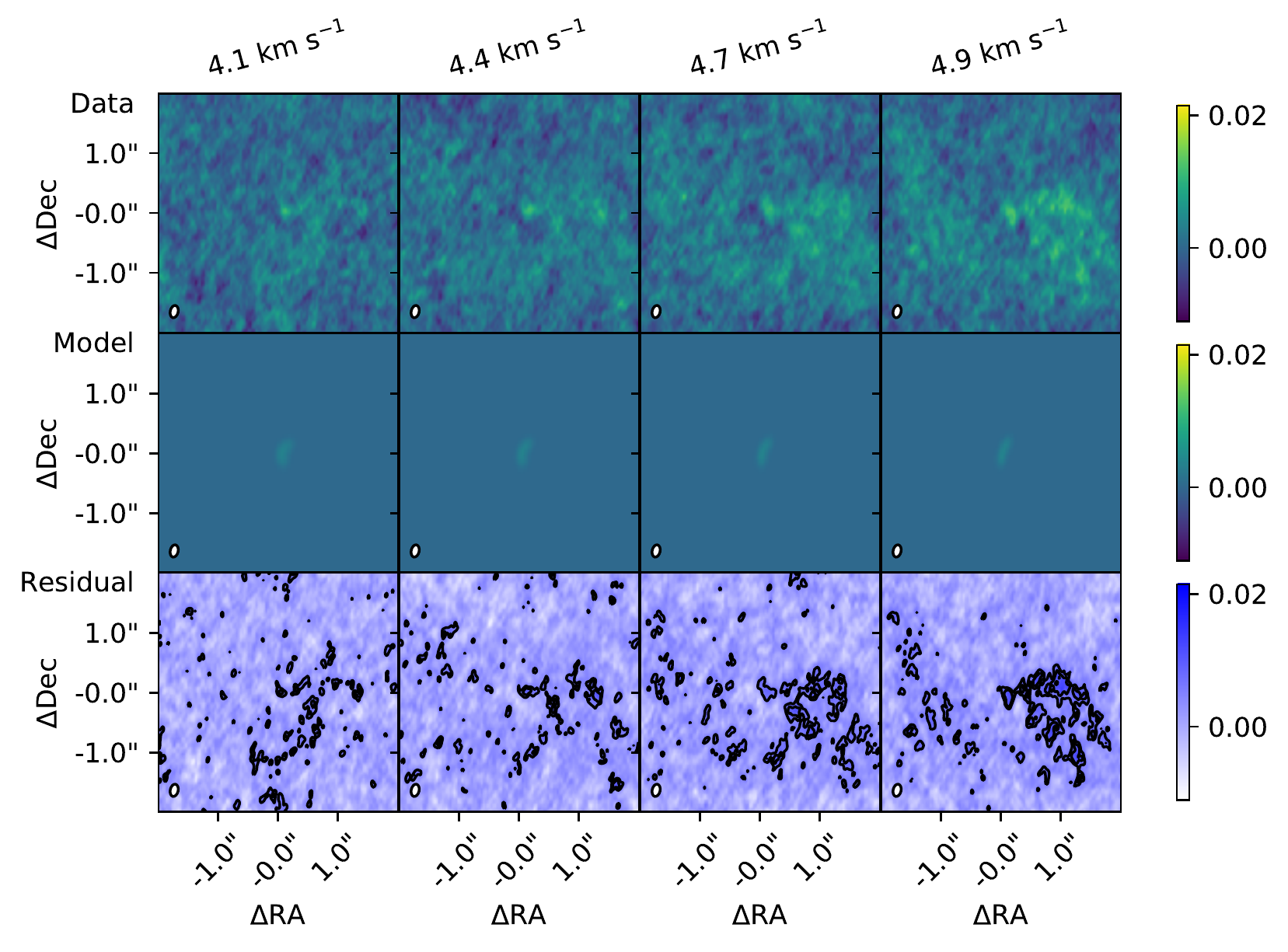}
\includegraphics[width=0.5\textwidth]{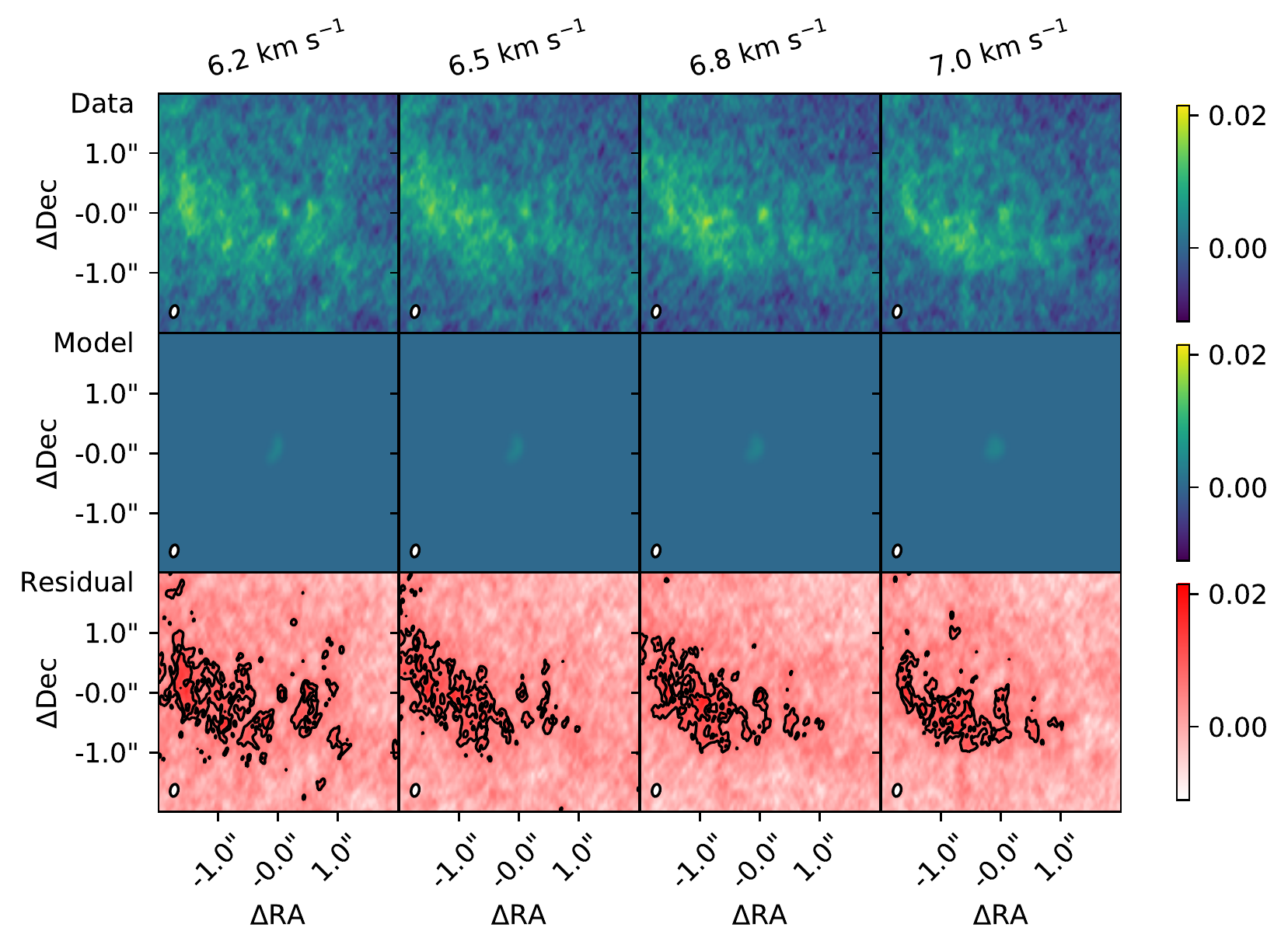}
\end{center}
\caption{IRS3A Kinematic Model comparison: A representative selection of channel maps that demonstrate the fit of the model to the data. The top figure is the blue Doppler shifted emission while the bottom figure is the red Doppler shifted emission. The first row contours are the model contours, generated at the 2, 3, 5, and 10$\sigma$  level overlaid the data channels selected at the same velocity to not overshadow the emission. The second row is the residual contours overlaid the same data channels. System velocity is \ab5.2~km~s$^{-1}$. There is residual emission at scales much larger than the continuum disk, especially prevalent near the system velocity, likely due to large scale emission from the cloud that is not included in the disk.}\label{fig:h13cn_res}
\end{figure}

\begin{figure}[H]
\begin{center}
\includegraphics[width=0.49\textwidth]{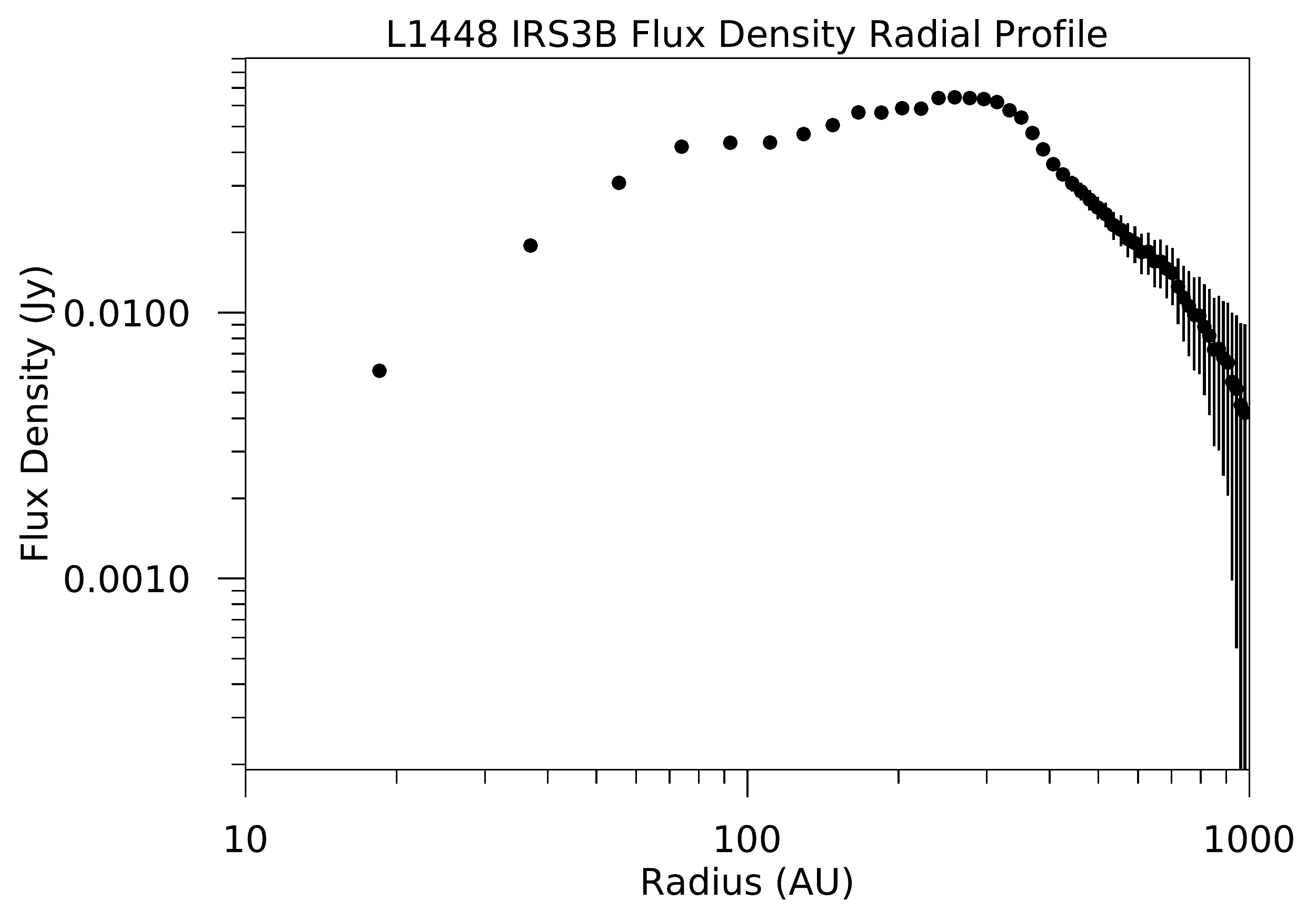}
\includegraphics[width=0.49\textwidth]{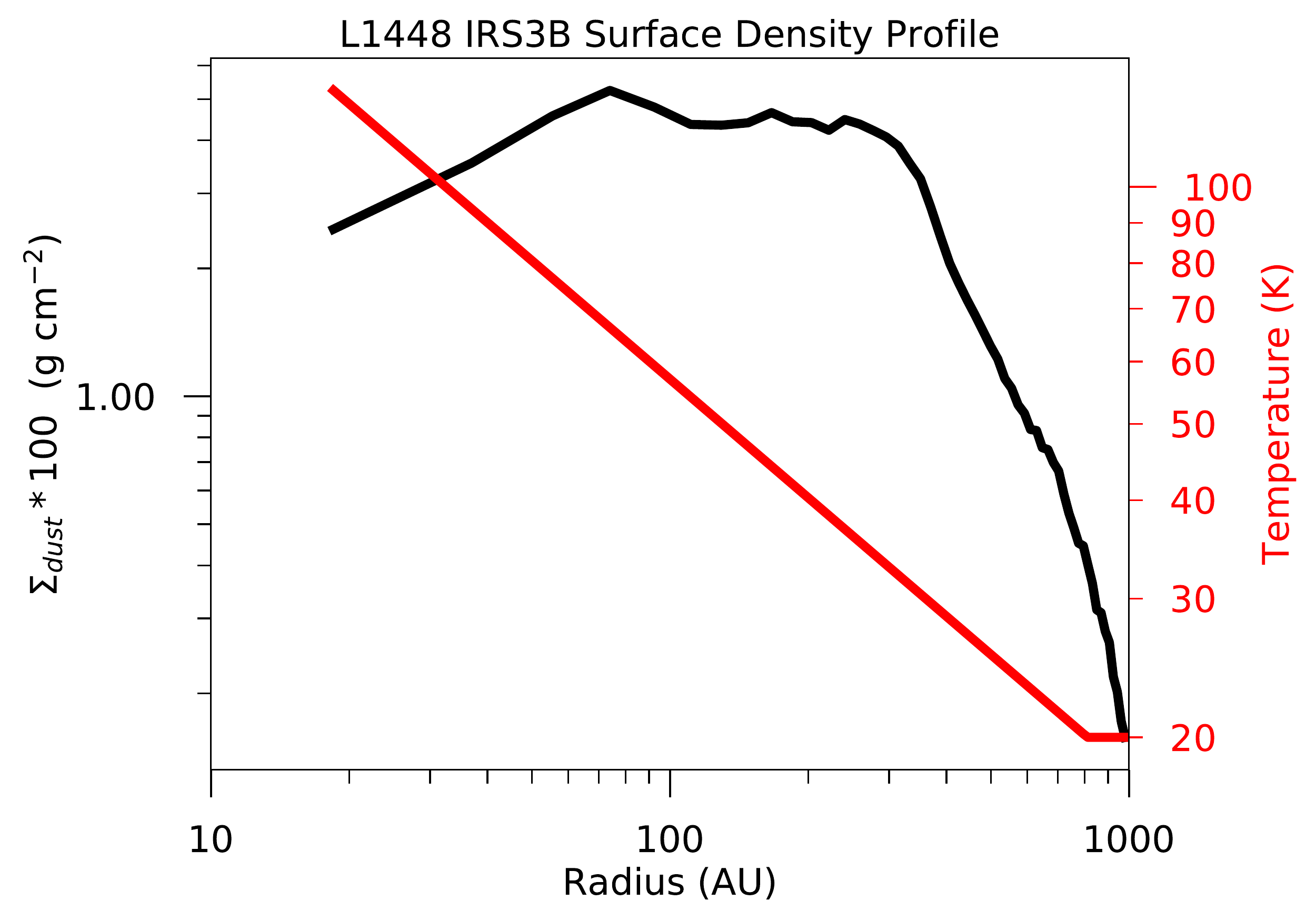}
\end{center}
\caption{The left plot is the continuum flux \added{density} radial profile of IRS3B. The right plot is the deprojected radial surface density profile of the dust continuum in black \deleted{points}, while the red line is the radial temperature profile of the disk. \replaced{The slope is -0.5 and the temperature is scaled such that the temperature of the disk}{The temperature profile is $\propto\text{R}^{-0.5}$ and is scaled such that} at 100~AU is described via (30~K)$\times(L_{*}$/\lsun)$^{0.25}\approx40.1$~K.}\label{fig:surfacedensity}
\end{figure}

\begin{figure}[H]
\begin{center}
\includegraphics[width=0.49\textwidth]{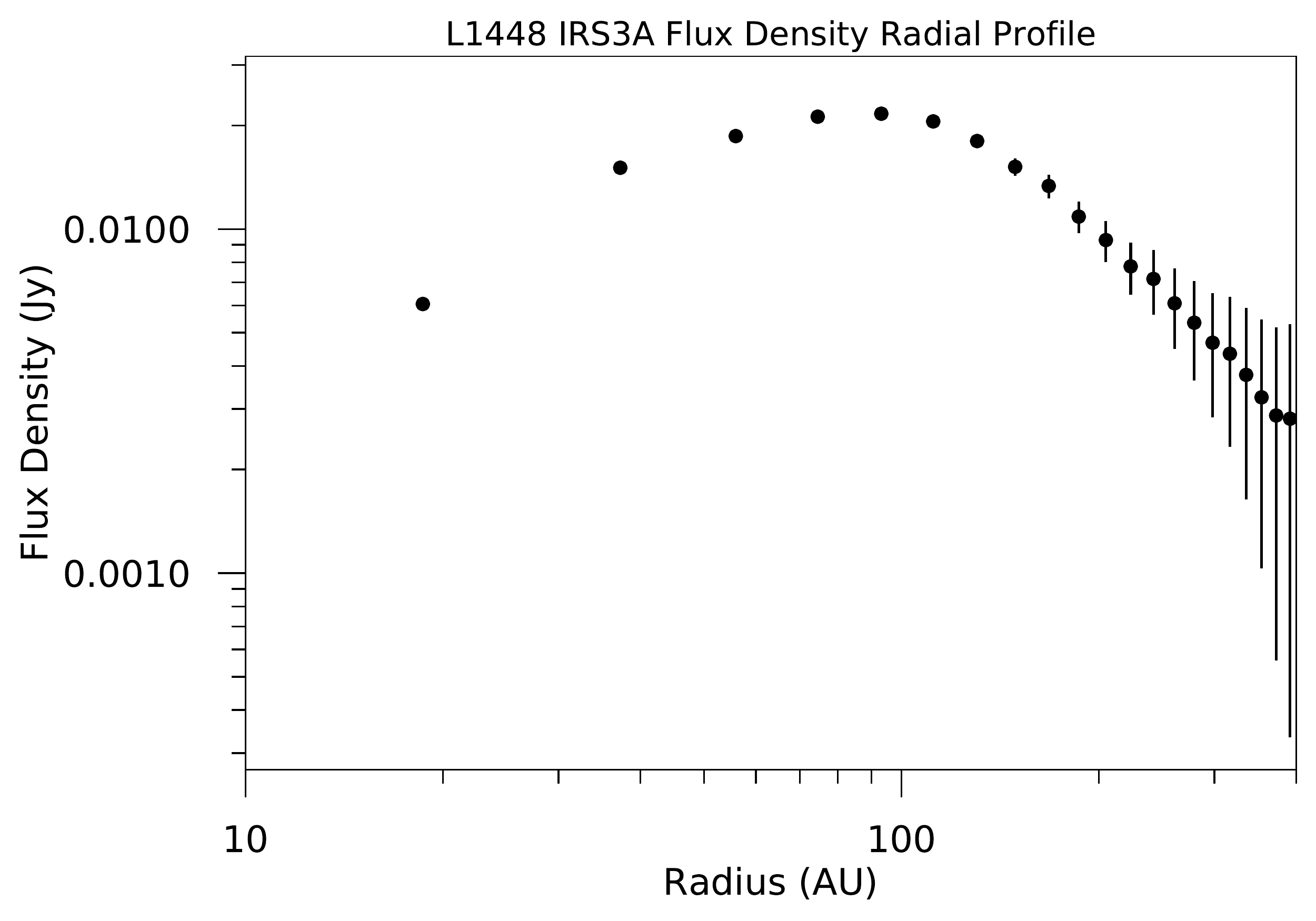}
\includegraphics[width=0.49\textwidth]{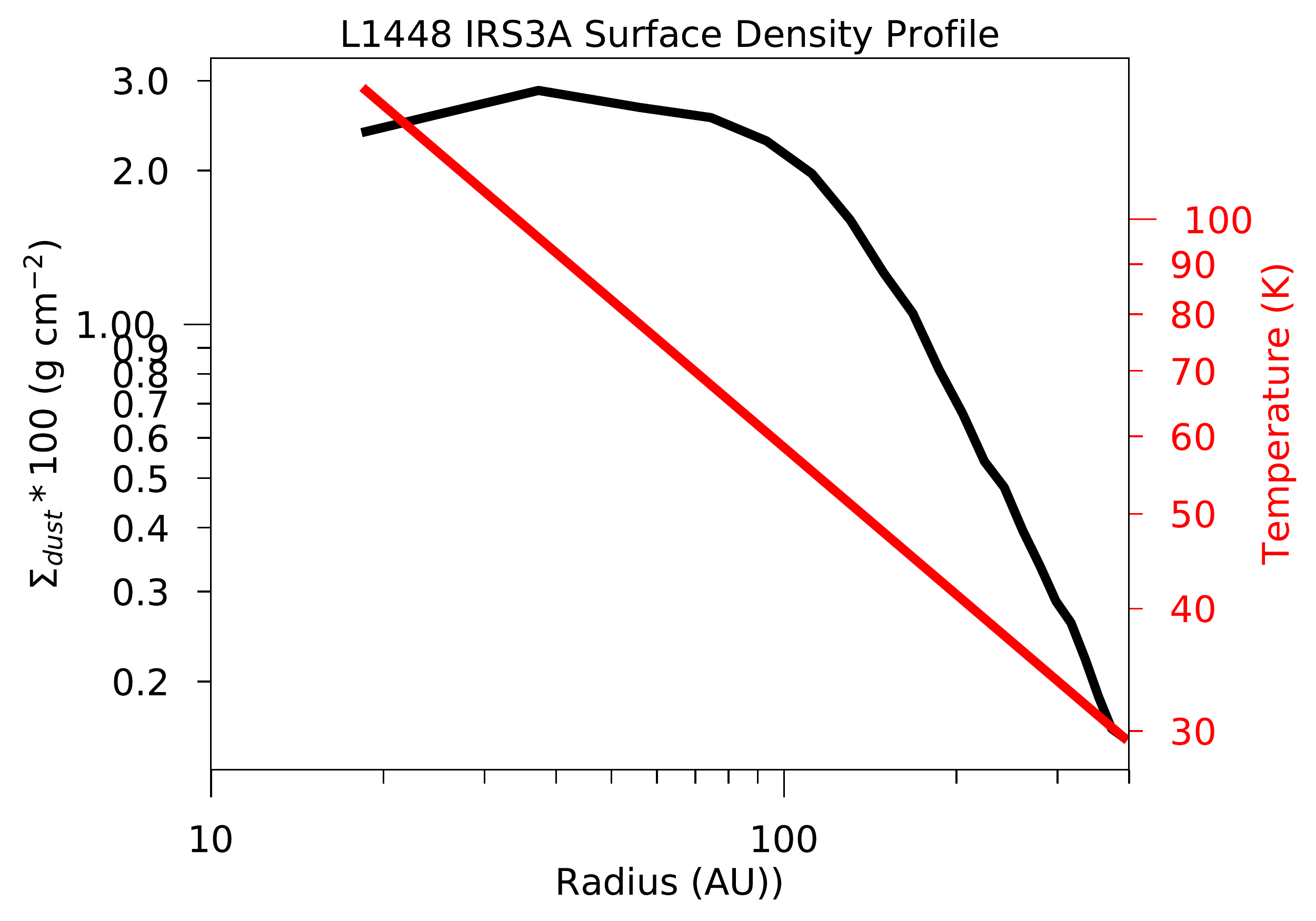}
\end{center}
\caption{The left plot is the continuum flux \added{density} radial profile of IRS3A. The right plot is the radial surface density profile of the dust continuum in black\deleted{points}, while the red line is the radial temperature profile of the disk. \replaced{The slope is -0.5 and the temperature is scaled such that the temperature of the disk}{The temperature profile is $\propto\text{R}^{-0.5}$ and is scaled such that} at 100~AU is described via (30~K)$\times(L_{*}$/\lsun)$^{0.25}\approx53.1$~K.}\label{fig:irs3asurfacedensity}
\end{figure}

\begin{figure}[H]
\begin{center}
\includegraphics[width=0.48\textwidth]{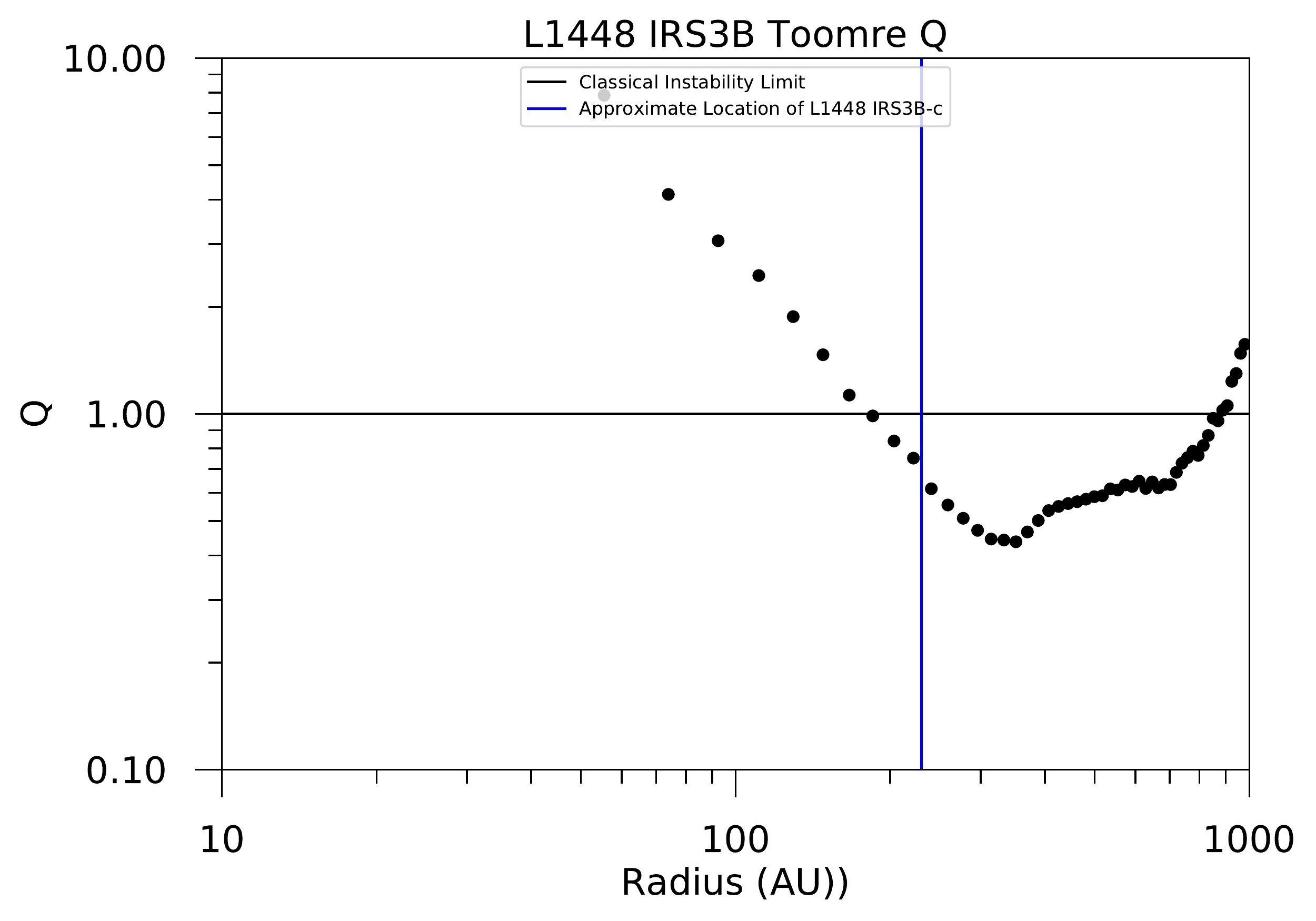}
\end{center}   
\caption{Toomre Q parameter plotted as a function of deprojected radius for IRS3B. The horizontal line indicates a Toomre Q parameter of one, at which the disk would be gravitationally unstable. As indicated, the disk Toomre Q parameter drops below 1 at a radius of \ab120~AU. The vertical line corresponds to the deprojected radius of IRS3B-c. The observed spiral arms also become most prominent at R $>$100~AU, where Toomre~Q approaches 1.}\label{fig:irs3btoomreq}
\end{figure}

\begin{figure}[H]
\begin{center}
\includegraphics[width=0.48\textwidth]{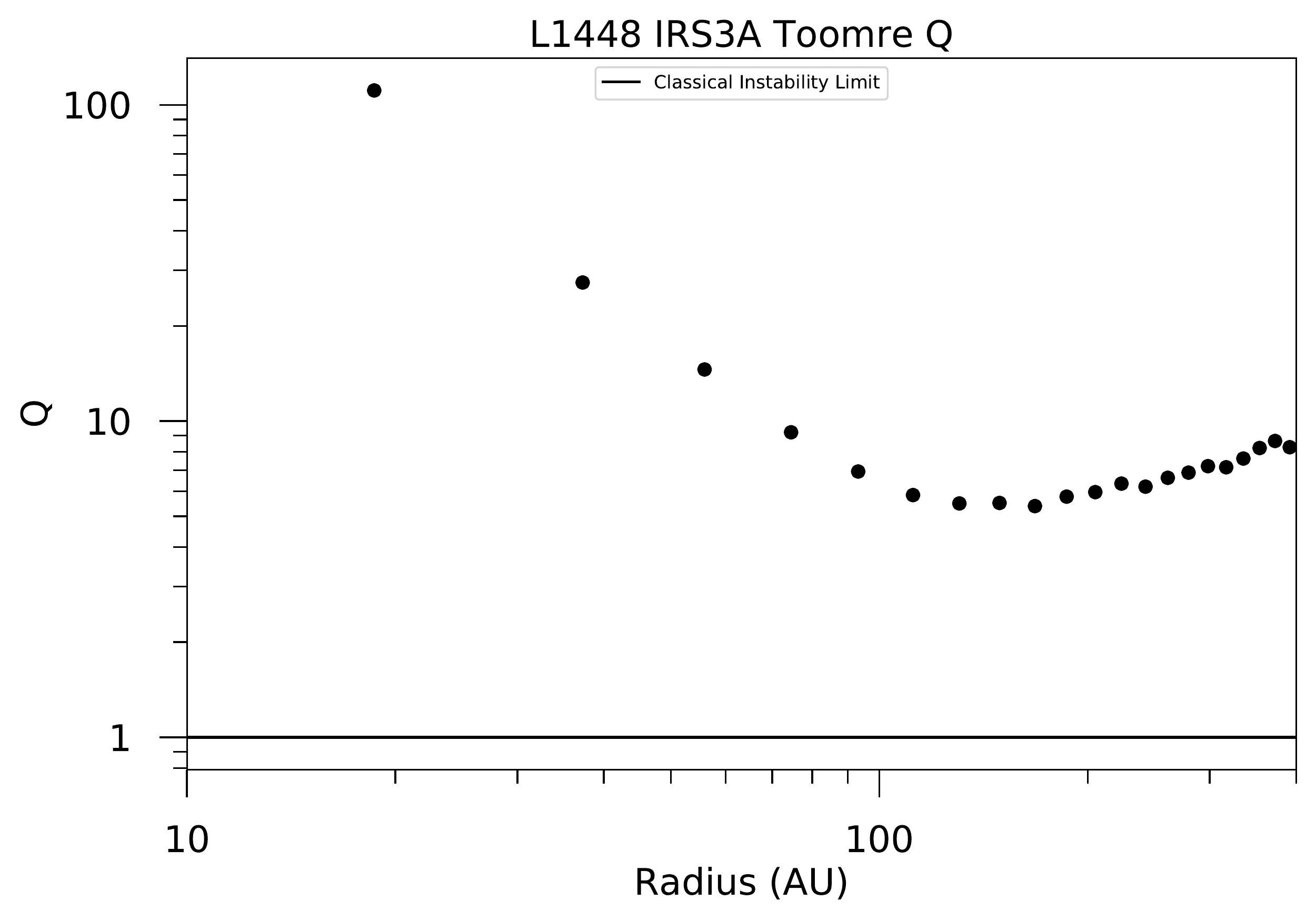}
\end{center}   
\caption{Toomre Q parameter plotted as a function of deprojected radius for IRS3A. The horizontal line indicates a Toomre Q parameter of one, at which the disk would be gravitationally unstable. The circumstellar disk of IRS3A is much less massive than IRS3B, coupled with a \replaced{more massive central gravitational source}{more massive protostar}, the disk is more stable against gravitational instabilities.}\label{fig:irs3atoomreq}
\end{figure}

\begin{figure}[H]
   \begin{center}
   \includegraphics[width=\textwidth]{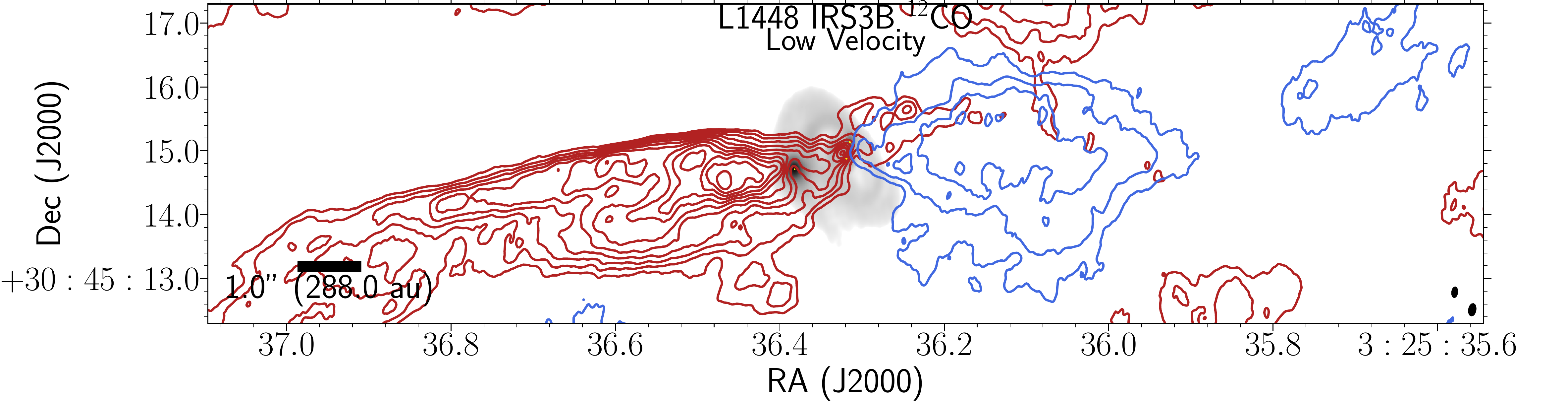}
   \includegraphics[width=\textwidth]{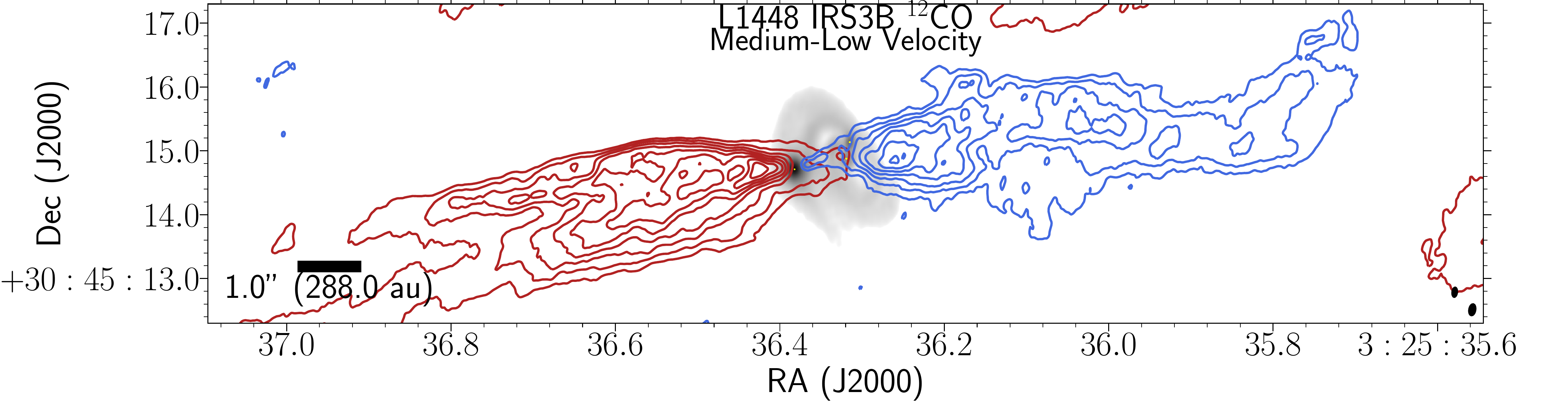}
   \includegraphics[width=\textwidth]{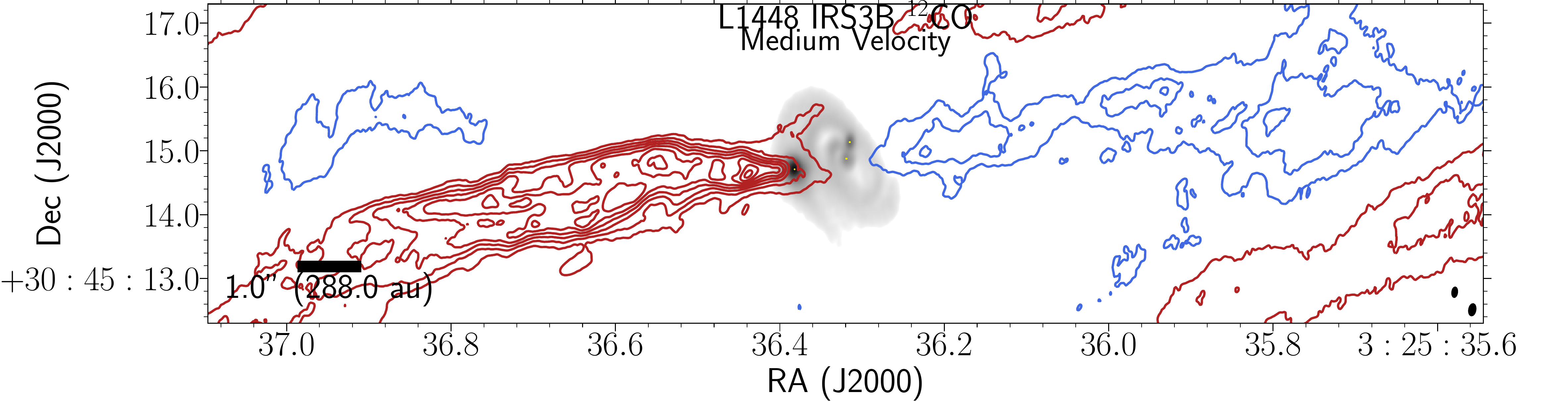}
   \caption{Moment 0 map (integrated intensity) of \co, overlaid on the continuum (grayscale) image from Figure~\ref{fig:contimage}, split up according to velocity ranges, providing exquisite detailing of the location and collimated of the IRS3B outflows. The central outflow from IRS3B extends 10\arcsec\space(2880~au)\added{, beyond the edge of the primary beam of ALMA at 879~micron,} from launch location on either side. The panels correspond to low, medium-low, and medium velocity ranges which are delineated as red(blue), respectively. \textbf{Low Velocity:} velocity ranges 5.5$\rightarrow$10.5~km~s$^{-1}$ (4$\rightarrow$-1~km~s$^{-1}$), contours start at 3(3)~$\sigma$ and iterate by 2(2)~$\sigma$ with the 1-$\sigma$~level starting at 0.1(0.1)~Jy~beam$^{-1}$ for the red(blue) channels respectively. \textbf{Medium-low Velocity:} velocity ranges 10.5$\rightarrow$15.5~km~s$^{-1}$ (-6$\rightarrow$-4~km~s$^{-1}$), contours start at 5(5)~$\sigma$ and iterate by 3(2)~$\sigma$ with the 1-$\sigma$~level starting at 0.04(0.004)~Jy~beam$^{-1}$ for the red(blue) channels respectively. \textbf{Medium Velocity:} velocity ranges 15.5$\rightarrow$20.5~km~s$^{-1}$ (-11$\rightarrow$-6~km~s$^{-1}$), contours start at 10(10)~$\sigma$ and iterate by 4(4)~$\sigma$ with the 1-$\sigma$~level starting at 0.02(0.02)~Jy~beam$^{-1}$ for the red(blue) channels respectively. The \co\space synthesized beam (\cobeam) is the bottom-right most overlay on each of the panels and the continuum synthesized beam (\contbeam) is offset diagonally.}\label{fig:comomentmap}
\end{center}
\end{figure}
\begin{figure}[H]
   \begin{center}
   \includegraphics[width=\textwidth]{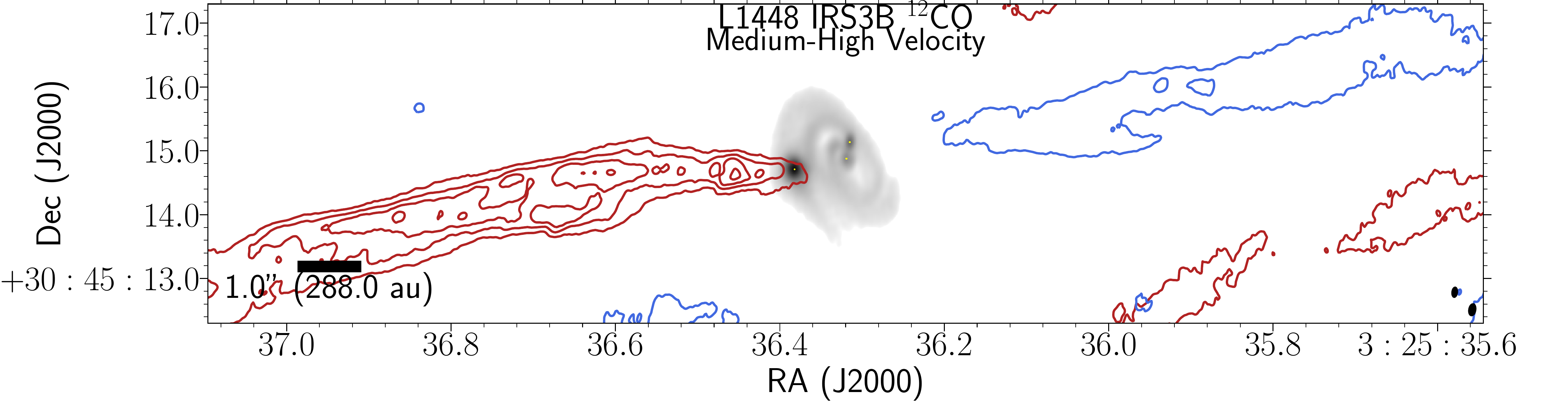}
   \includegraphics[width=\textwidth]{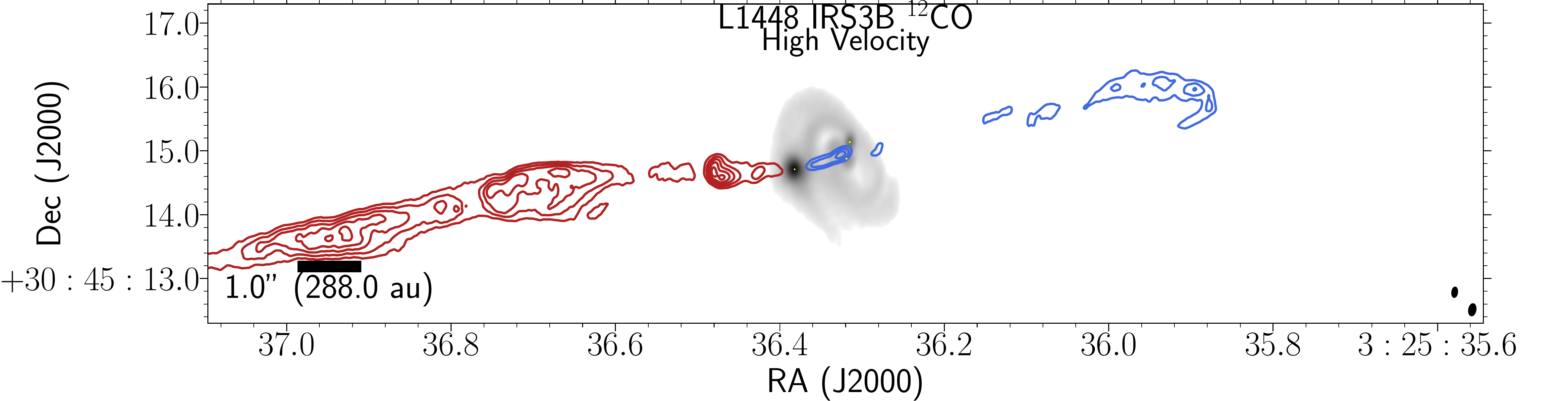}
   \caption{Same as Figure~\ref{fig:comomentmap} but for the \textbf{Medium-high Velocity:} velocity ranges 20.5$\rightarrow$25.5~km~s$^{-1}$ (-16$\rightarrow$-11~km~s$^{-1}$), contours start at 3(3)~$\sigma$ and iterate by 4(4)~$\sigma$ with the 1-$\sigma$~level starting at 0.04(0.04)~Jy~beam$^{-1}$ for the red(blue) channels respectively. \textbf{High Velocity:} velocity ranges 25.5$\rightarrow$30.5~km~s$^{-1}$ (-21$\rightarrow$-16~km~s$^{-1}$), contours start at 5(5)~$\sigma$ and iterate by 2(2)~$\sigma$ with the 1-$\sigma$~level starting at 0.04(0.04)~Jy~beam$^{-1}$ for the red(blue) channels respectively. The \co\space synthesized beam (\cobeam) is the bottom-right most overlay on each of the panels and the continuum synthesized beam (\contbeam) is offset diagonally.}\label{fig:comomentmap2}
\end{center}
\end{figure}

\begin{figure}[H]
   \begin{center}

\includegraphics[width=0.45\textwidth]{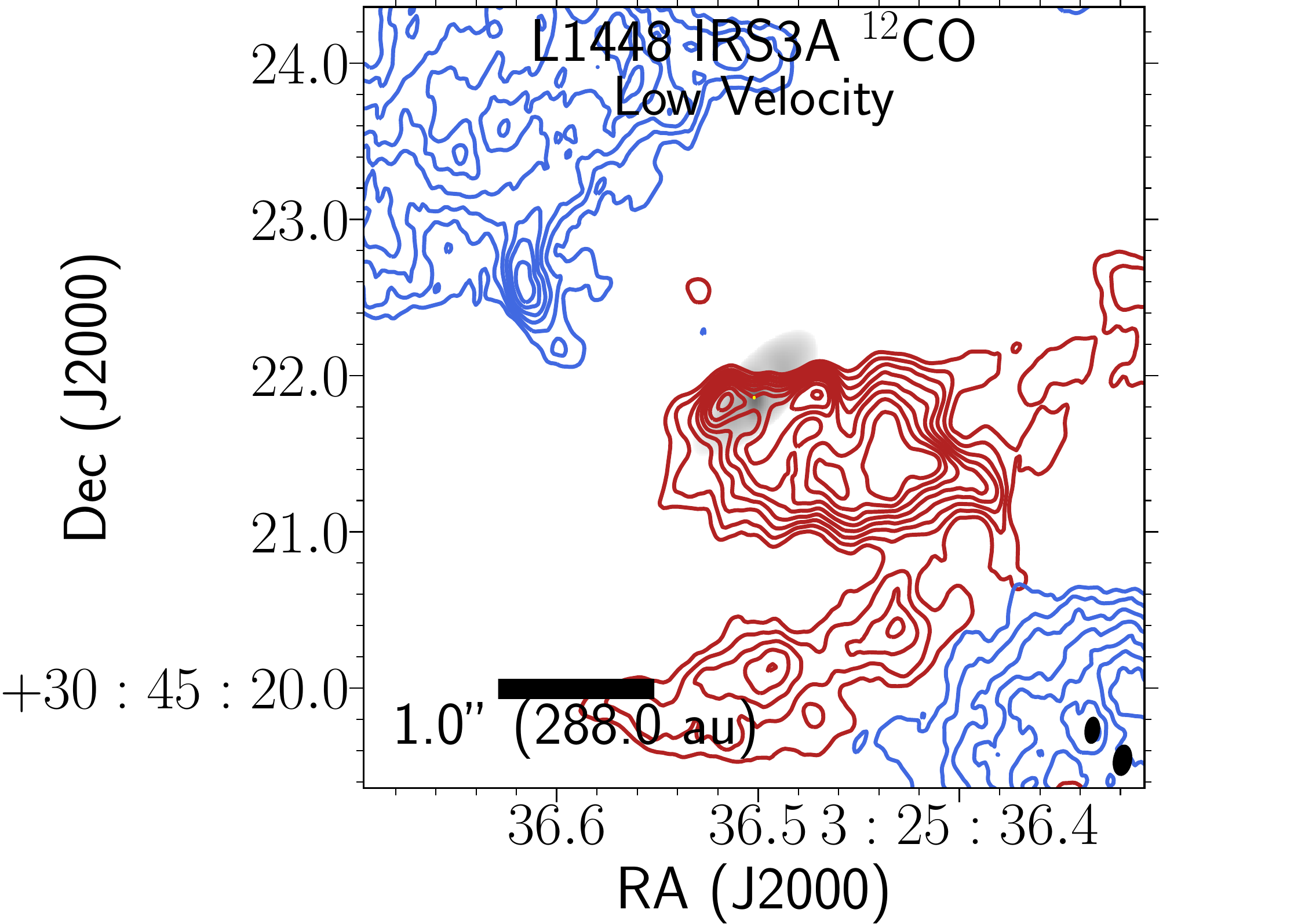}
\includegraphics[width=0.45\textwidth]{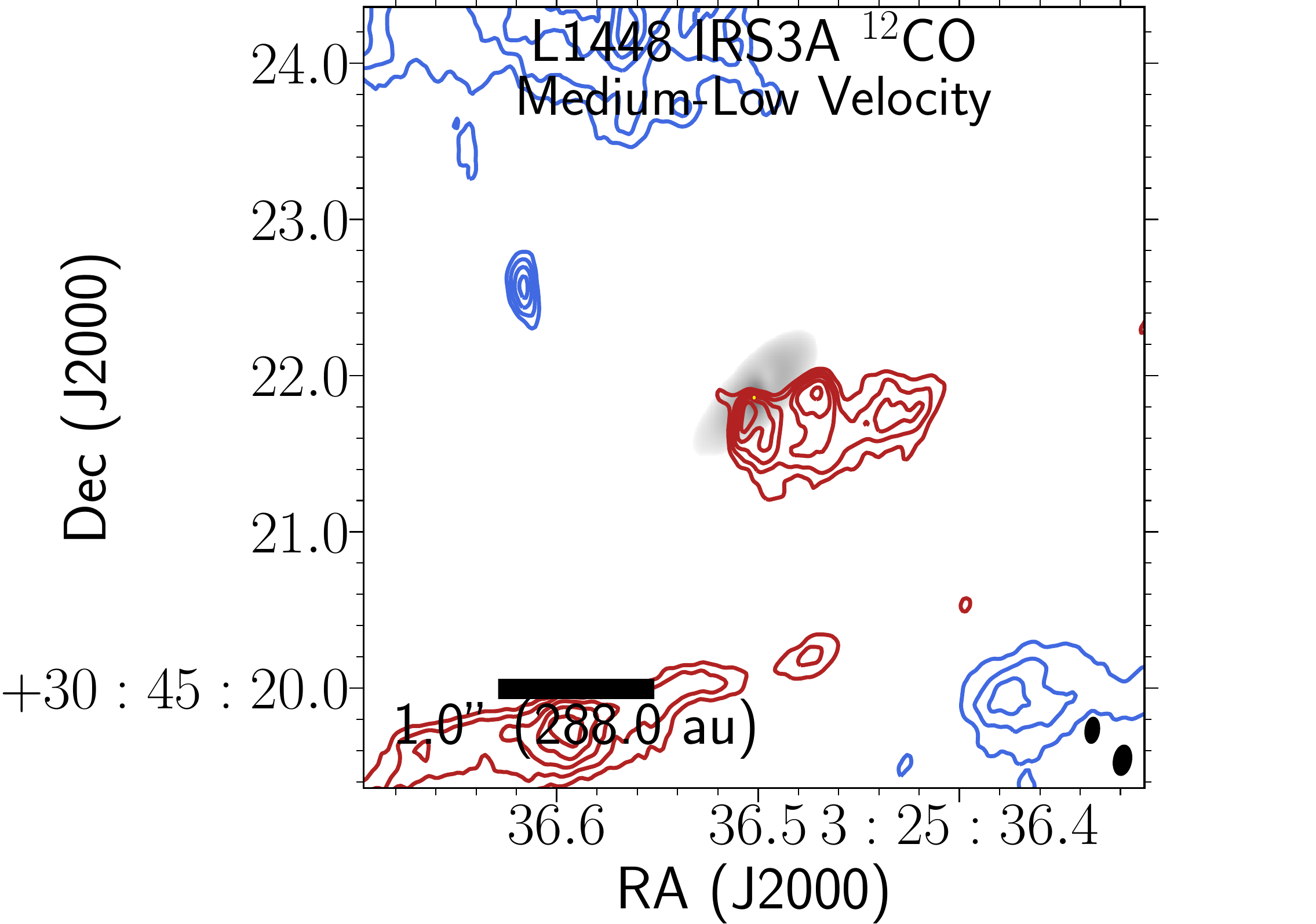}
\includegraphics[width=0.45\textwidth]{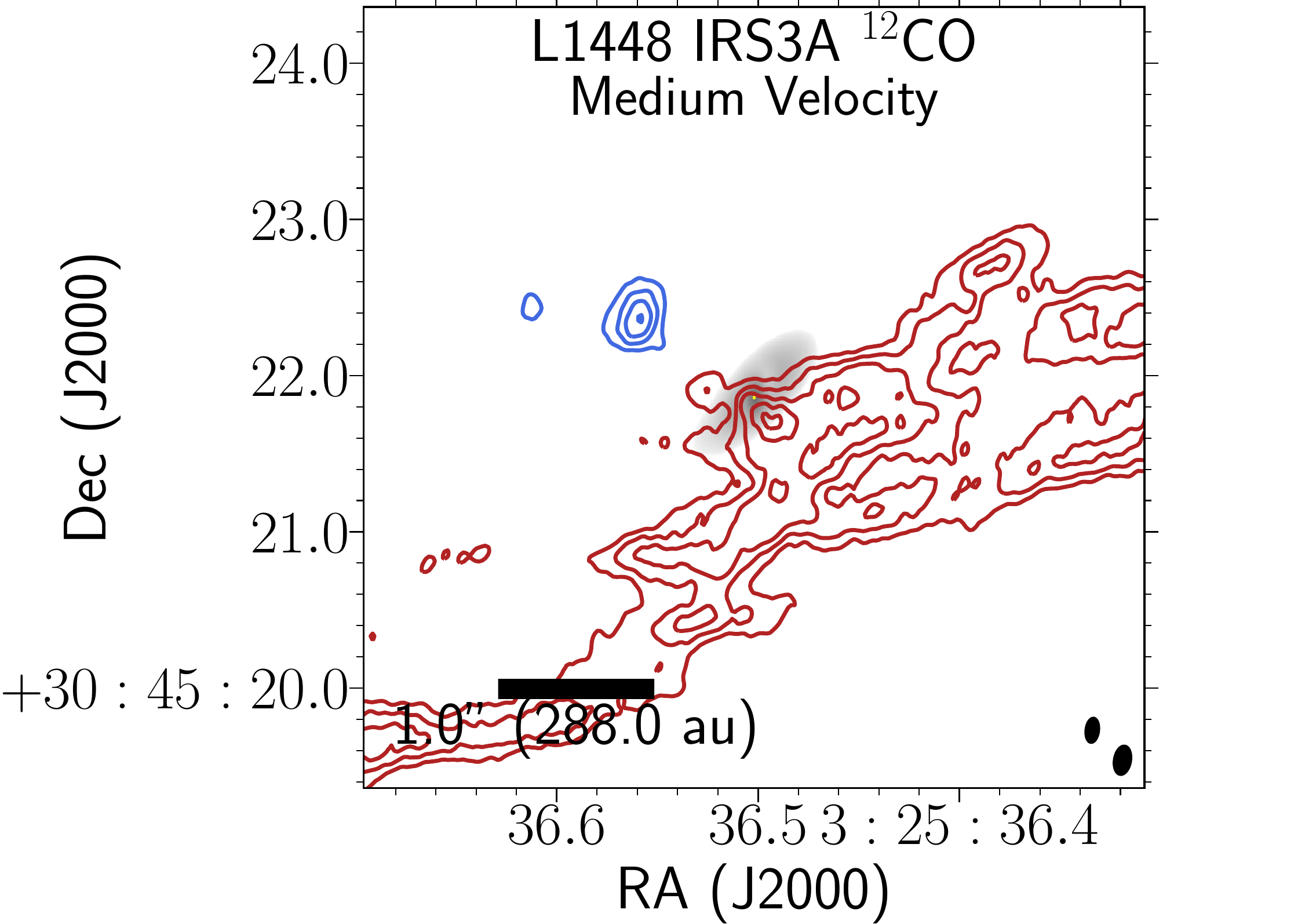}
\includegraphics[width=0.45\textwidth]{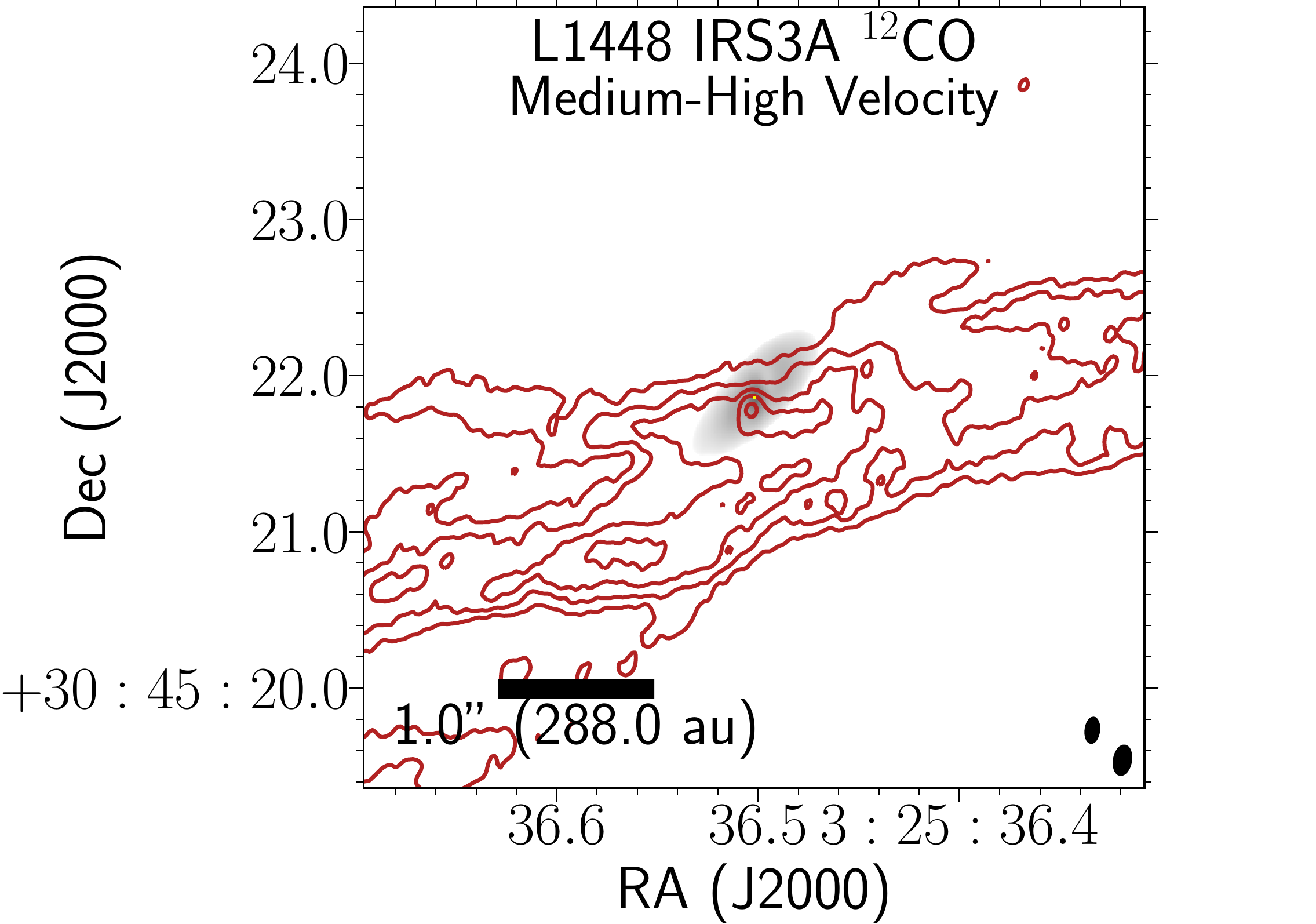}
\includegraphics[width=0.45\textwidth]{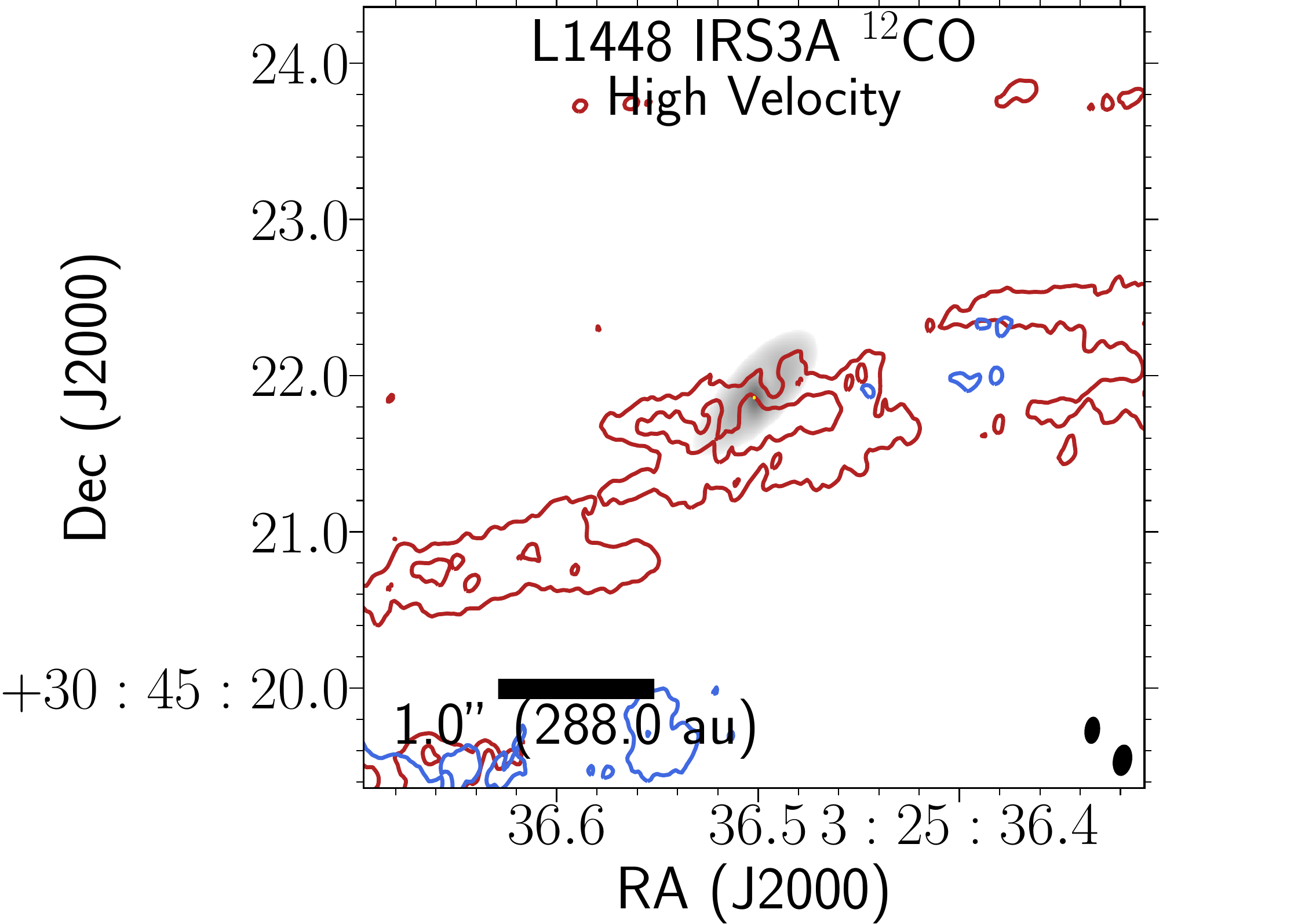}
   \caption{Similar to Figures~\ref{fig:comomentmap}~and~\ref{fig:comomentmap2} towards the IRS3A source with the same velocity ranges. \textbf{Low Velocity:} velocity ranges 5.5$\rightarrow$10.5~km~s$^{-1}$ (4$\rightarrow$-1~km~s$^{-1}$), contours start at 5(5)~$\sigma$ and iterate by 4(2)~$\sigma$ with the 1-$\sigma$~level starting at 0.1(0.1)~Jy~beam$^{-1}$ for the red(blue) channels respectively. \textbf{Medium-low Velocity:} velocity ranges 10.5$\rightarrow$15.5~km~s$^{-1}$ (-6$\rightarrow$-4~km~s$^{-1}$), contours start at 5(5)~$\sigma$ and iterate by 2(2)~$\sigma$ with the 1-$\sigma$~level starting at 0.04(0.004)~Jy~beam$^{-1}$ for the red(blue) channels respectively. \textbf{Medium Velocity:} velocity ranges 15.5$\rightarrow$20.5~km~s$^{-1}$ (-11$\rightarrow$-6~km~s$^{-1}$), contours start at 5(5)~$\sigma$ and iterate by 2(2)~$\sigma$ with the 1-$\sigma$~level starting at 0.02(0.02)~Jy~beam$^{-1}$ for the red(blue) channels respectively. \textbf{Medium-high Velocity:} velocity ranges 20.5$\rightarrow$25.5~km~s$^{-1}$ (-16$\rightarrow$-11~km~s$^{-1}$), contours start at 3(3)~$\sigma$ and iterate by 2(2)~$\sigma$ with the 1-$\sigma$~level starting at 0.04(0.04)~Jy~beam$^{-1}$ for the red(blue) channels respectively. \textbf{High Velocity:} velocity ranges 25.5$\rightarrow$30.5~km~s$^{-1}$ (-21$\rightarrow$-16~km~s$^{-1}$), contours start at 3(3)~$\sigma$ and iterate by 2(2)~$\sigma$ with the 1-$\sigma$~level starting at 0.04(0.04)~Jy~beam$^{-1}$ for the red(blue) channels respectively. The \co\space synthesized beam (\cobeam) is the bottom-right most overlay on each of the panels and the continuum synthesized beam (\contbeam) is offset diagonally.}\label{fig:comomentmapirs3a}
\end{center}
\end{figure}

\begin{figure}[H]
   \begin{center}
\includegraphics[width=\textwidth]{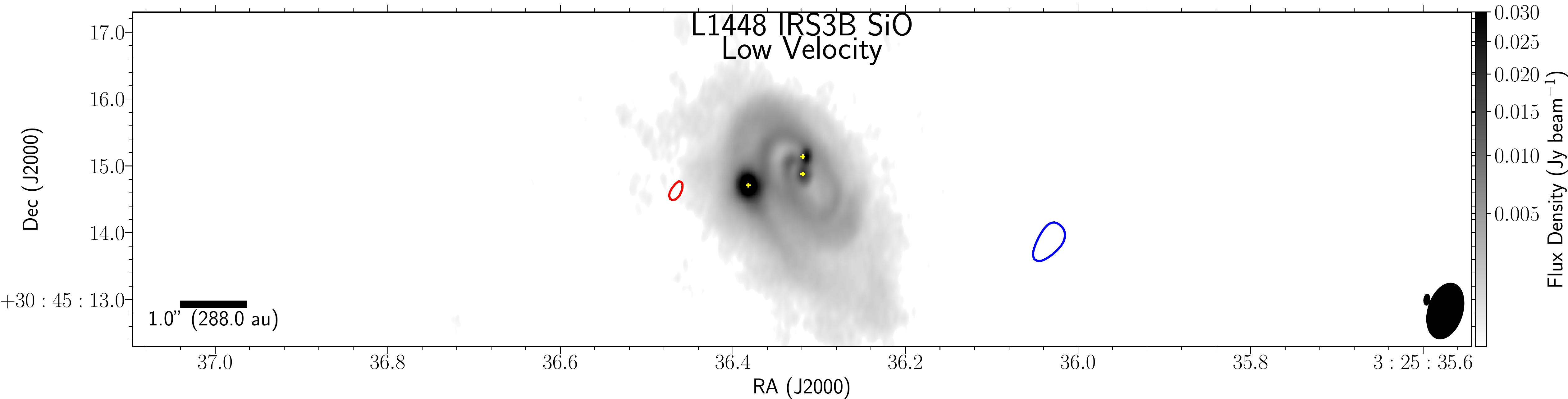}
\includegraphics[width=\textwidth]{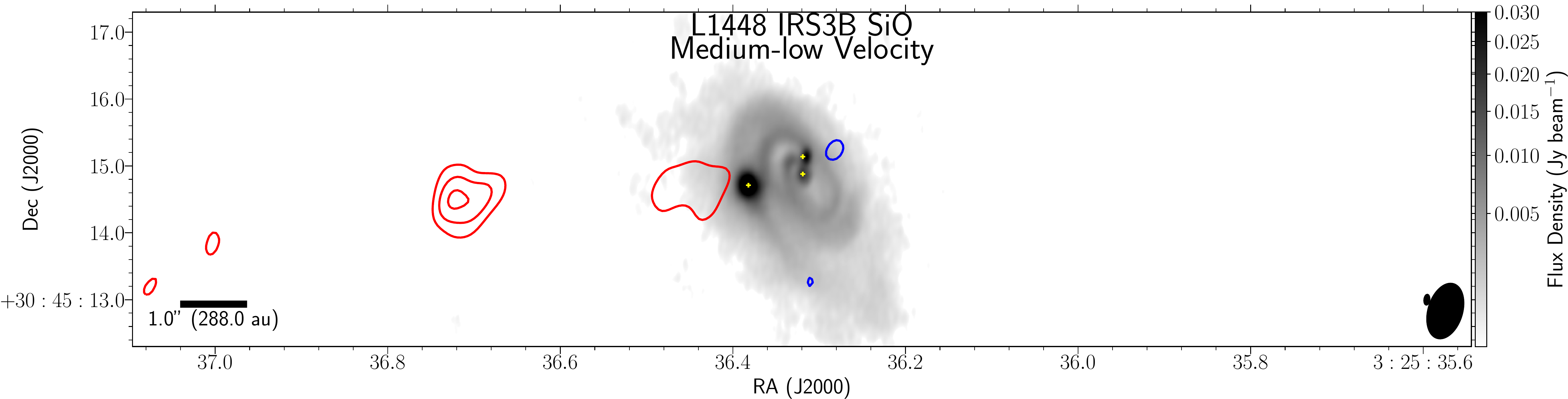}
\includegraphics[width=\textwidth]{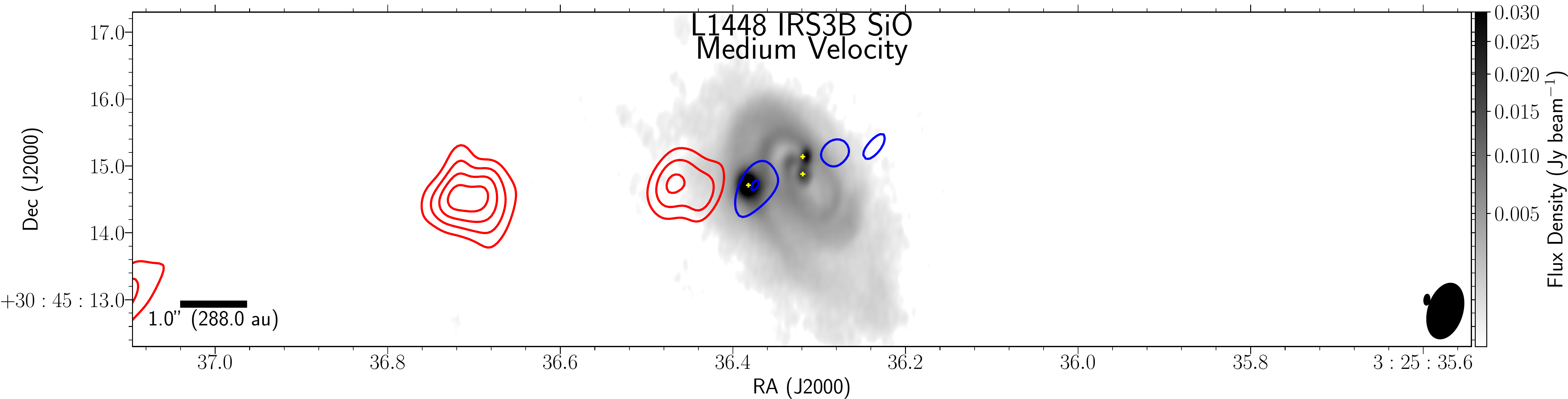}
   \end{center}
   \caption{Moment 0 map (integrated intensity) of \sio, overlaid on the continuum (grayscale) image from Figure~\ref{fig:contimage}. \sio\space shows locations of shocked gas fronts. \deleted{There is significant blue-shifted emission on the eastern side of the image, in the same location as the red-shifted outflow, which is coming from the L1448-C outflow, located \ab3\arcmin south of L1448 IRS3B.} The panels correspond to low, medium-low, and medium velocity ranges which are delineated as red(blue), respectively. \textbf{Low Velocity:} velocity ranges 5.5$\rightarrow$10.5~km~s$^{-1}$ (4$\rightarrow$-1~km~s$^{-1}$), contours start at 5(5)~$\sigma$ and iterate by 3(3)~$\sigma$ with the 1-$\sigma$~level starting at 0.11(0.09)~Jy~beam$^{-1}$ for the red(blue) channels respectively. \textbf{Medium-low Velocity:} velocity ranges 10.5$\rightarrow$15.5~km~s$^{-1}$ (-6$\rightarrow$-4~km~s$^{-1}$), contours start at 5(5)~$\sigma$ and iterate by 3(3)~$\sigma$ with the 1-$\sigma$~level starting at 0.01(0.01)~Jy~beam$^{-1}$ for the red(blue) channels respectively. \textbf{Medium Velocity:} velocity ranges 15.5$\rightarrow$20.5~km~s$^{-1}$ (-11$\rightarrow$-6~km~s$^{-1}$), contours start at 5(5)~$\sigma$ and iterate by 3(3)~$\sigma$ with the 1-$\sigma$~level starting at 0.009(0.012)~Jy~beam$^{-1}$ for the red(blue) channels respectively. The \sio\space synthesized beam (\siobeam) is the bottom-right most overlay on each of the panels and the continuum synthesized beam (\contbeam) is offset diagonally.}\label{fig:siomomentmap}
\end{figure}

\begin{figure}[H]
   \begin{center}
\includegraphics[width=\textwidth]{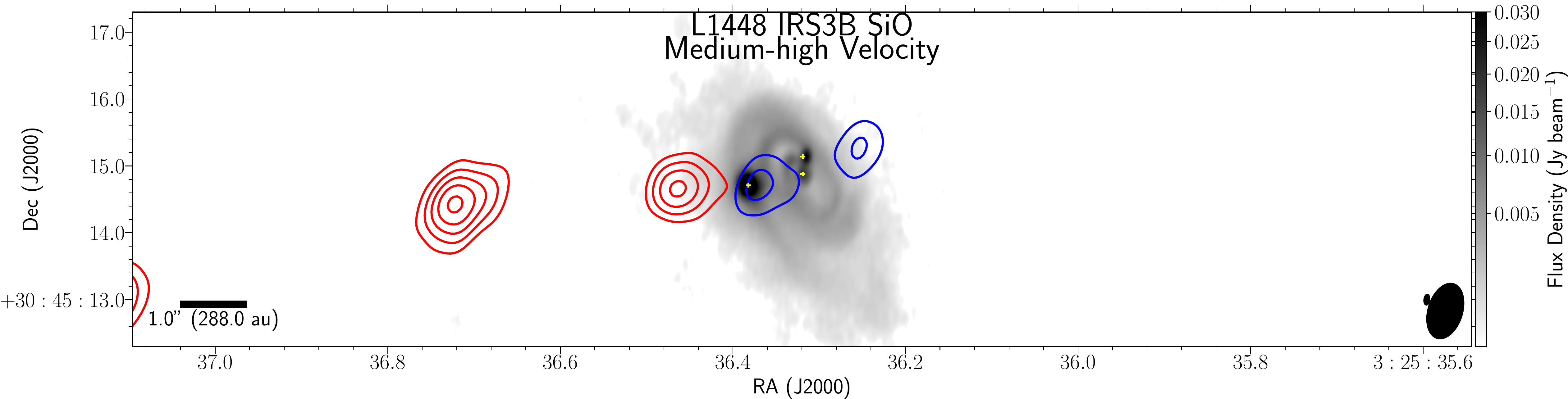}
\includegraphics[width=\textwidth]{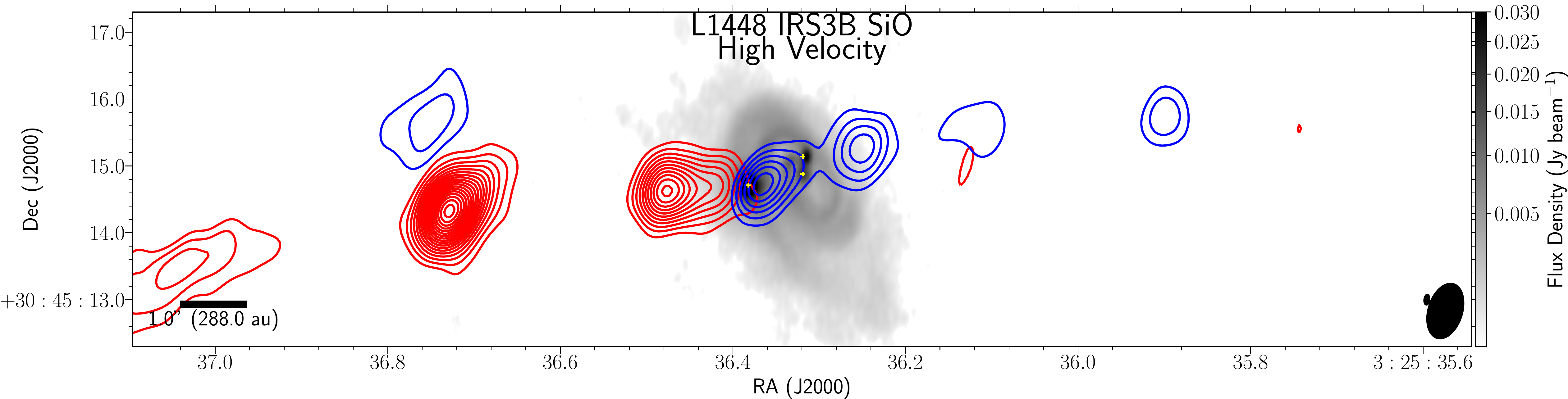}
\includegraphics[width=\textwidth]{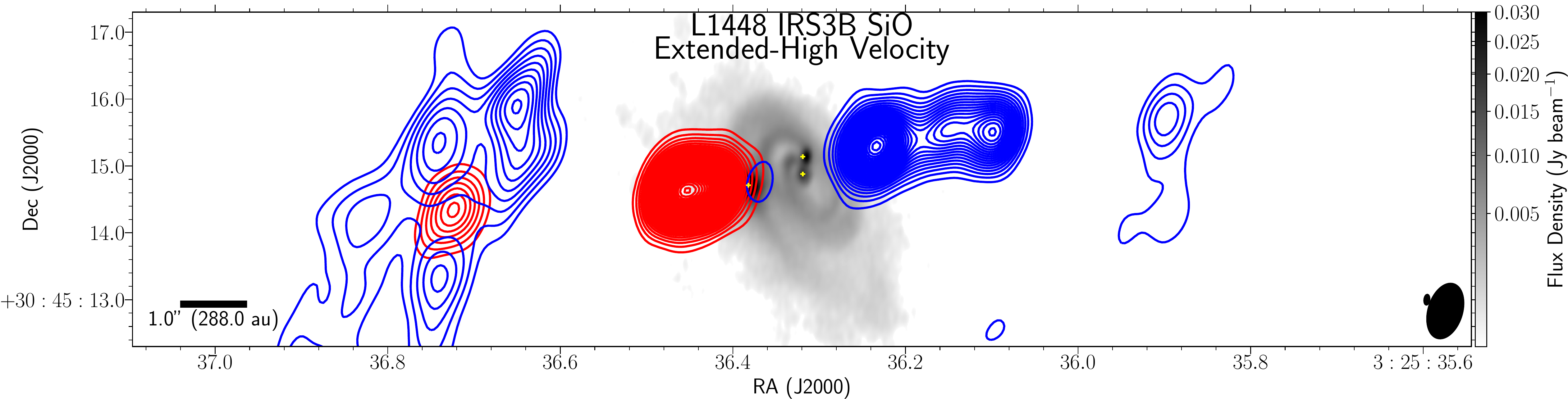}
   \end{center} 
   \caption{Similar to Figure~\ref{fig:siomomentmap} but for the \textbf{Medium-high Velocity:} velocity ranges 20.5$\rightarrow$25.5~km~s$^{-1}$ (-16$\rightarrow$-11~km~s$^{-1}$), contours start at 5(5)~$\sigma$ and iterate by 3(3)~$\sigma$ with the 1-$\sigma$~level starting at 0.012(0.015)~Jy~beam$^{-1}$ for the red(blue) channels respectively. \textbf{High Velocity:} velocity ranges 25.5$\rightarrow$30.5~km~s$^{-1}$ (-21$\rightarrow$-16~km~s$^{-1}$), contours start at 5(5)~$\sigma$ and iterate by 3(3)~$\sigma$ with the 1-$\sigma$~level starting at 0.008(0.015)~Jy~beam$^{-1}$ for the red(blue) channels respectively. \textbf{Extended-High Velocity:} velocity ranges 30.5$\rightarrow$50~km~s$^{-1}$ (-40$\rightarrow$-21~km~s$^{-1}$), contours start at 5(5)~$\sigma$ and iterate by 3(3)~$\sigma$ with the 1-$\sigma$~level starting at 0.025(0.025)~Jy~beam$^{-1}$ for the red(blue) channels respectively. \added{ There is significant blue-shifted emission on the eastern side of the image, in the same location as the red-shifted outflow, which is coming from the L1448-C outflow, located \ab3\arcmin south of L1448 IRS3B. }The \sio\space has additional emission well beyond the velocity range of the emission in \co\space and is presented as an additional panel (``extended-high velocity'') which only features the red-shifted emission. The \sio\space synthesized beam (\siobeam) is the bottom-right most overlay on each of the panels and the continuum synthesized beam (\contbeam) is offset diagonally. }\label{fig:siomomentmap2}
\end{figure}

\appendix

\section{Observations}\label{sec:appobs}

The \added{ALMA} correlator was configured to observe \lco, \lcso, \lhtcop, \lhtcn, \lsio, and a broad 2~GHz continuum band centered at 335.5~GHz (894~\micron). A summary of the correlator setup is provided in Table~\ref{table:obssummary2}. The raw data were reduced using the Cycle 4 ALMA pipeline within the \textit{Common Astronomy Software Application} (CASA) \citep{2007ASPC..376..127M} version 4.7.0. All further processing was done using the CASA version 4.7.2 and the reduction sequence is described here. To maximize the sensitivity of the continuum observations, emission free channels from the higher resolution windows were added to the continuum after appropriate flagging of line emission. The total bandwidth recovered from this method was \ab1.2~GHz, which, in conjunction with the continuum spectral window bandwidth of 1.875~GHz, yields \ab3~GHz of aggregate continuum bandwidth with an average frequency center of 341.0~GHz.
\added{
Given the high signal-to-noise (S/N) of the sources, we performed self-calibration (summary to reproduce in Table~\ref{table:selfcal}) on the separate configurations (C40-3 and C40-6) to further increase the S/N by correcting short timescale phase and amplitude fluctuations. During the phase-only self-calibration, the solution intervals for each additional iteration were: ``inf'' (The entire scan length, dictated by the time on a single pointing), 30.25~seconds (5 integrations), and 12.1~seconds (2 integrations). During the amplitude self-calibration, the solution interval of ``inf'' was used. The final self-calibrated measurement sets from the two configurations were concatenated using the CASA task \textit{concat}. 
}
The resulting images were generated from this concatenated dataset using \textit{Briggs weighting} with a \textit{robust} parameter of 0.5~(Figure~\ref{fig:contimage}). The beamsize of the combined continuum image is \contbeam\space (32$\times$15~au). We achieved 69~$\mu$Jy~beam$^{-1}$\space sensitivity for the aggregate continuum data and the full list of frequencies, bandwidths, beamsizes, sensitivities, and tapering for the suite of molecules is provided in Table~\ref{table:obssummary2}.

During the high-resolution execution, the central frequency of the SiO spectral window was set to 347.01~GHz with a bandwidth of 469~MHz, falling outside of the emission range for the target molecule. However, for the C40-3 configuration, the spectral setup was corrected and SiO was observed.

\section{Optimal Disk tracing Molecular lines}
To infer properties about the central potential from the circumstellar disk characteristics, we must disentangle the envelope and disk kinematics from the molecular line emission. Previous observations conducted by \citet{2016Natur.538..483T}\space of IRS3B included molecular lines \ceo\space and \tco. However, while emission from \ceo\space spatially coincides with the disk and is optically thin, it can have resolved-out emission towards the molecular line center, under-representing the underlying gas structure and reducing the fidelity of the tracer \citep{2020MNRAS.493L.108B}. Furthermore, \tco\space is a poor kinematic tracer for embedded Class 0 disks because it is a more abundant molecule that will have a high optical depth (and subsequently a larger degree of spatial filtering which limits the velocity range it is sensitive to) and confusion with the outflow. This tracer is better suited towards more evolved Class I sources (possibly IRS3A). Combining all previous observations of the sources, we find \cso\space is possibly the best tracer for Class 0 disks, being the least abundant molecule and thus experiencing the least amount of spatial filtering, both of which allow for accurate emission reconstruction near the line center. However, due to the low abundance, this molecule requires substantial integration time and is not suited for Class I disks.

\section{Application of Radiative Transfer Models}\label{sec:apppdspy}
We generate a set of \textit{priors} for the protostellar parameters based on the observational constraints. These \textit{priors} are then sampled via a uniform distribution and fed into \textit{emcee} to generate the samples, each sample describing a unique set of model parameters. These parameters are used to generate synthetic channel maps for the lines of interest, computed with RADMC-3D. These synthetic data cubes are Fourier transformed to recover a synthetic visibility dataset. These are re-gridded and subsequently cross-compared with the observed data in the uv-plane. The likelihood of the parameters for this comparison is then updated internally, the MCMC either probabilistically accepts the sample and migrates to this new point, or does not accept it by comparing the new likelihood to the previous sample. The whole process is repeated until convergence.

\deleted{The Affine Invariant MCMC algorithm (\textit{emcee}) utilizes Bayesian statistics at its core which provides a way to marginalize over nuisance parameters (e.g., variance of the priors), map out the posterior for the model, and provide inferences on the parameters of interest. MCMC itself, provides a way to sample a large, degenerate sample of parameter space and move towards regions of higher likelihoods and samples the posterior distributions.}

We assume the kinematic rotation of the disk is described by a Keplerian orbit, with an azimuthal velocity (in cylindrical coordinates) of $V(R) = \sqrt{GM_{\*}/R}$. We assume the molecular line emission comes from a flared disk geometry as motivated by viscous and irradiated disk evolution, where the mass density profile is described, in cylindrical coordinates with the origin at the gravitational source, by the equation:
\begin{equation}
\rho\left(R, z\right) = \frac{\Sigma(R)}{\sqrt{2\pi}h(R)}exp\left(-0.5\left(\frac{z}{h(R)}\right)^2\right)
\end{equation}
where R is the distance in the radial direction in cylindrical coordinates, $\Sigma$\space is the surface mass density of each molecule species,  and \textit{h} is the disk scale height. We assume the disk can be described via a power-law surface mass density profile that is truncated at some outer radius, of the form:
\begin{equation}
\Sigma\left(R\right) = \Sigma_{0}\times{R}^{-\gamma}.
\end{equation}
We also define
\begin{equation}
\Sigma_{0} = \frac{(2-\gamma)M_{disk}}{2\pi\left(R_{out}^{2-\gamma} - R_{in}^{2-\gamma}\right)}
\end{equation}
where R$_{out}$\space is the outer cutoff radius, R$_{in}$\space is the inner cutoff radius, and $\gamma$\space is the surface density power law exponent.

Another assumption we make is that the vertical structure of the disk is set by Local Hydro-static Equilibrium (LHSE) with a vertically isothermal temperature profile and a radial power-law temperature profile of the form:
\begin{equation}
T\left(R\right) = T_{0}\left(\frac{R}{1~au}\right)^{-q}
\end{equation}
which then sets the scale height of the disk, under the balance of thermal pressure and gravity, to be 
\begin{equation}
h\left(R\right) = \left(\frac{k_{b}R^{3}T(R)}{GM_{\*}\mu m_{H}}\right)^{1/2}
\end{equation}
where $k_{b}$ is the Boltzmann constant, G is the gravitational constant, $m_{H}$ is the mass of hydrogen, and $\mu$\space is the mean molecular weight \citep[assuming classic protostellar mean molecular weight, $\mu\approx2.37$; ][]{2003ApJ...591.1220L}. Additionally, chemical variations such as gas freeze-out onto dust grains towards the midplane and outer disk are excluded from the models.

Combining the aforementioned parameters that describe the disk structure plus the inclusion of disk geometric orientations, we have the following free parameters:  position angle (p.a.), inclination (inc.), temperature (T$_0$), stellar mass (M$_{*}$), disk radius (R$_D$), disk mass (M$_{disk}$), surface density power law ($\gamma$), system source velocity (V$_{sys}$), and uniform microturbulent line broadening ($\alpha$) (Table~\ref{table:pdspykinematic}). Furthermore, we have a number of fixed parameters that are used throughout the models but are not fit: molecular gas-to-H$_{2}$ abundance ratio (for IRS3B \cso\space=5.88$\times10^{-8}$; for IRS3A \htcn\space=2.04$\times10^{-7}$), inner disk cutoff radius (R$_{in}$ = 0.1~au), and the temperature power law index (q = 0.35).

\added{The combined fitting is computationally expensive, requiring on order 10$^{4}$~core-hours to reach convergence. A bulk of the computation time (up to 10 minutes per individual model) is used when RADMC-3D attempts to ray-trace massive disks.}

\section{Outflows}\label{sec:outflow}
\subsection{\co\space Line Emission}\label{sec:coemission}
The second most abundant molecule to H$_{2}$, $^{12}$CO, is shown as moment 0 maps in Figures~\ref{fig:comomentmap}, \ref{fig:comomentmap2}, and \ref{fig:comomentmapirs3a}. The \co\space integrated intensity maps towards IRS3B (Figures~\ref{fig:comomentmap}~and~\ref{fig:comomentmap2}) show clear signs of a collimated outflow originating from a region near IRS3B-ab and IRS3B-c that extends to \ab20\arcsec. Outflows are thought to be a signature of stellar birth with the highest velocity outflows ($>$20~km~s$^{-1}$) and high collimation are frequently found toward Class 0 protostars \citep[][]{1993ApJ...406..122A}. We observe asymmetric emission of the \co\space outflows with excess red-shifted emission dominating the data cube. The low velocity outflows appear to originate from IRS3B-ab while the high velocity jets appear to originate from both IRS3B-ab and -c. The outflows from IRS3B-ab and IRS3B-c are highly entangled at the lower velocity emission ($<10$~km~s$^{-1}$) but become more easily separated at higher velocity emission ($>20$~km~s$^{-1}$). The outflows of IRS3B-ab and IRS3B-c appear aligned within the wide opening angle (\ab45\deg) of the IRS3B-ab emission. However, both of these sources are marginally misaligned from the IRS3B-ab continuum disk minor axis ($<10$\deg). In the blue-shifted emission, there appears a faint but very wide opening angle (\ab65\deg) for the outflows which is resolved out in these observations but more clear in \citet{2016Natur.538..483T}. Additionally, there is a crescent shaped over-density along the blue-shifted emission, which could be due to orbital movement of the tertiary and/or precessions of the outflows. In the red-shifted emission there are 3 main over densities that occur along the line of the outflow, possibly indicative of irregular, high accretion events in the past. \co\space integrated intensity maps towards IRS3A (Figure~\ref{fig:comomentmapirs3a}) show low velocity, wide angle outflows towards line center, unlike the collimated outflows towards IRS3B.

\subsection{\sio\space Line Emission}\label{sec:sioemission}
The \sio\space emission (Figures~\ref{fig:siomomentmap}\space~and~ \ref{fig:siomomentmap2}) corresponds to shocks along the outflow. \sio\space most probably forms via dust grain sputtering which can inject either silicon atoms or \sio\space molecules into the gas \citep[][]{1997AA...322..296C}. This happens from neutral particle impacts on charged grains in addition to grain-grain collisions at sufficient velocities \citep[25-35~km~$s^{-1}$;][]{1997AA...322..296C}. Furthermore, we observe a relatively high asymmetry in the emission intensity between the red- and blue-shifted, while the radial extent (distance from launch location) is more symmetric about the outflow launch origin. Unlike the \co\space emission, the outflow launch location from \sio\space seems to coincide with IRS3B-c for both the high and low velocity emission rather than IRS3B-ab. However, the lower resolution leaves some ambiguity as to the true launch location.

\section{Molecular Line Spectra}\label{sec:spectra}
In order to visualize the structure and dynamics in 3D datacubes,  we construct moment 0 maps and PV diagrams to reduce the number of axis by either integrating along the frequency axis or along slices across the minor axis, respectively. We can also construct spectra, centered on the sources, and integrated radially outwards in annuli.

\begin{figure}[H]
  \begin{center}
   \includegraphics[width=\textwidth]{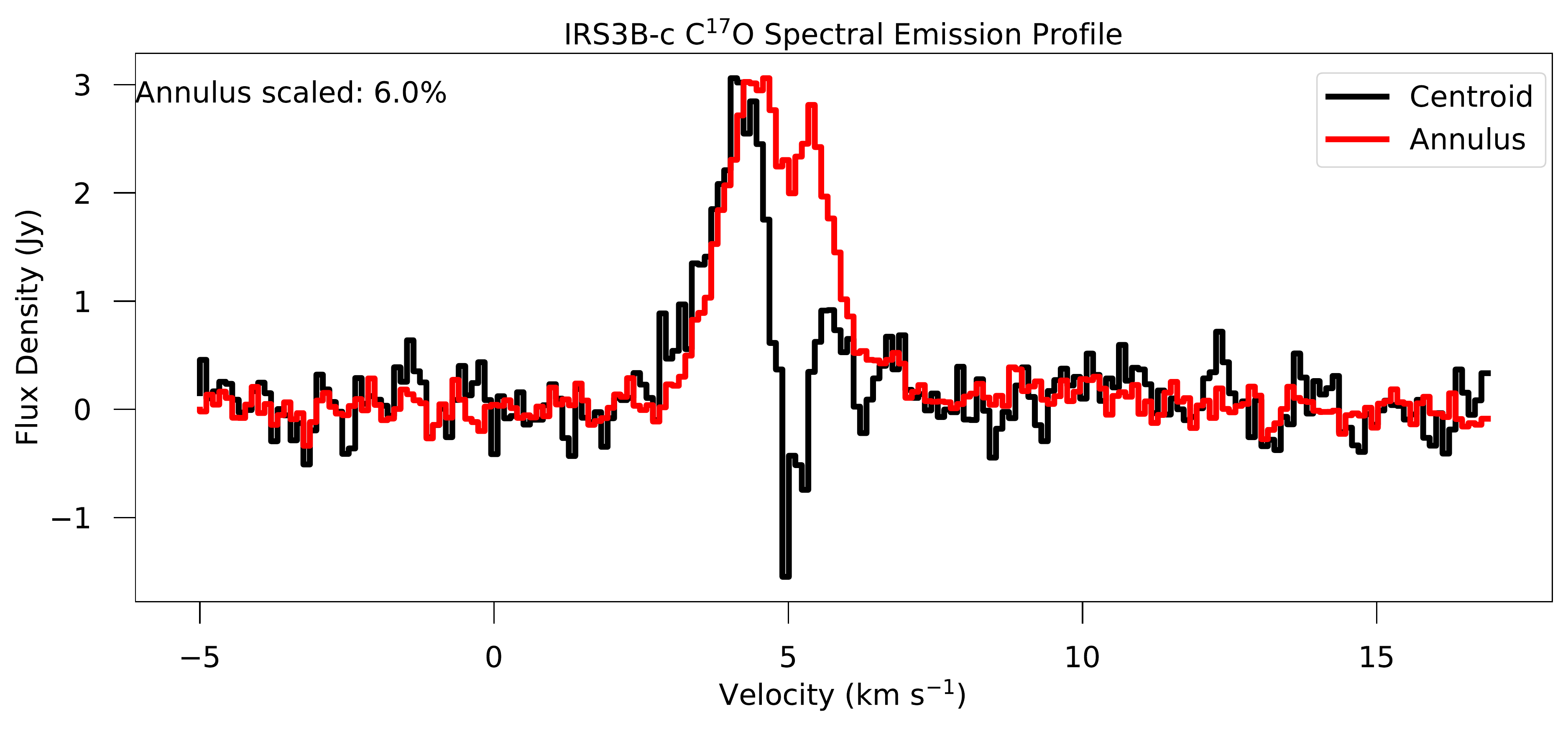}
   \end{center}
   \caption{\cso\space integrated spectral emission profile of IRS3B-c, set to the rest frequency of \cso. The profiles were extracted by integrating the emission within an annulus, where the co-center of the annuli is set to the center point of IRS3B-c, while the inclination and position angle of the annuli is set to the IRS3B-ab parameters. The ``black'' profile is extracted from a central ellipse 2 times the size of the restoring beam, while the ``red'' profile is extracted from an annulus with the same width as the average restoring beam, three beam widths off of the source. The central emission features a deficit of emission towards line center. The profiles are normalized to highlight the emission profiles rather than the actual values of the emission.}\label{fig:irs3bspec}
\end{figure}

\begin{figure}[H]
  \begin{center}
   \includegraphics[width=\textwidth]{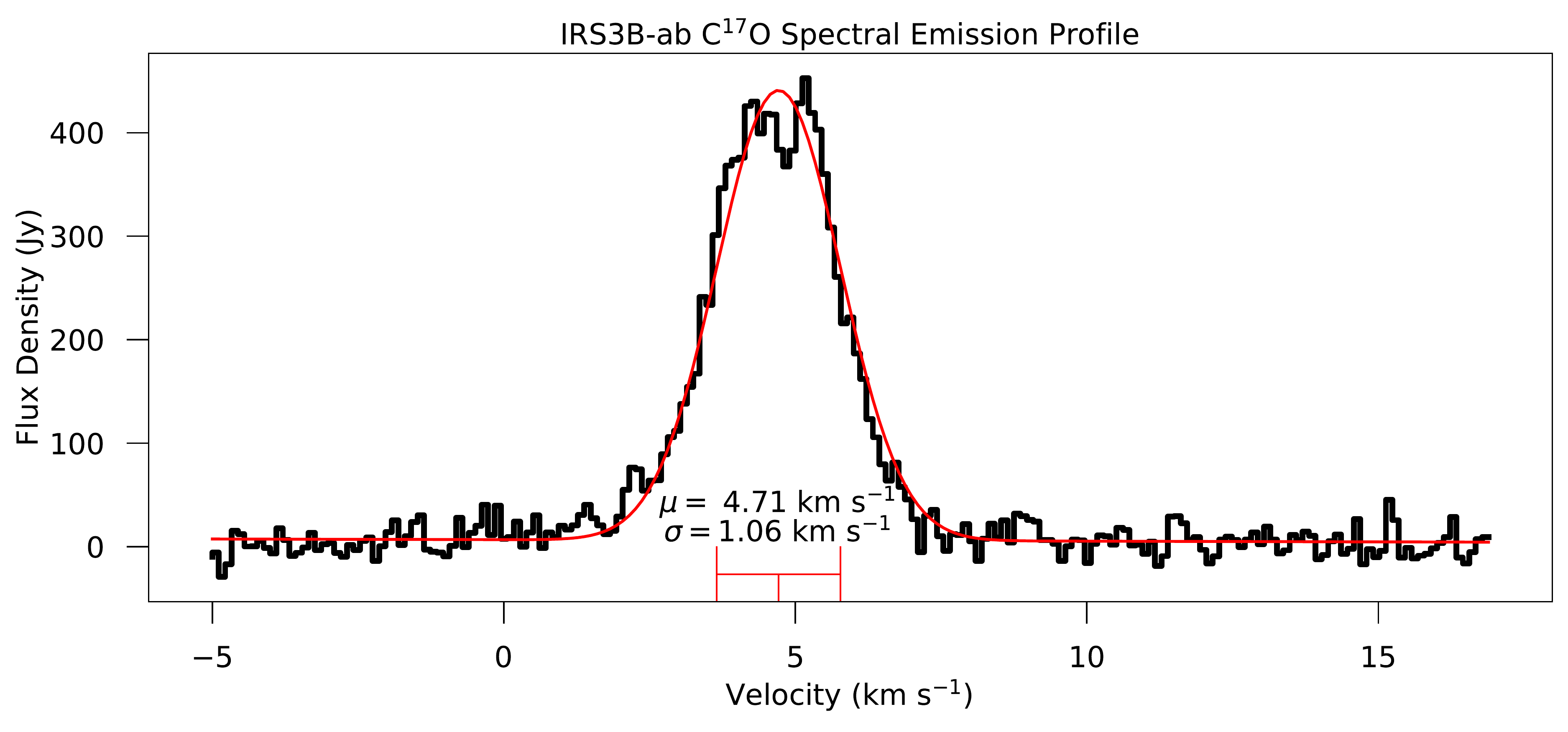}
   \end{center}
   \caption{\cso\space integrated spectral emission profile of IRS3B-ab, set to the rest frequency of \cso. The profile is extracted by integrating the emission within an ellipse, where the center, inclination, and position angle are set to the center point of IRS3B-ab. The ``black'' profile is extracted from a central ellipse the same size as the gaseous disk in Table~\ref{table:obssummary3}. The red line is a Gaussian fit to the spectra, with parameters $\mu=$4.71$^{+0.02}_{-0.02}$~\kms\space and $\sigma=$1.06$^{+0.02}_{-0.02}$~\kms.}\label{fig:irs3babspec}
\end{figure}

\begin{figure}[H]
  \begin{center}
   \includegraphics[width=\textwidth]{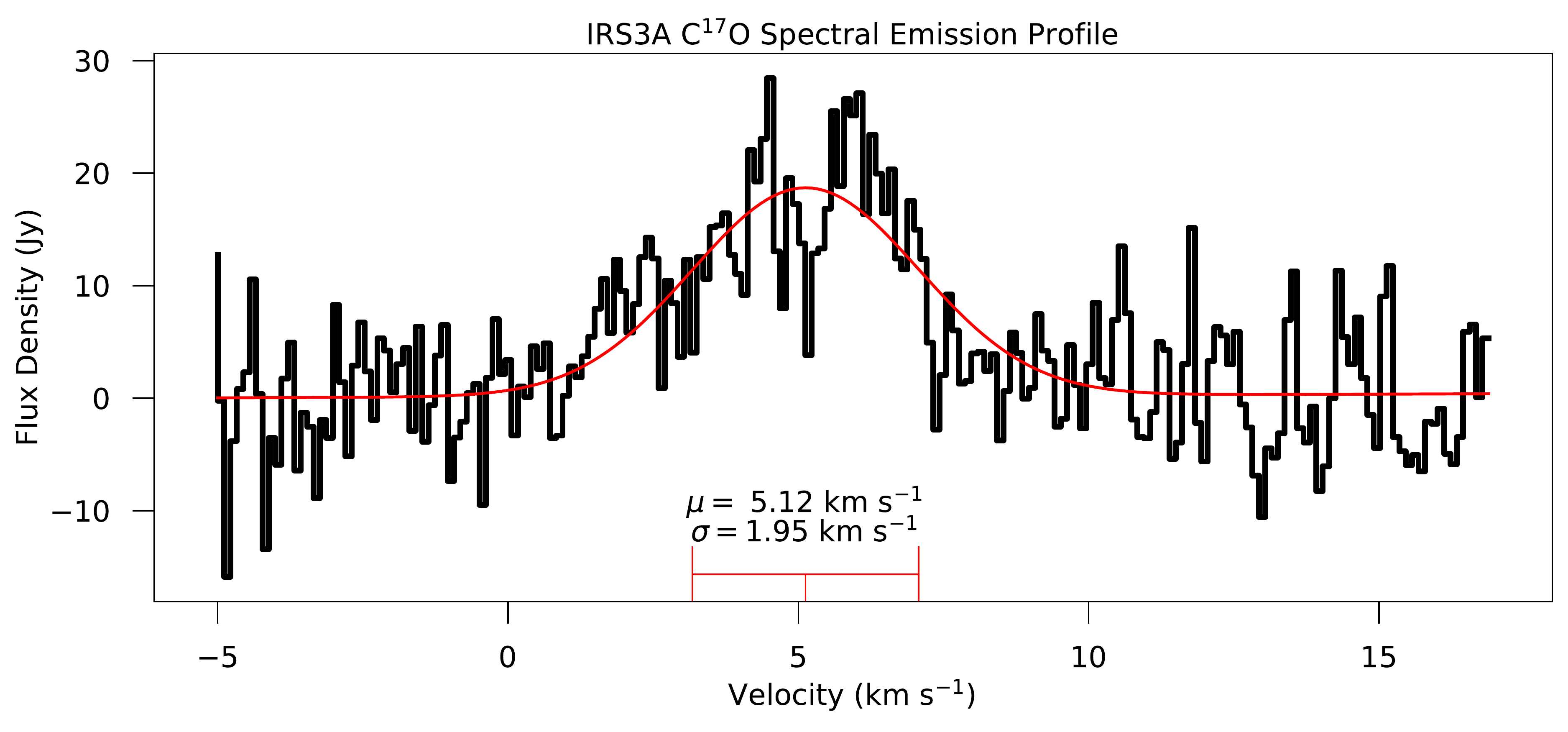}
   \end{center}
   \caption{\cso\space integrated spectral emission profile of IRS3A, set to the rest frequency of \cso. The profiles were extracted by integrating the emission within an ellipse, where the center, inclination, and position angle are set to the center point of IRS3A. The ``black'' profile is extracted from a central ellipse the same size as the gaseous disk in Table~\ref{table:obssummary3}. The \cso\space emission towards this source is fainter than the emission from other dense gas tracers, thought to trace disk kinematics like that of \htcn.  The red line is a Gaussian fit to the spectra, with parameters $\mu=$5.12$^{+0.15}_{-0.15}$~\kms\space and $\sigma=$1.95$^{+0.75}_{-0.17}$~\kms.}\label{fig:irs3aspec}
\end{figure}

Figure~\ref{fig:irs3bspec}\space is the \cso\space spectra for the IRS3B-c system. We extract the emission within an ellipse centered on IRS3B-c to define the main core of the IRS3B-c spectra in ``red'' and an annulus just outside of this ellipse to define the comparative IRS3B-ab disk spectra in ``red''. The IRS3B-c spectra features a deficit of emission towards line center due to the high optical depths towards this clump. Figure~\ref{fig:irs3babspec}\space is the \cso\space spectra for the IRS3B system. This spectra is centered on the kinematic center of the disk (Table~\ref{table:pvtable}) and is integrated out to the size of the gaseous disk (Table~\ref{table:obssummary3}). Figure~\ref{fig:irs3aspec}\space is the \cso\space emission towards IRS3A which is faint in these observations, making it not a suitable molecule for tracing disk kinematics.


\section{Tertiary Subtraction and Gaussian Fitting}\label{sec:tertsub}
\begin{figure}[H]
  \begin{center}
   \includegraphics[width=0.48\textwidth]{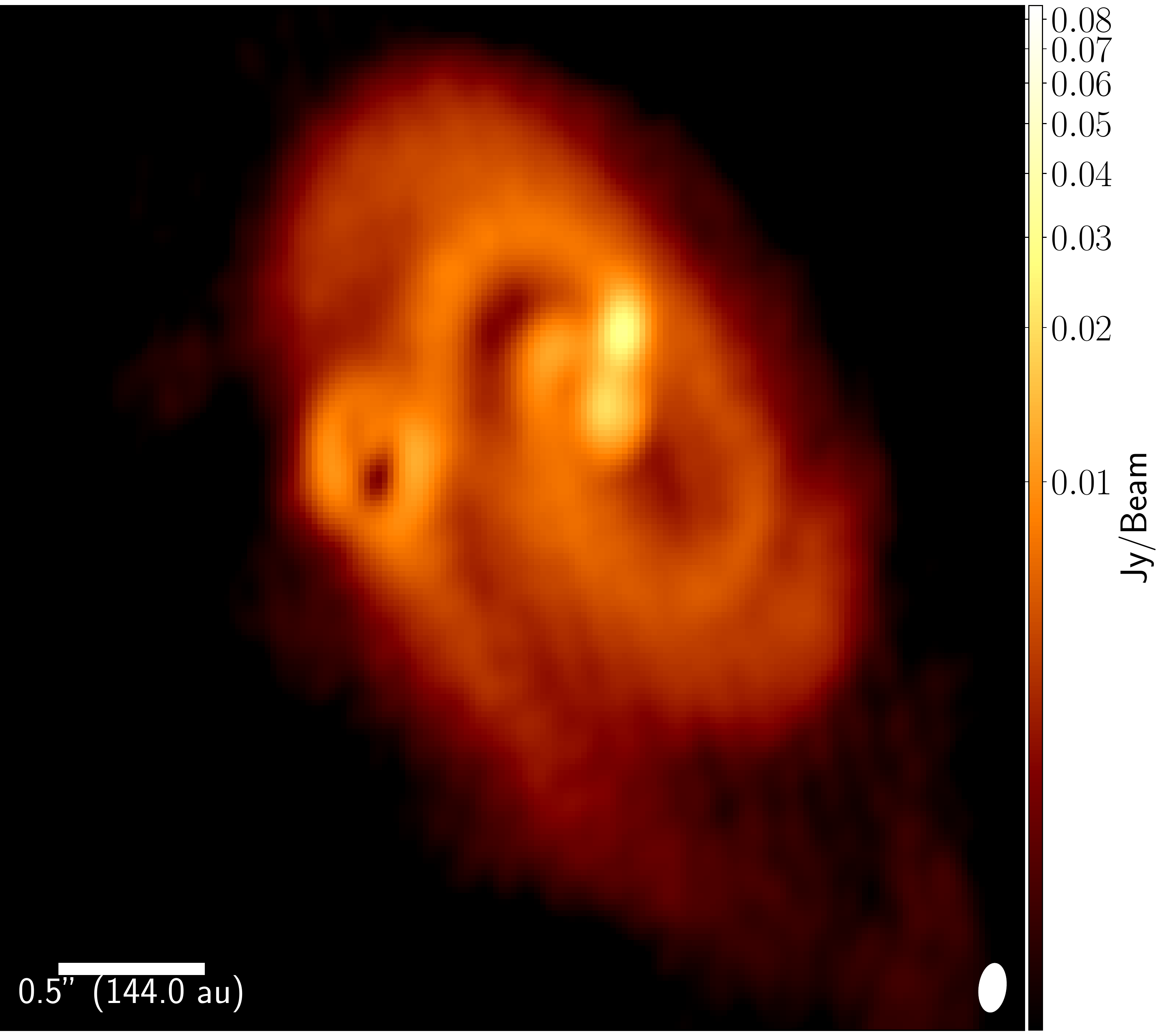}
   \includegraphics[width=0.48\textwidth]{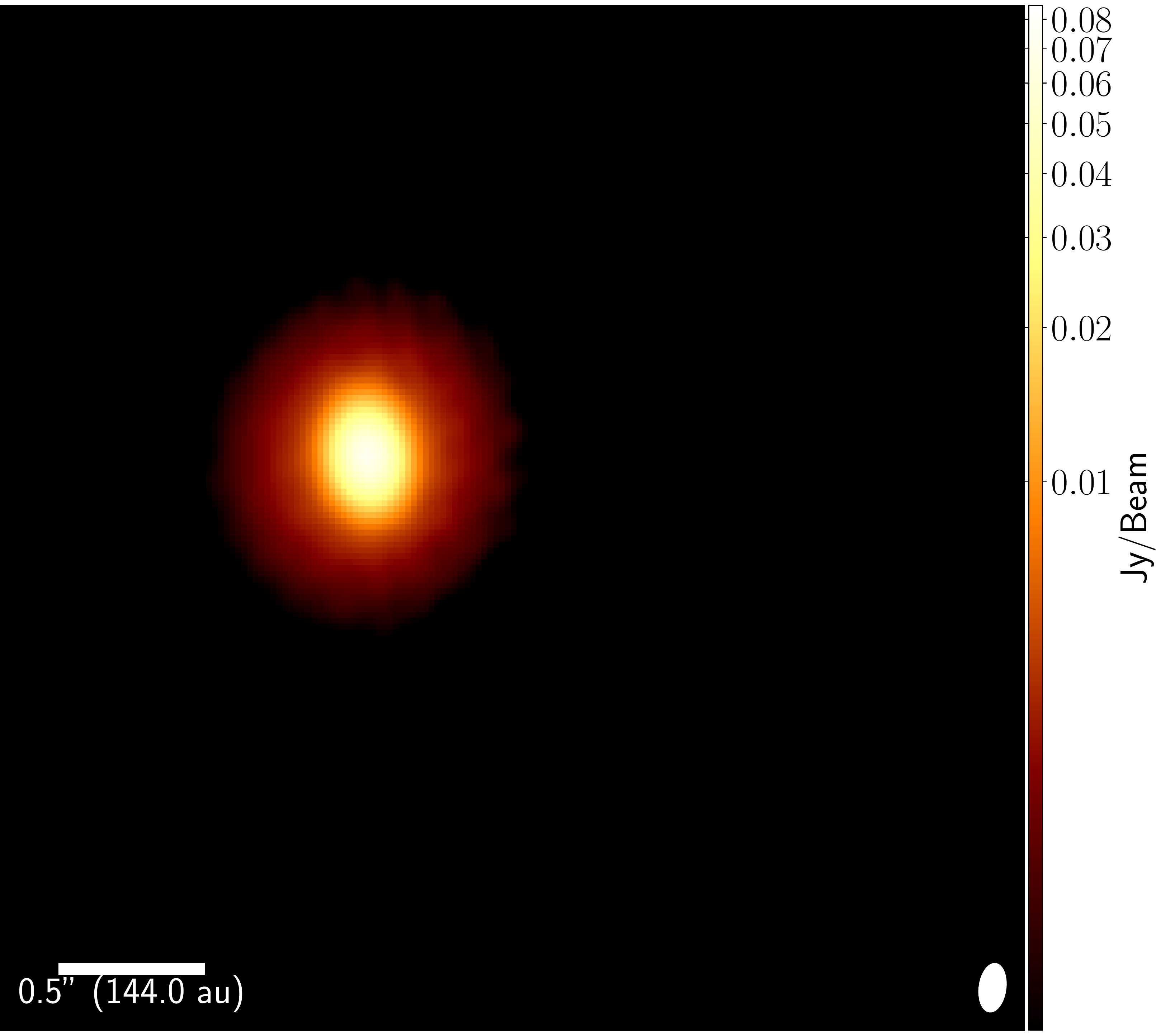}
   \end{center}
   \caption{Continuum (879~\micron) images of IRS3B with the tertiary clump removed (left image) for analysis and the model of the tertiary clump (right image). The tertiary model was constructed using two 2D Gaussians with a zero-level offset in order to properly restore the underlying disk emission without introducing additional features. The left image was used to exclude the embedded tertiary mass from the dust component of the circumstellar disk while the right image image was then used for analysis of the compact dust emission around the tertiary.} \label{fig:subclump}
\end{figure}

\begin{figure}[H]
  \begin{center}
   \includegraphics[width=0.48\textwidth]{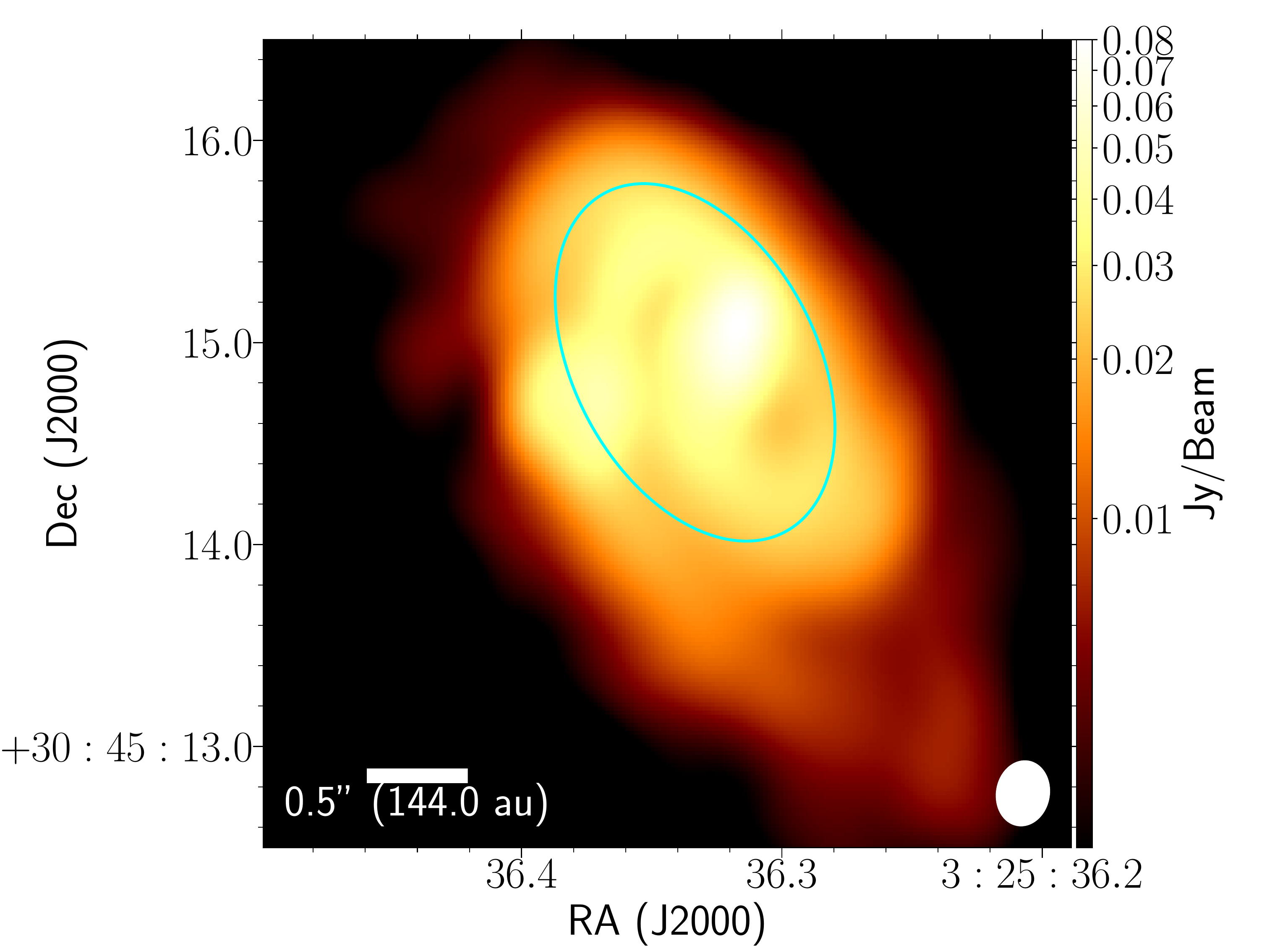}
   \end{center}
   \caption{Continuum (879~\micron) image of IRS3B with the tertiary removed, reconstructed with Briggs weighing robust parameter of 2 and tapered to 500k$\lambda$. This smooths over the substructure of the continuum disk to enable fitting of the disk with a single 2-D Gaussian profile, without over-fitting the substructure. The cyan line is the Gaussian fit of the circum-multiple disk of IRS3B, with the major and minor axis of the ellipses defined by the FWHM major and minor axis of the 2-D Gaussian fit.} \label{fig:subclumptaper}
\end{figure}

The continuum emission of the bright, embedded source, IRS3B-c, biases the analysis of the radial disk structure and circumstellar disk mass estimate of the IRS3B system. By removing this source, we can independently examine the disk and the tertiary source in order to characterize their properties separately~(Figure~\ref{fig:subclump}). In order to remove the tertiary source, we fit two Gaussians with a zero-level offset to the position of the source using the \textit{imfit} task in CASA (a point source and single Gaussian did not provide adequate fit while preserving the underlying disk emission). The offset serves to preserve the emission from the underlying IRS3B-ab disk emission. We also restricted the imfit task to a 0\farcs8$\times$0\farcs7\space ellipse around the source such that the fit does not extend into the surrounding emission from the spiral arms. With these parameters generated from the imfit task, we then constructed a model image of the tertiary. We used the CASA task \textit{setjy} to Fourier transform the model image and fill the model column of the measurement set with the model visibility data. We then use the task \textit{uvsub} to subtract this model from the data, producing the residual visibilities without the tertiary. A tertiary subtracted image is generated from this residual dataset and shown in Figure~\ref{fig:subclump} along with the model of the tertiary used to construct this dataset. The masses generated from this fit is \ab0.07~\solm, as described in Section~\ref{sec:dcont} and provided in Table~\ref{table:pvtable}. We then are able to reconstruct and taper the resulting visibilities to smooth over the substructure of the disk, in order to better fit the circum-multiple disk. The image (Figure~\ref{fig:subclumptaper}) is fit with a 2-D Gaussian using the \textit{imfit} in CASA and the results of the fit are provided in Table~\ref{table:obssummary3}.

\end{document}